\documentclass{aa}  

\usepackage{nicefrac}
\usepackage{thesis_defs}
\usepackage{blindtext}
\usepackage{subcaption}
\usepackage{tikz}
\usepackage{graphics}
\usepackage[utf8]{inputenc}
\usepackage{graphicx}
\usepackage[export]{adjustbox}
\usepackage{appendix}
\usepackage{stfloats}
\usepackage{cuted}
\usepackage{graphicx}
\usepackage{txfonts}
\usepackage{natbib}
\bibpunct{(}{)}{;}{a}{}{,}
\usepackage[final]{hyperref}
\usepackage{enumitem}
\begin{document} 

\title{Ray-Tracing in Relativistic Jet Simulations: A Polarimetric Study of Magnetic Field Morphology and Electron Scaling Relations}

   \author{Joana A. Kramer
          \inst{1,*}
          \and
          Nicholas R. MacDonald\inst{1}
          }

   \institute{\inst{1}Max-Planck-Institut für Radioastronomie, Auf dem Hügel 69, D- 
   				53121 Bonn, Germany\\
   			\inst{*}\email{jkramer@mpifr-bonn.mpg.de}
             }

   \date{}

 \abstract{The jets emanating from the centers of active galactic nuclei (AGN) are among the most energetic objects in the universe. Investigating how the morphology of the jet's synchrotron emission depends on the magnetic nature of the jet's relativistic plasma is fundamental to the comparison between numerical simulations and the observed polarization of relativistic jets.
 }{Through the use of 3D relativistic magnetohydrodynamic (RMHD) jet simulations (computed using the \emph{PLUTO} code) we study how the jet's synchrotron emission depends upon the morphology of the jet's magnetic field structure. Through the application of polarized radiative transfer and ray-tracing (via the \emph{RADMC-3D} code) we create synthetic radio maps of the jet's total intensity as well as the linearly and circularly polarized intensity for each jet simulation. }{ In particular, we create synthetic ray-traced images of the jet's polarized synchrotron emission when the jet carries a predominantly poloidal, helical, and toroidal magnetic field. We also explore several scaling relations in which the underlying electron power-law distribution is set proportional to: (i) the jet's thermal plasma density, (ii) the jet's internal energy density, and (iii) the jet's magnetic energy density.}{We find that: (i) the jet emission is edge brightened when the magnetic field is toroidal in nature and a spine brightened when the magnetic field is poloidal in nature, (ii) the circularly polarized emission exhibits both negative and positive sign for the toroidal magnetic field morphology at an inclination of $i=45^\circ$ as well as $i=5^\circ$, and (iii) the relativistic jet's emission is largely independent of different emission scaling relations when the ambient medium is excluded. }{} 

   
  \authorrunning{J.~A.~Kramer \& N.~R.~MacDonald}
  \titlerunning{Ray-tracing in RMHD Jet Simulations}
 \maketitle
%

\section{Introduction}
Collimated supersonic flows of plasma are a characteristic of many 
astrophysical objects. These phenomena are known as jets and emanate from 
compact systems (e.g., proto-stars) as well as from supermassive black holes 
(SMBHs). They are among the most energetic objects in the universe and commonly emanate from the centers of active galaxies. The class of radio loud active galactic nuclei (AGN) exhibit jet emission. They are mostly embedded in massive elliptical galaxies and only account for less than $10\,\%$ of observed AGN. Launched from a central engine such as a SMBH, the jets can be 
accelerated to highly relativistic speeds and remain collimated up to kilo-parsec (kpc) scales. 
AGN emit radiation across the electromagnetic spectrum, 
and observations of the jet emission reveal a featureless power-law spectrum.
Together with the high level of 
linear polarization (up to $60$-\SI{70}{\percent}), the physical process of 
synchrotron radiation can explain the emission as well as the optical 
flux~\citep{Troja}. From the radio emission, the presence of a magnetic field can be 
inferred and is commonly thought to play a key role in the launching and collimation process of the jet. AGN jets can extend to 
hundreds of kilo-parsecs even though the jet launching region occurs on 
scales of a few gravitational radii from the black hole (BH).
On larger scales, jets are thought to be kinetically dominated, and therefore, 
contain relatively weak magnetic fields. Within the collimation region, 
however, jets 
are thought to be magnetically dominated.
Theoretical studies of jet formation suggest that strong magnetic fields
are an essential mechanism for 
launching jets~\citep{BlandfordZnajek}. One of the main 
conclusions of a number of relativistic jet simulations 
is that the jet transitions from being magnetically dominated to 
being kinetically dominated as it propagates~\citep[e.g.,][]{Marti97}.

Very-long-baseline interferometric (VLBI) imaging of the synchrotron emission 
emanating from jets commonly reveal a bright central feature (referred to as the radio 
core) and a series of components which separate from the core over time 
(i.e., blobs/plasmoids). There are also (in some sources) features downstream 
of the 
core that, in contrast to the plasmoids, appear to be stationary relative to 
the radio core~\citep[e.g.,][]{Ojha2010, Fromm2013}. These standing features within 
the jet are 
commonly interpreted as recollimation shocks within the 
jet flow~\citep{DalyMarscher}.

A continuum approximation of the plasma nature of the jet (i.e., relativistic 
magnetohydrodynamics - RMHD) can be implemented based on the 
assumption that the jet radius, $R_j$, is much larger than the 
Debye-Length\footnote{The Debye-Length is the length scale in a plasma over 
	which the charge of a plasma is shielded by intervening electrons.} and 
gyroradius\footnote{The gyroradius (also referred to as the Larmor radius) is 
	the radius about which an electron rotates about a magnetic field line.} of 
the jet's 
plasma~\citep{Hawley2015}.
Theoretical models of AGN jets postulate that the jet plasma is likely 
magnetized with 
a large-scale helical morphology related to the launching of the 
jet by the rotation of the central black hole and accretion 
disk~\citep{BlandfordZnajek, BlandfordPayne, Hardee2007}. Recent observational 
evidence indicates that a large fraction of parsec-scale jets do
indeed exhibit polarization signatures of helical magnetic field components. 
This is based on the detection of 
statistically significant transverse Faraday rotation measure
(RM) gradients across the jet on parsec scales~\citep{Gabuzda2008}.

The presence of a helical/toroidal magnetic field within the jet can (in theory) produce current driven instabilities within the jet flow~\citep[][]{Kadowaki}. These instabilities can then produce sites of magnetic reconnection within the jet plasma which in turn can result in particle acceleration~\citep[][]{Singh2016, Striani2016}. Recent particle-in-cell (PIC) simulations~\citep[e.g.,][]{ Sironi2021} have indeed shown that magnetic reconnection events within relativistically jetted plasma can efficiently generate power-law distributions of electrons~\citep[see also,][]{Sironi2014, Guo2015, Werner2016, Guo2019, Matthews2020}. Many of these PIC calculations, however, lack sufficient grid sizes to model the length scales of astrophysical jets.

In this paper we set about carrying out a 
systematic study of how the fractional levels and morphology of both linearly 
and circularly 
polarized synchrotron emission depend on the underlying magnetic field morphology of the jet as well as various fluid scalings for the underlying electron power-law distribution. This 
study is executed with fully 3D relativistic magnetohydrodynamic jet 
simulations coupled 
with full Stokes polarized radiative transfer via ray-tracing.

RMHD simulations are unable to reproduce the kinetic scale physics 
of 
the jet (i.e., self-consistently generating the non-thermal distribution of electrons responsible for
the observed synchrotron emission). We therefore rely on a purely macroscopic 
model 
of the jet that simulates the large scale dynamics of the thermal
plasma within the jet flow. We explore various emission 
\emph{recipes} for 
mapping from the thermal fluid variables to the non-thermal distribution of 
electrons~\citep[see, e.g., ][]{Porth2011}. 
This mapping is carried out as a post-process step. In particular, we apply 
three scaling relations in which the non-thermal distribution of electrons
is assumed to be proportional to the plasma's: \emph{(i)} density, \emph{(ii)} 
thermal pressure, and 
\emph{(iii)} magnetic energy density.
We also examine the effect that different magnetic field morphologies within 
the jet (namely; poloidal, helical, and toroidal) have on the dynamics of the 
jet 
as well as the resultant polarized emission.

Our current jet simulations, while applicable to parsec scale jets, lack sufficient micro physics (such as magnetic reconnection) to self-consistently generate power-law distributions of electrons. Our emission calculations, however, provide: \emph{(i)} an important bridge between the micro physical scales of reconnecting current sheets and parsec scale jets, and \emph{(ii)} a valuable point of comparison for the next generation of synthetic synchrotron emission maps to be produced via hybrid fluid particle schemes~\citep[see, e.g.,][]{Vaidya2018}.

This paper is structured as follows: \sect{sec:num} gives an 
introduction to the principles of relativistic 
magnetohydrodynamics and polarized radiative transfer and introduces the \emph{PLUTO}~\citep{Mignone2007} and \emph{RADMC-3D}~\citep{RADMC3D} codes. We perform a full Stokes analysis with an emphasis on studying the jet's circularly polarized synchrotron emission. For this, the dependence 
of the jet's polarization on the magnetic field morphology, i.e., poloidal, 
helical, and toroidal, is investigated in \sect{3}. \sect{4} explores the effect that different thermal fluid 
to non-thermal electron emission scaling relations have on the resulting jet emission. \sect{variations} outlines different numerical approaches, i.e., when no jet tracer is included or the lower energy cutoff is computed differently. In~\sect{discuss} we compare our numerical results to recent observations of jets.
Finally, our conclusions are 
summarized in \sect{con}.

\section{Numerical Methods}\label{sec:num}
\subsection{Principles of the Relativistic Magnetohydrodynamics in the \emph{PLUTO} Code}\label{pluto}
To model magnetized fluid flows, the \emph{PLUTO} code integrates a system of conservation laws which can be expressed in general 
as:
\begin{equation}\label{cons}
\partial_t U^k + \sum_{i \in \{x,y,z\}} \partial_i T^{ik} = 0 ,
\end{equation}
where $U^k$ is a state vector of $k$ conservative quantities and
$T^{ik}$ is a rank 2 tensor. Moreover, $\partial_i$ is the four-gradient.
The explicit form depends on the physical module selected within the code.

\emph{PLUTO} solves a time-dependent non-linear system of special relativistic 
conservation laws, which in general have the form of \eq{cons}. To account for 
the motion of an ideal relativistic magnetized fluid, i.e., in relativistic 
magnetohydrodynamics, mass and energy-momentum are conserved.
The solution to the specified problem of relativistic magnetohydrodynamics and therefore the conservative variables and 
respective fluxes for RMHD are 
expressed as:

\begin{align}\label{sol}
U^{k} = 
\begin{pmatrix}
D\\ m_x\\ m_y\\ m_z\\ B_x\\ B_y\\ B_z\\ E 
\end{pmatrix}_k
=
\begin{pmatrix}
D\\
\VEC{m}\\
\VEC{B}\\
E
\end{pmatrix}_k,
\quad
T^{ik} = 
\begin{pmatrix}
D v_i \\
p \hat{e}^x_i  - \frac{b_x}{\gamma} B_i+m_x v_i \\
p \hat{e}^y_i - \frac{b_y}{\gamma} B_i+m_y v_i\\
p\hat{e}^z_i - \frac{b_z}{\gamma} B_i +m_z v_i \\
B_x v_i-v_x B_i \\
B_y v_i-v_y B_i \\
B_z v_i - v_z B_i \\
m_i
\end{pmatrix}_k
\end{align}
where $\VEC{v}=\left(v_x, v_y, v_z\right)^T$ is the fluid's velocity, $\VEC{\hat{e}}^j$ ($j\in \{x,y,z\}$) is the unit vector in the direction of the 
$i$th 
axis of a 3D Cartesian coordinate system, and $b_i$ are the spatial components of the covariant magnetic field vector.
The quantities in \eq{sol} are defined as, e.g., 
~\citet{Mignone2006}:
\begin{align}\label{231}
\begin{split}
D &= \rho\gamma, \\
\VEC{m} &= \left(\rho h \gamma^2 + \VEC{B}^2\right)\VEC{v} - 
\left(\VEC{v}\cdot\VEC{B}\right)\VEC{B},\\
E &= \rho h\gamma^2 - p_g +\frac{\VEC{B}^2}{2} + \frac{\VEC{v}^2\VEC{B}^2 - 
   	\left(\VEC{v}\cdot\VEC{B}\right)^2}{2}.
\end{split}
\end{align}
Hence, the components of $U^k$, resulting from the conservation laws, are 
the laboratory density $D$, the three 
components of both momentum $\VEC{m}$ and magnetic field $\VEC{B}$, 
and the 
total energy density $E$, respectively.

For a proper solution of \eq{cons} an additional equation 
of state (EoS) is specified, i.e., defining the specific enthalpy ~\citep[see][]{Mignone2005a}:
\begin{align}
h = \frac{5}{2}\Theta + \sqrt{\frac{9}{4}\Theta +1},
\end{align}
which approximates a single-special relativistic perfect gas.
Here, $\Theta$ is the 
ratio of pressure to density, i.e., $\Theta=\nicefrac{p}{\rho}$.

\subsection{Scaling}\label{scaling}
The \emph{PLUTO} simulations are computed in dimensionless grid 
units, therefore, the thermal quantities of the RMHD jet flow must be properly 
scaled into physical units as a post process step. Computing dimensionless 
quantities has the advantage of 
avoiding either extremely small or large numbers at run time. A physical 
scaling is necessary whenever specific scales of length, time, and 
energy are included into the problem. The physical scaling of 
our RMHD jet simulations requires the definition of 
three fundamental units\footnote{For further details see the \emph{PLUTO} Code (\url{http://plutocode.ph.unito.it})}:
\begin{align}
\begin{split}
&\text{unit density } \quad \rho_0 \left[\SI{}{\gram\,\centi\metre^{-3}}\right]\\
&\text{unit length    }\,\, \quad L_0 \left[\SI{}{\centi\meter}\right]\\
&\text{unit velocity} \quad v_0 \left[\SI{}{\centi\meter\per\second}\right].
\end{split}
\end{align}
From these unit values, other quantities such as the timescale of the code 
$t_0\left[\text{s}\right]$ can be computed: $t_0 = L_O/v_0$. Similarly, the 
pressure 
and magnetic field scale factors can be computed from these unit values: $p_0=\rho_0 v_0^2$ and $B_0 = 
\sqrt{4\pi\rho_0v_0^2}$. To then scale the dimensionless fluid variables into cgs units we apply the scale factors:
\begin{align}
\begin{split}
\rho_\text{cgs} &= \rho \cdot \rho_0,\\
p_\text{cgs} &= p \cdot \rho_0v_0^2,\\
\VEC{B}_\text{cgs} &= \VEC{B}\cdot \sqrt{4\pi\rho_0v_0^2},
\end{split}
\end{align}
where $\rho$, $p$, and $\VEC{B}$ are dimensionless grid values. For the 
simulations 
presented in this work we specify the following unit values: $\rho_0 
\simeq 
1.67\cdot\SI{e-22}{\gram\,\centi\meter^{-3}}$, $L_0 \simeq 
1.23\cdot\SI{e16}{\centi\meter}$, and $v_0\simeq 
3.0\cdot\SI{e10}{\centi\meter\per\second}$. With this choice of values the magnetic field strength along the jet will be of the order of gauss (G) to mG.

\subsection{Mapping the Non-Thermal onto the Thermal}\label{emission1}
An additional scaling relation is required in order to model the resultant non-thermal 
jet 
emission from our numerical simulations. The non-thermal quantities (i.e., 
electron number density and power-law energy cutoff) are inferred from the scaled thermal fluid
variables of the simulation (i.e., plasma density, pressure, and magnetic field). 
This thermal to non-thermal mapping of our 3D RMHD jet simulations is
used in the calculation of synchrotron emission maps. 
In particular, we initially assume an energy distribution of non-thermal relativistic electrons (NTEs) 
$n_e$ which follows a global power law distribution. This approach is justified 
by 
observations as well as theoretical expectations for particle acceleration within jets. 
In particular, we adopt a power-law distribution in electron energy $\gamma$ 
(where $E=\gamma m_e c^2$):
\begin{align}\label{ne}
n_e(\gamma) = n_0 \left(\frac{\gamma}{\gamma_\text{min}}\right)^{-s} \quad 
\text{for} \quad \gamma_\text{min} \leq \gamma \leq \gamma_\text{max},
\end{align}
where $n_e(\gamma)$ is the differential number of NTEs and $ 
\gamma_\text{min}$, $\gamma_\text{max}$ are the power-law cutoffs. The 
term $n_0$ is a normalization constant, and the power-law index $s$ is 
related to the spectral index $\alpha = \left(s-1\right)/2$.

To solve for the unknowns of the electron power-law ($\gamma_\text{min}$ and $n_0$), see \eq{ne}, we map the total number density of non-thermal electrons (NTEs) 
onto the thermal fluid variables~\citep[similar to][]{Fromm2016}.
First, we assume the number density of the injected NTEs is proportional to 
the thermal fluid number density $\rho$:
\begin{align}\label{32}
\int_{\gamma_\text{min}}^{\gamma_\text{max}} \text{d}\gamma n_e(\gamma) = 
\zeta_e \frac{\rho}{m_p},
\end{align}
where $\zeta_e$ is the ratio of non-thermal to thermal 
particles~\citep[see][]{Mimica2012}. Second,  we assume that the total energy 
density of 
the NTEs is 
proportional to the thermal pressure $p$. Here, we connect the fluid's pressure 
to the fluid's internal energy density $\epsilon$ via the equation of state $p 
= \left(\hat{\gamma}-1\right)\epsilon$, where $\hat{\gamma}$ is the adiabatic 
index. Consequently, the energy density becomes proportional to the internal 
energy density: 
\begin{align}\label{33}
\int_{\gamma_\text{min}}^{\gamma_\text{max}}\text{d}\gamma n_e(\gamma) m_ec^2 = 
\epsilon_e \epsilon = \epsilon_e \left(\hat{\gamma}-1\right) p,
\end{align}
where $\epsilon_e$ is the ratio between the energy stored in non-thermal 
particles to the thermal particles. We set the thermal to non-thermal 
conversion factors to $\zeta_e 
= 1.0$ and $\epsilon_e=0.5$.

Assuming that $\gamma_\text{max} \gg 
\gamma_\text{min}$ and $s>2$ (we set $s=2.3$), we solve this system of two equations (\eqs{32}{33}) for two unknowns ($\gamma_\text{min}$ and $n_0$), which yields:
\begin{align}\label{34}
\begin{split}
\gamma_\text{min}  =& \frac{pm_p \epsilon_e 
	(s-2)}{\rho m_ec^2\zeta_e 
	\left(s-1\right)\left(\hat{\gamma}-1\right)}\\\\
n_0 =& \frac{\zeta_e^2 \rho^2 m_e c^2 
	\left(1-2\right)^2\left(\hat{\gamma}-1\right)}{m_p^2 p \epsilon_e 
	\left(2-s\right)}.
\end{split}
\end{align}

\subsection{Polarized Radiative Transfer and Ray-tracing via \emph{RADMC-3D}}\label{sec:ray}
For our ray-tracing calculations we use the code \emph{RADMC-3D}, which is a well tested and documented ray-tracing software for computing astrophysical radiative transfer in 3D geometries. Our ray-tracing calculations are carried out in Cartesian coordinates and in the co-moving frame of the plasma after which the resultant fluxes are Doppler boosted to obtain the jet flux in the observer's frame.
The code reads in \emph{PLUTO} output files that have been scaled into physical units (namely; the 3D distributions of $B$, $n_e$, and $\gamma_\text{min}$). \emph{RADMC-3D} produces 2D Fits images containing full Stokes polarization maps. 
The radiative transfer is implemented in our plasma simulations through the use of 
transport coefficients for synchrotron absorption $\left(\kappa_I, \kappa_Q, \kappa_U, 
\kappa_V\right)$, synchrotron emissivity
$\left(\eta_I, \eta_Q, \eta_U, \eta_V\right)$, Faraday rotation ($\kappa_V^*$), and Faraday conversion ($\kappa_Q^*$ and $\kappa_U^*$).
Along individual rays our modified version of \emph{RADMC-3D} solves the following transfer matrix: 
\begin{align}\label{326}
\begin{pmatrix}
\dfrac{\text{d}}{\text{d}l} + \kappa_{I} &\kappa_{Q} &
\kappa_{U} & \kappa_{V} \\
\kappa_{Q} &  \dfrac{\text{d}}{\text{d}l} + \kappa_{I}  &
\kappa^{*}_{~ V} & -\kappa^{*}_{~ U} \\
\kappa_{U} & -\kappa^{*}_{~ V} & \dfrac{\text{d}}{\text{d}l} + 
\kappa_{I}  & \kappa^{*}_{~ Q} \\
\kappa_{V} &  \kappa^{*}_{~ U} & -\kappa^{*}_{~ Q} &
\dfrac{\text{d}}{\text{d}l} + \kappa_{I}
\end{pmatrix}
\begin{pmatrix}
I_{\nu} \vphantom{\left(\dfrac{\text{d}}{\text{d}l}+\kappa_I\right)}\\
Q_{\nu} \vphantom{\left(\dfrac{\text{d}}{\text{d}l}+\kappa_I\right)}\\
U_{\nu} \vphantom{\left(\dfrac{\text{d}}{\text{d}l}+\kappa_I\right)}\\
V_{\nu}  \vphantom{\left(\dfrac{\text{d}}{\text{d}l}+\kappa_I\right)}
\end{pmatrix} = 
\begin{pmatrix}
\eta_I  \vphantom{\left(\dfrac{\text{d}}{\text{d}l}+\kappa_I\right)}\\
\eta_Q  \vphantom{\left(\dfrac{\text{d}}{\text{d}l}+\kappa_I\right)}\\
\eta_U  \vphantom{\left(\dfrac{\text{d}}{\text{d}l}+\kappa_I\right)}\\
\eta_V  \vphantom{\left(\dfrac{\text{d}}{\text{d}l}+\kappa_I\right)},
\end{pmatrix}
\end{align}
to obtain linear and circular polarization as a function of optical 
depth, i.e., $\text{d}\tau = \kappa_I \text{d}l$. The code applies an 
analytical 
solution to \eq{326} presented in~\citep{JonesOdell} and which is summarized in~\citep{MacDonald2018}. The analytical solution is a function of the 
normalization constant $n_0$, the low-energy cutoff 
$\gamma_\text{min}$ of the power-law distribution in \eq{ne}, the strength of 
the magnetic field, and its orientation to our line of sight.

\section{Magnetic Field Morphology Study}\label{3}

\subsection{Magnetic Field Prescriptions}\label{mag}
Based on the physical scaling (presented in \sect{scaling}) and our NTE scaling relations (see~\eq{34}), we proceed to study the impact that different magnetic field 
morphologies within the jet (i.e., poloidal, toroidal, and helical) have on the jet's
polarized synchrotron emission.
To produce different magnetic field morphologies we implement (in Cartesian 
coordinates) expressions for the poloidal and toroidal 
components of the jet's magnetic field at the jet injection point (i.e., orifice) in our simulations.
The jet is oriented along the z-axis in our ray-tracing calculations. The $x$ and $y$ components of the jet's magnetic field's toroidal component are given by~\citep{Nishikawa2019}:
\begin{align}\label{36}
\begin{split}
B_{x}=\frac{\left(\frac{y-y_{c}}{a}\right)\cdot 
	b_m}{1+\left(\frac{r}{a}\right)^2}, \qquad
B_{y}=-\frac{\left(\frac{x-x_{c}}{a}\right)\cdot 
	b_m}{1+\left(\frac{r}{a}\right)^2},
\end{split}
\end{align}
where $\left(x_{c},y_{c}\right)$ is 
the location of the center of the jet, which in our simulation is set to $(0,0)$.
The variable $r$ defines the jet's radius while $a$ represents a magnetization 
radius. The parametrization constant of the magnetic field ($b_m$) is 
given by:
\begin{equation}
b_m=\sqrt{\frac{-4p_j\sigma_\phi}{a^2\left(2\sigma_\phi-1+4\log(a)\right)}},
\end{equation}
where $\sigma_\phi$ is the magnetization parameter for the toroidal component. 
\begin{figure}
	\centering
	\setlength{\fboxsep}{1pt}
	\setlength{\fboxrule}{1pt}
	\fbox{\includegraphics[width=0.42\textwidth]{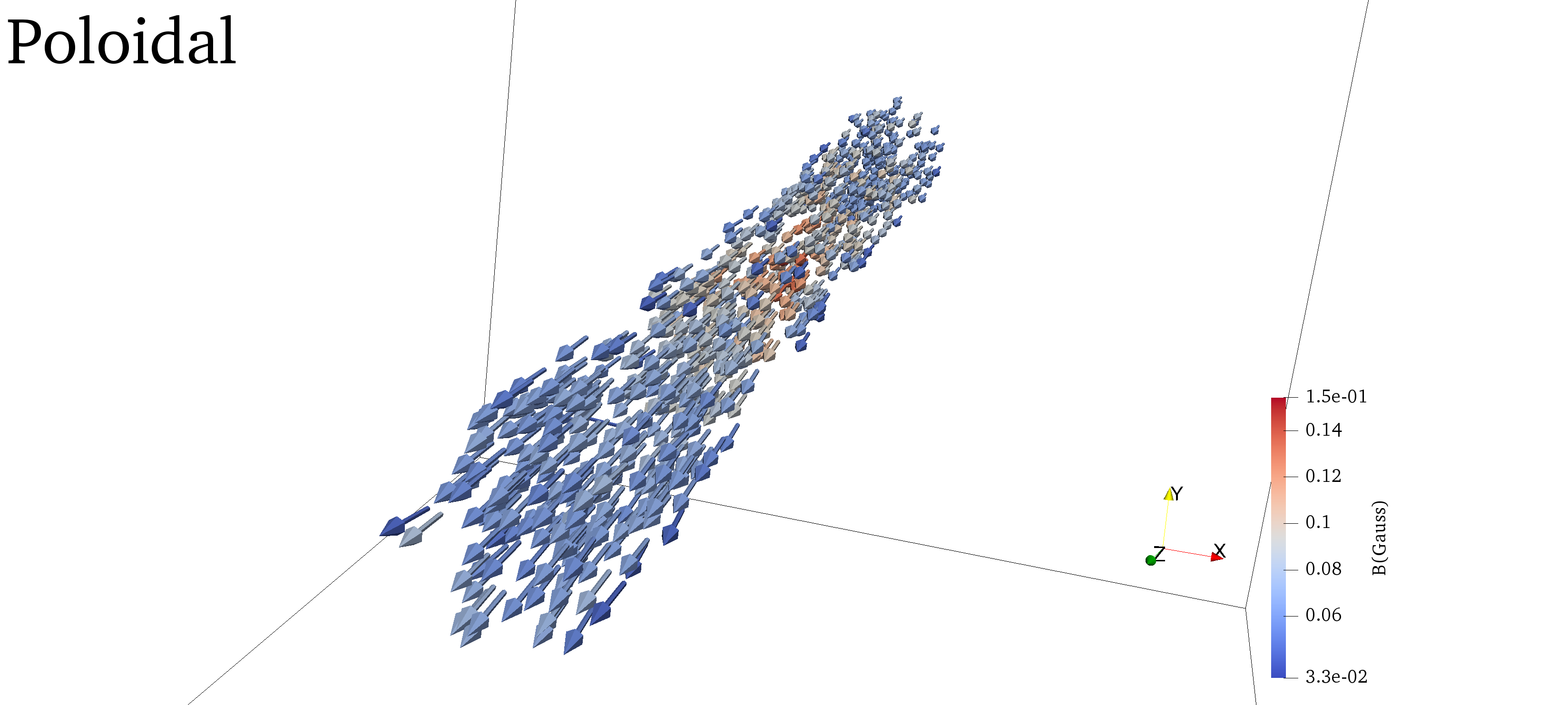}}
	\fbox{\includegraphics[width=0.42\textwidth]{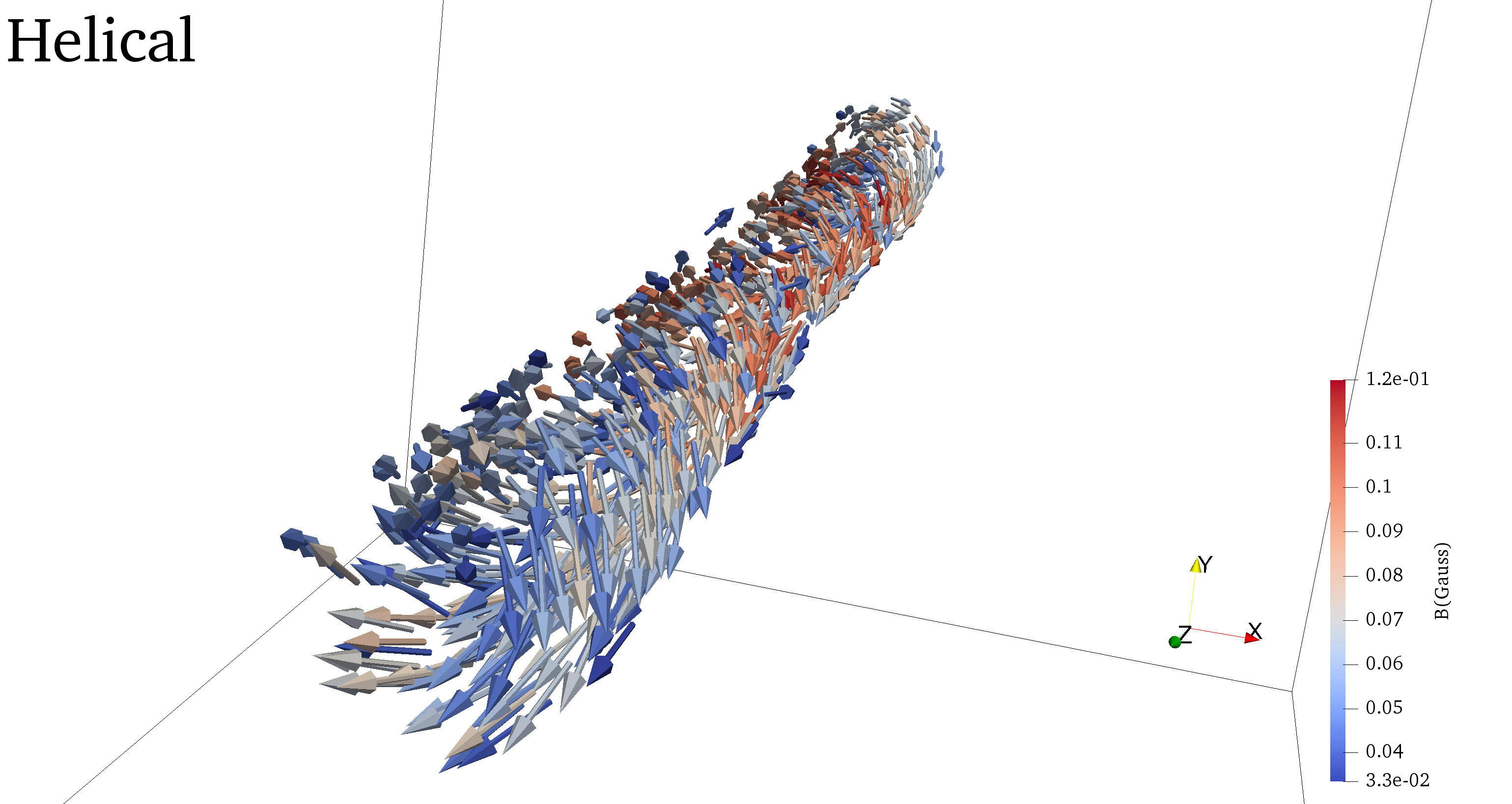}}
	\fbox{\includegraphics[width=0.42\textwidth]{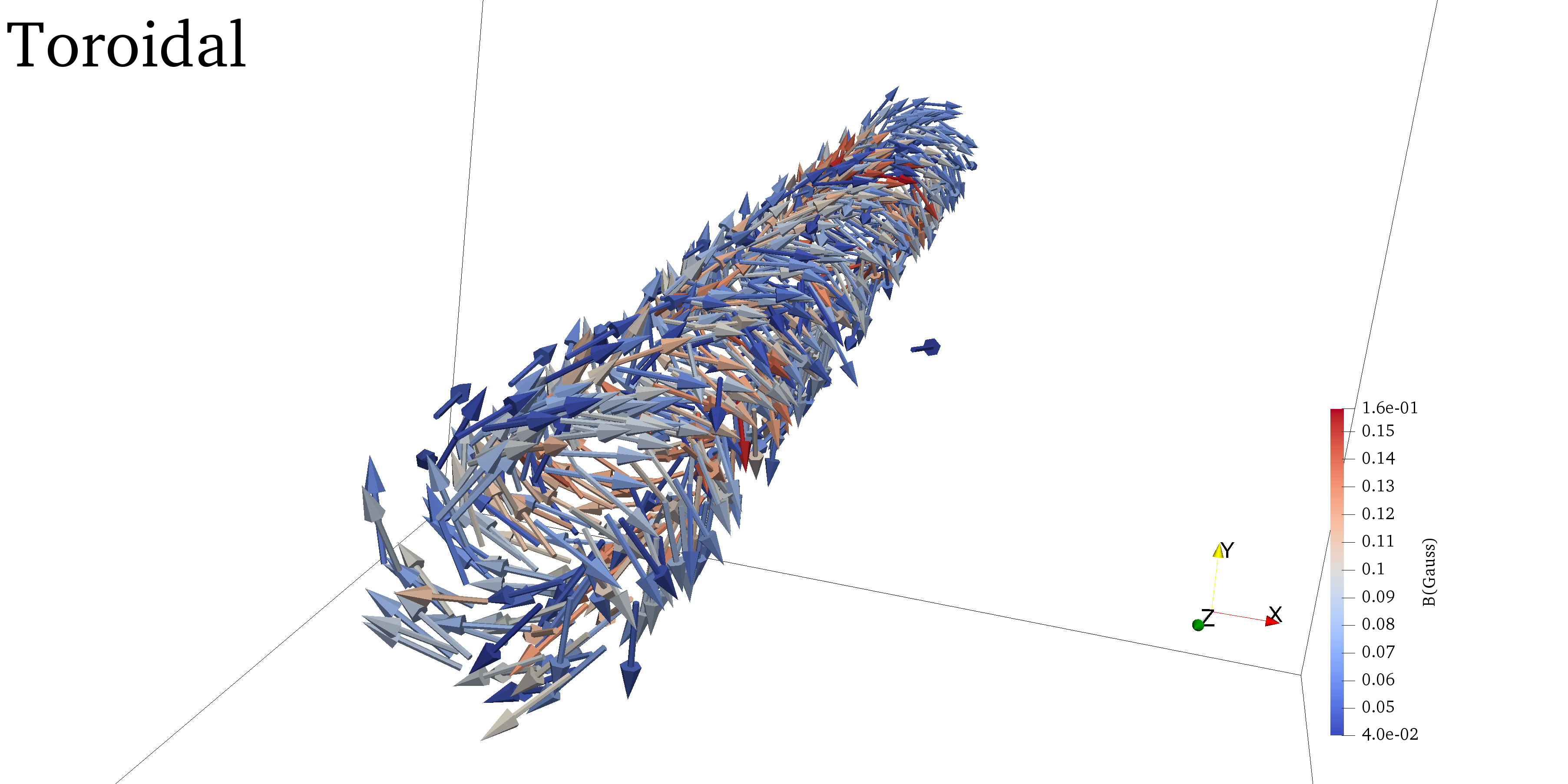}}
	\caption[Illustration of three different magnetic field morphologies 
	within 
	the 3D RMHD jet]{Illustration of the three different magnetic field 
		morphologies 
		within 
		the 3D RMHD jet simulations. The jet is streaming in 
		$z$-direction. The vectors 
		represent the magnetic field strength (and orientation) in gauss (see the color bar) within each computational cell. 
		From top to bottom: poloidal, helical, and toroidal magnetic field morphologies.}
	\label{figmag}
\end{figure}
The constant poloidal term $B_z$, threading the jet, is 
written as:
\begin{equation}\label{37}
B_{z}=\sqrt{\sigma_z\left(b_m^2a^2+2p_j\right)},
\end{equation}
where $\sigma_z$ is the magnetization parameter of the poloidal component. We 
choose $\sigma_z=1$ for a purely poloidal magnetic field and $\sigma_\phi=1$ for a purely toroidal field~\citep{Nishikawa2019, Mignone2009}. A helical field is produced by setting $\sigma_z = \sigma_\phi = 0.5$.
In \eq{37}  the variable $p_j$ is the jet
pressure. In particular, $p_j$ is determined from our simulated jet Mach number 
$M=2.7$ and bulk Lorentz factor $\Gamma=7$ (i.e., $M = v_j \cdot \sqrt{\rho_j/\left(\Gamma p_j\right) + 1/\left(\Gamma - 1\right)}$ ), where $\rho_j$ is the jet density. The sum 
of both magnetization parameters is set to 1 to enforce an 
equipartition between the magnetic pressure and thermal pressure within the jet 
plasma. We point out that the magnetization radius $a$ equals the jet radius 
$r_j=\sqrt{\left(x-x_c\right)^2+\left(y-y_c\right)^2}$ as it reaches its 
maximum 
value.

\Fig{figmag} illustrates the three different simulated magnetic field morphologies 
within the jet (from top to bottom: poloidal, helical, toroidal). Here, the 
poloidal magnetic field vectors are streaming along the jet in 
$z$-direction while the vectors for the toroidal components are predominantly
perpendicular to the jet axis. The helical magnetic field vectors are rotated about $\sim 45^\circ$ in the jet's direction.
2D slices through the jet that illustrate how the injected field morphologies persist down the jet axis are included in Appendix~\ref{appendix}.

\subsection{Results}\label{resfromm}
\begin{figure*}
	\begin{tikzpicture}
	\small
	\node[anchor=south west,inner sep=0] (Fig1) at (0,0) 
	{\includegraphics[trim={    1.4cm     0.4cm      3.4cm       0.0cm      }, clip,width=0.16\textwidth]{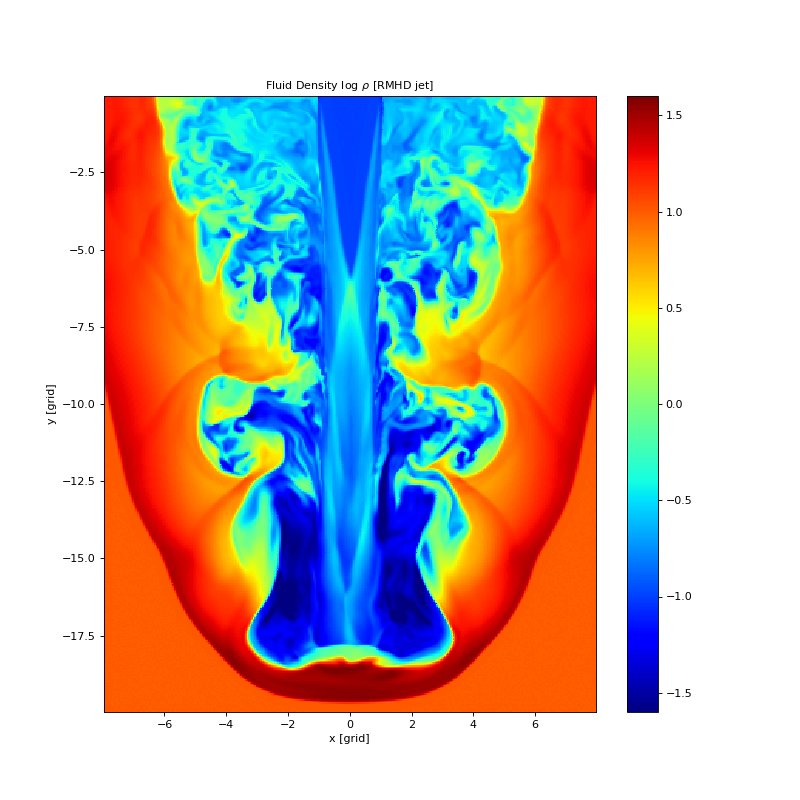}};
	\hspace{5pt}
	\node[anchor=south west, inner sep=0] (Fig2) at (Fig1.south east) 
	{\includegraphics[trim={    0.9cm     0.0cm      2.1cm       0.0cm      }, clip,width=0.21\textwidth]{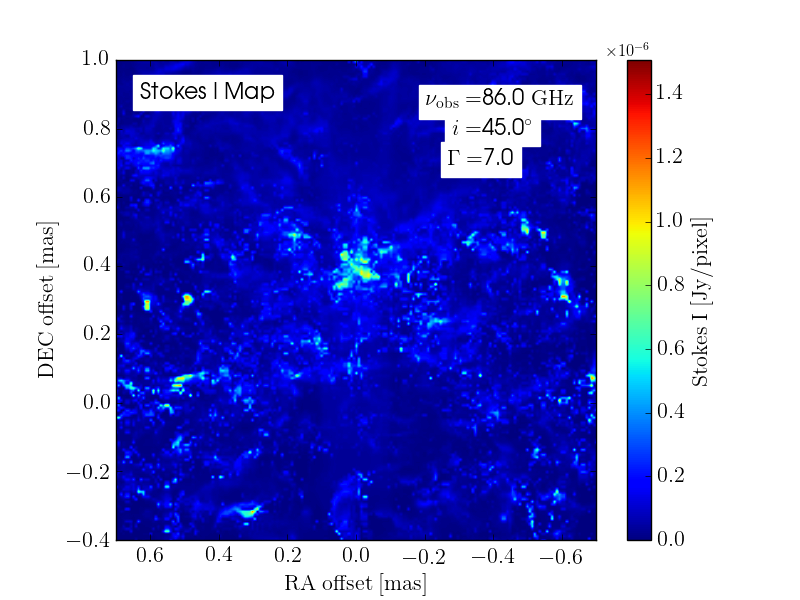}};
	\node[anchor=south west, inner sep=0] (Fig3) at (Fig2.south east) 
	{\includegraphics[trim={    0.9cm     0.0cm      2.1cm       0.0cm      }, clip,width=0.21\textwidth]{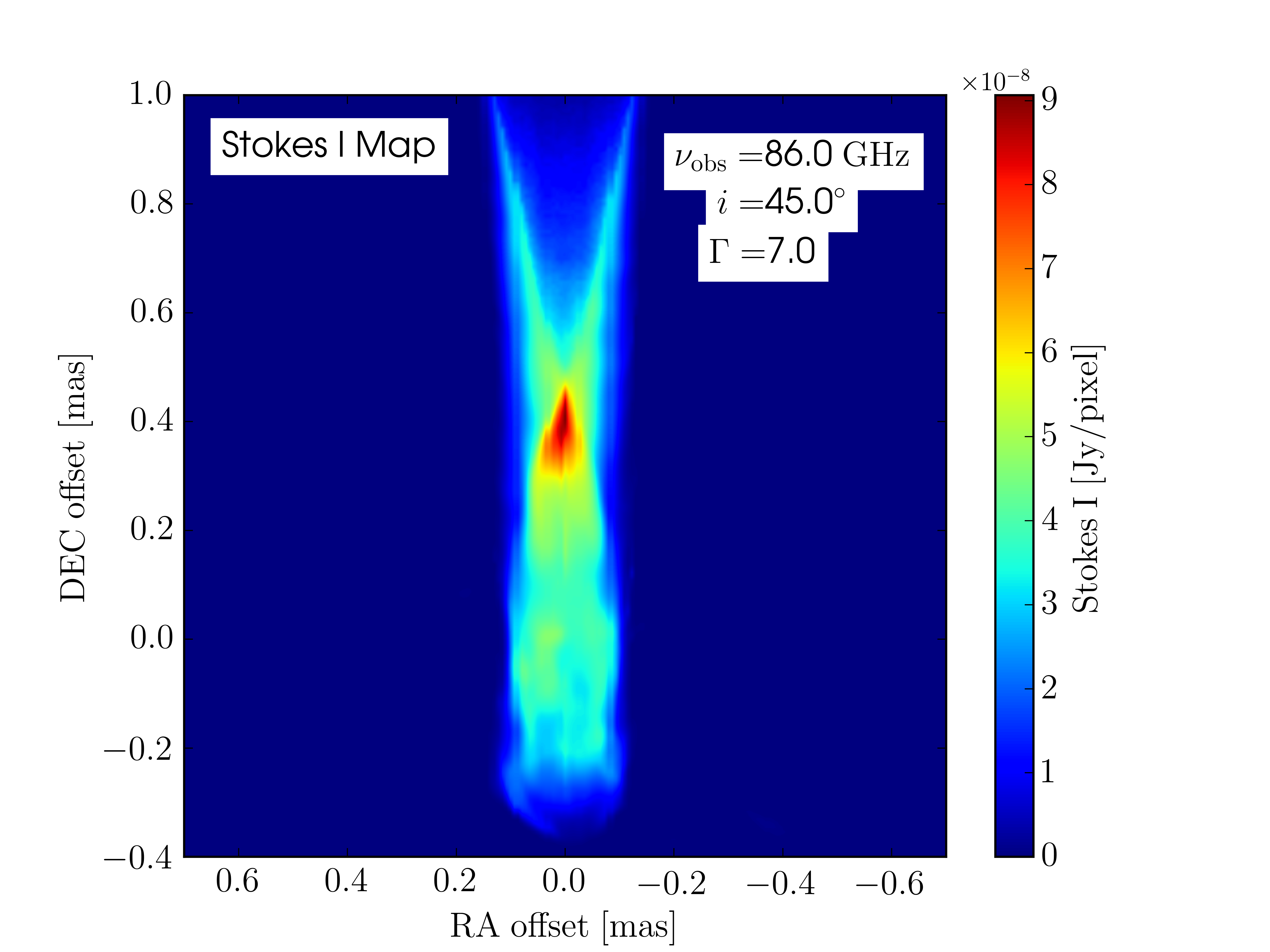}};
	\node[anchor=south west, inner sep=0] (Fig5) at (Fig3.south east) 
	{\includegraphics[trim={    0.9cm     0.0cm      2.1cm       0.0cm      }, clip,width=0.21\textwidth]{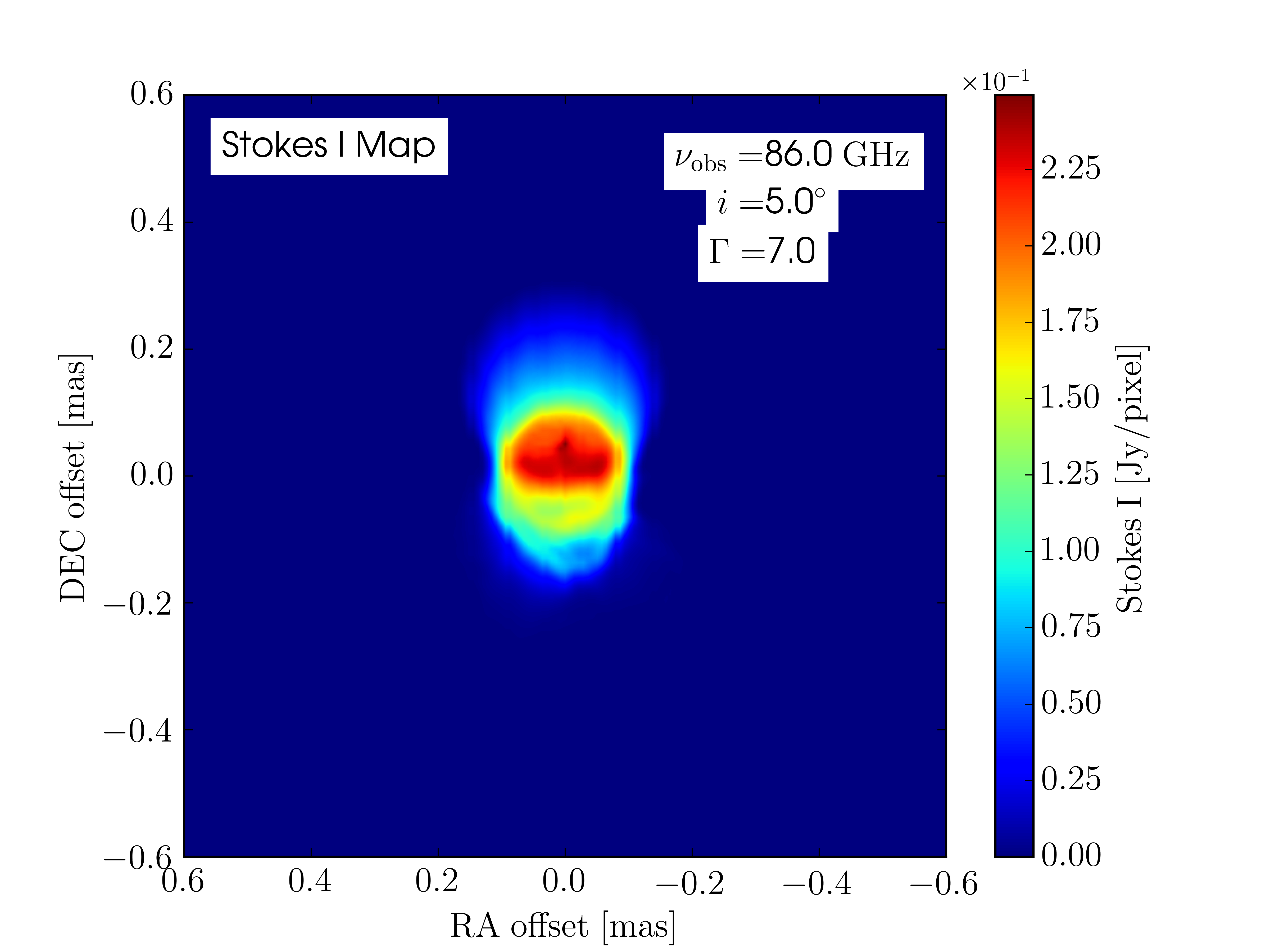}};
	\node[anchor=south west, inner sep=0] (Fig6) at (Fig5.south east) 
	{\includegraphics[trim={    0.9cm     0.0cm      2.1cm       0.0cm      }, clip,width=0.21\textwidth]{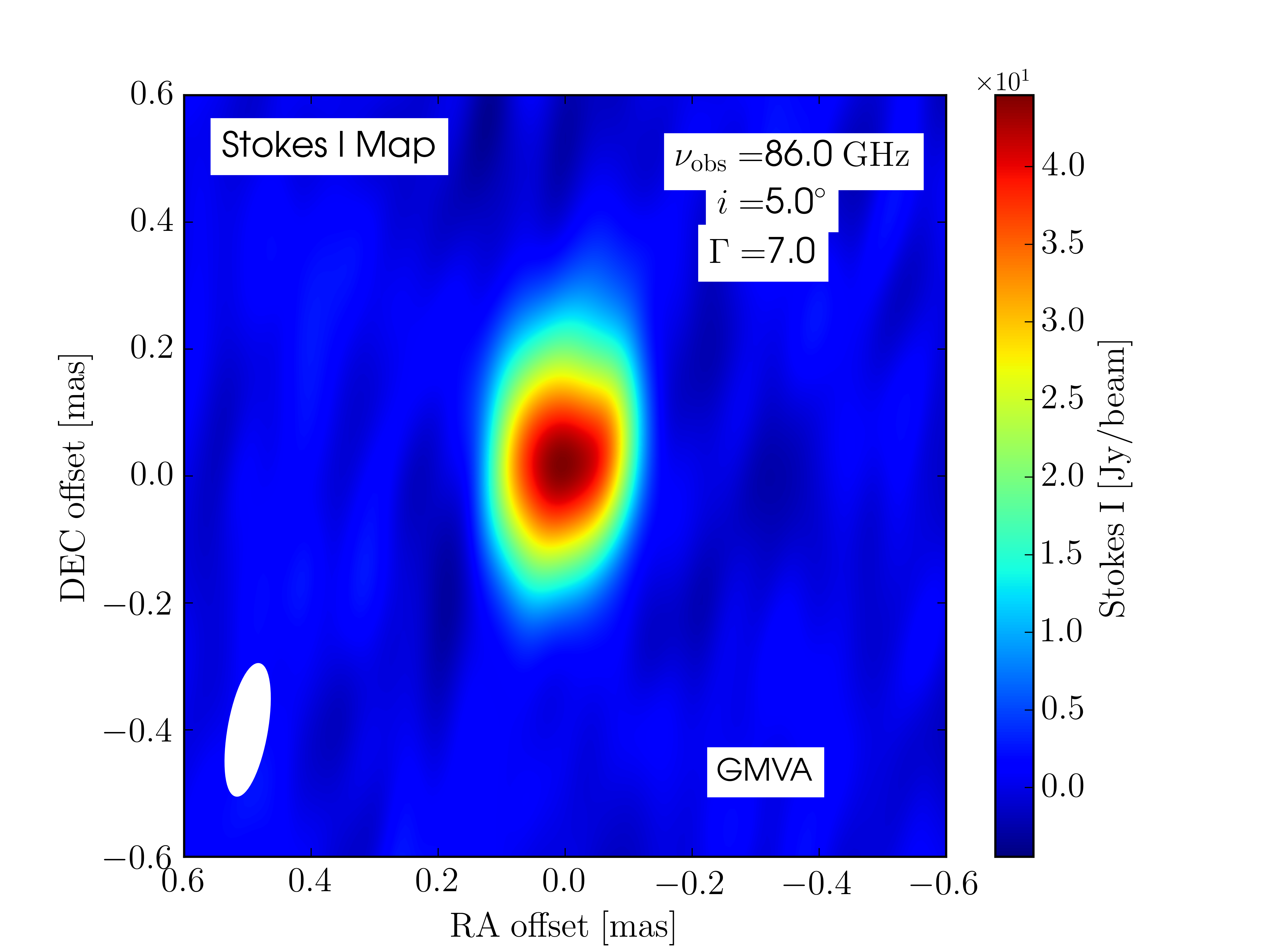}};
	
	\draw[semithick, violet] (0.32,3.13) -- 
	node[anchor=north, rotate=90] {} (2.1,3.13);
	\draw[semithick, violet] (0.32,1.5) -- 
	node[anchor=north, rotate=90] {} (2.1,1.5);
	\draw[semithick, violet] (0.32,1.5) -- 
	node[anchor=north, rotate=90] {} (0.32,3.13);
	\draw[semithick, violet] (2.1,1.5) -- 
	node[anchor=north, rotate=90] {} (2.1,3.13);
	
	\draw[semithick, violet] (0.32,3.13) -- 
	node[anchor=north, rotate=90] {} (3.43,3.059);
	\draw[semithick, violet] (0.32,1.5) -- 
	node[anchor=north, rotate=90] {} (3.43,0.349);	

	\draw[semithick, violet] (3.43,3.059) -- 
	node[anchor=north, rotate=90] {} (6.15,3.059);
	\draw[semithick, violet] (3.43,0.349) -- 
	node[anchor=north, rotate=90] {} (6.15,0.349);
	\draw[semithick, violet] (3.43,0.349) -- 
	node[anchor=north, rotate=90] {} (3.43,3.059);
	\draw[semithick, violet] (6.15,0.349) -- 
	node[anchor=north, rotate=90] {} (6.15,3.059);	
	
	\hspace{-6pt}\node[anchor=north,inner sep =0, label=below:$(a)$] (a) at (Fig1.south) {};
	\hspace{3pt}\node[anchor=north,inner sep =0, label=below:$(b)$] (b) at (Fig2.south) {};
	\node[anchor=north,inner sep =0, label=below:$(c)$] (c) at (Fig3.south) {};
	\node[anchor=north,inner sep =0, label=below:$(d)$] (d) at (Fig5.south) {};
	\node[anchor=north,inner sep =0, label=below:$(e)$] (e) at (Fig6.south) {};
	\end{tikzpicture}
	
	\caption[Synthetic emission maps: Total intensity for three 
	different magnetic fields and emission recipes]{
		From left to right - a demonstration of our synthetic imaging pipeline: panel (a) starts with a 2D cut through our 3D RMHD jet colored by thermal density. Panels (b) through (e) are zoomed into an unobscured region of the jet's standing shock (demarcated with a purple box in panel (a)) and show: (b) the ray-traced synchrotron emission without the use of a jet tracer, (c) the ray-traced synchrotron emission with the use of a jet tracer to exclude the ambient medium, (d) the same simulation epoch but rotated to a viewing angle of $i=5^\circ$, and (e) the ray-traced image convolved with a Gaussian beam indicative of the resolution of the GMVA and with a Gaussian noise floor (to mimic array sensitivity) of $10^{-1}\,\text{Jy/beam}$. }
	\label{fig:41}
\end{figure*}
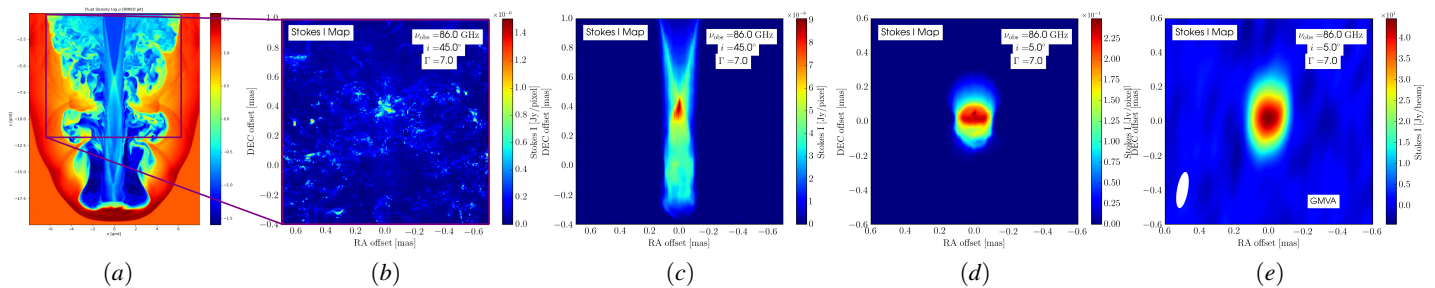

\Fig{fig:41} presents a summary of the various steps in our synthetic imaging pipeline. In particular, we are interested in studying the polarized properties of the jet's recollimation shock.
We image an intermediate epoch of each jet simulation, i.e., when the jet's hot spot/terminal shock has not propagated off the grid yet. Through the use of a jet tracer we extract the region of plasma around the standing shock (demarcated with a purple box in the first panel of~\fig{fig:41}) in order to yield an unobscured view of the jet's central spine thus allowing us to focus on the jet flow upstream of the termination shock. These initial images were created using the poloidal magnetic field simulation.
Panel (a) in~\fig{fig:41} shows a 2D slice through the 3D jet simulation and displays the jet's density in dimensionless grid units. Panels (b) and (c) illustrate the zoomed-in ray-traced total intensity maps of the resulting synchrotron emission when the jet is resolved without and with the use of a jet tracer to exclude the ambient medium, respectively. The jet is viewed at an angle to the jet-axis of $i=45^\circ$ and propagates from top to bottom. In the absence of radiative cooling (i.e., synchrotron losses) and larger simulation sizes we remove the bow shock from our ray-tracing calculations arbitrarily.
Panel (d) displays the same snap-shot but rotated to a viewing angle of $i=5^\circ$. Finally, panel (e) is convolved with a Gaussian beam, i.e., indicative of the resolution of the Global mm VLBI Array (GMVA), and with an added Gaussian noise level (to mimic array sensitivity) of $10^{-1}\,\text{Jy/beam}$. These final images show a bright radio core associated with the standing shock in our simulations.
In all our RMHD jet simulations (in which each computational box consists of $320\times320\times400$ zones) we choose to view the source at a luminosity distance of $100\,\text{Mpc}$. The individual scaled cell size is $0.004\,\text{pc}$. All images in this paper are generated at an observing frequency of $\nu_{\rm{obs}} = 86\, \rm{GHz}$. 

\begin{figure*}
	\centering
	\begin{tikzpicture}
	\small
	\node[anchor=south west,inner sep=0] (Fig1) at (0,0) 
	{\includegraphics[width=0.31\textwidth]{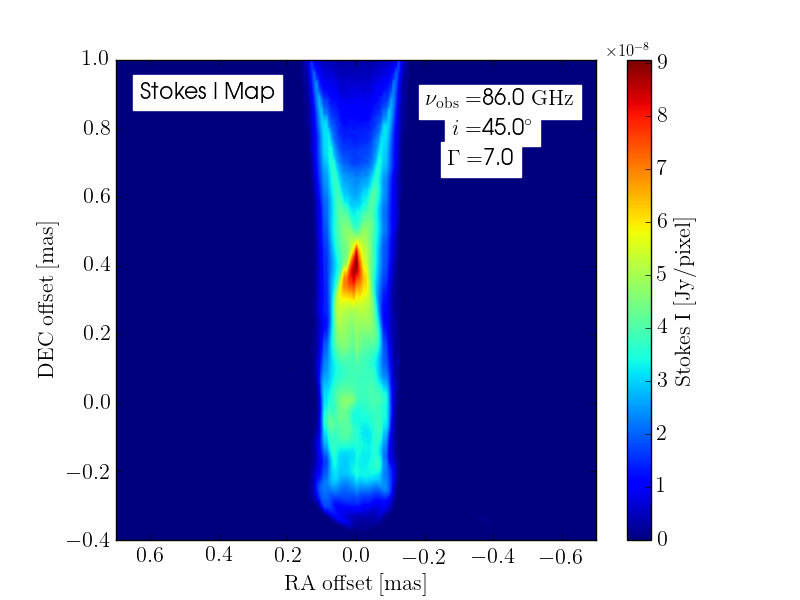}};
	\node[anchor=south west, inner sep=0] (Fig2) at (Fig1.south east) 
	{\includegraphics[width=0.31\textwidth]{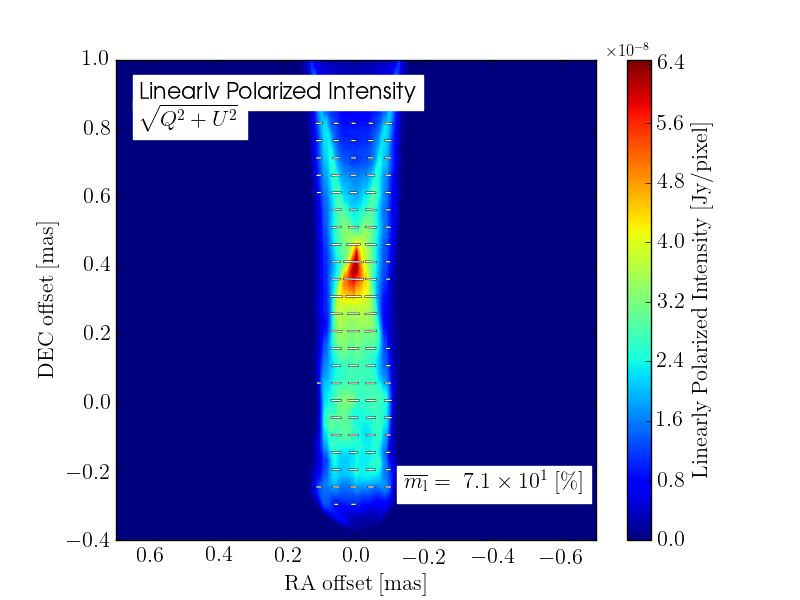}};
	\node[anchor=south west, inner sep=0] (Fig3) at (Fig2.south east) 
	{\includegraphics[width=0.31\textwidth]{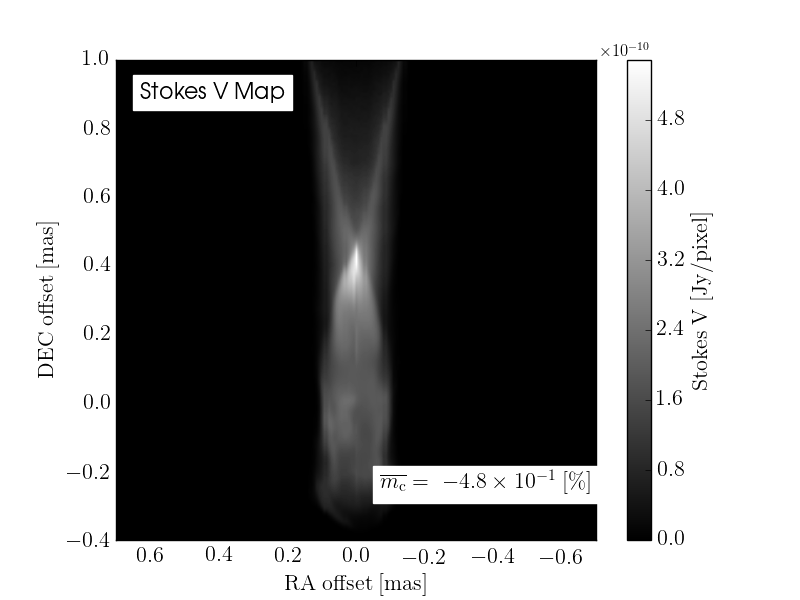}};
	
	\node[anchor=north west, inner sep=0] (Fig4) at (Fig1.south west) 
	{\includegraphics[width=0.31\textwidth]{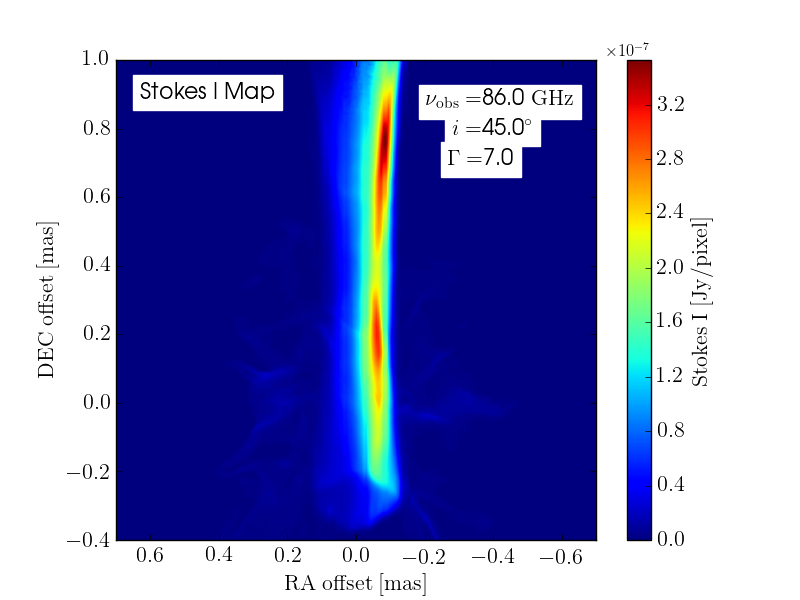}};
	\node[anchor=south west, inner sep=0] (Fig5) at (Fig4.south east) 
	{\includegraphics[width=0.31\textwidth]{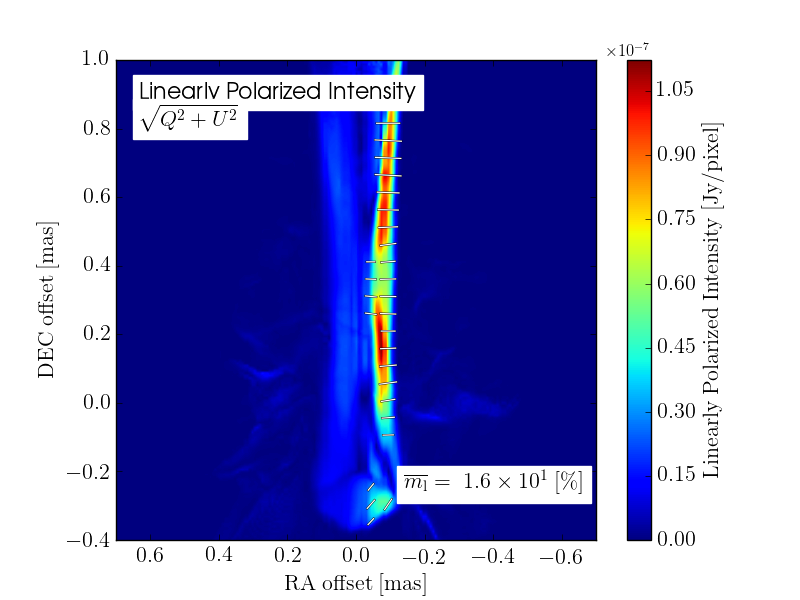}};
	\node[anchor=south west, inner sep=0] (Fig6) at (Fig5.south east) 
	{\includegraphics[width=0.31\textwidth]{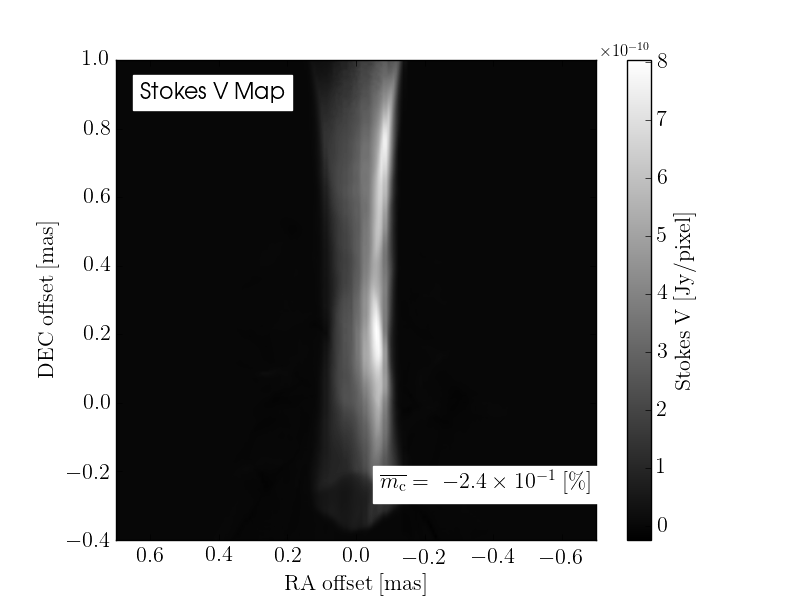}};
	
	\node[anchor=north west, inner sep=0] (Fig7) at (Fig4.south west) 
	{\includegraphics[width=0.31\textwidth]{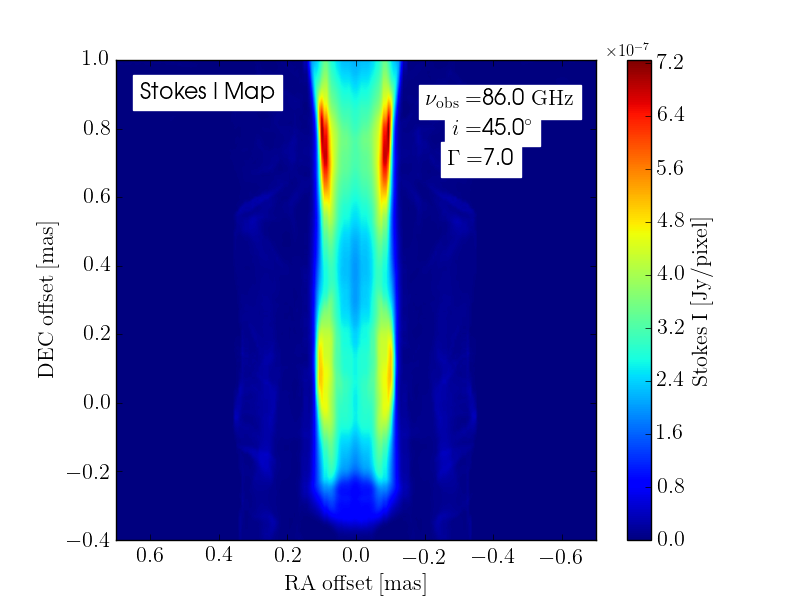}};
	\node[anchor=south west, inner sep=0] (Fig8) at (Fig7.south east) 
	{\includegraphics[width=0.31\textwidth]{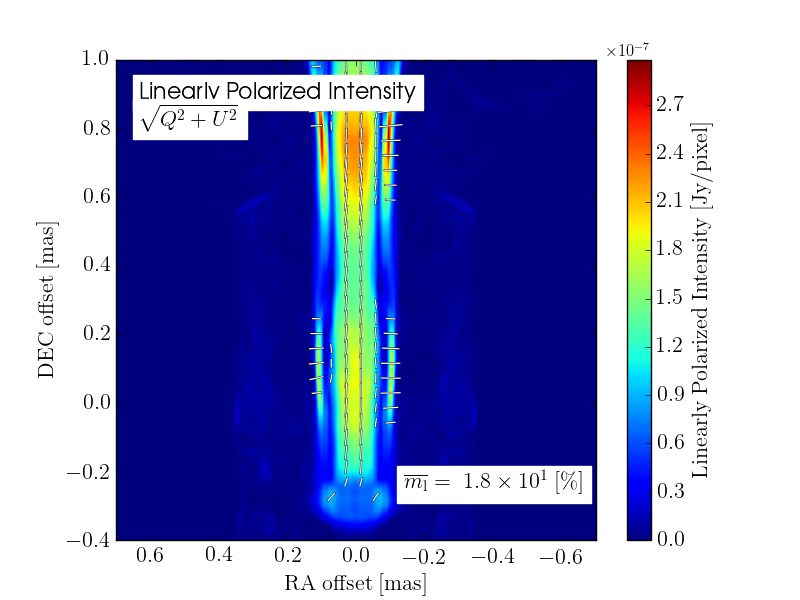}};
	\node[anchor=south west, inner sep=0] (Fig9) at (Fig8.south east) 
	{\includegraphics[width=0.31\textwidth]{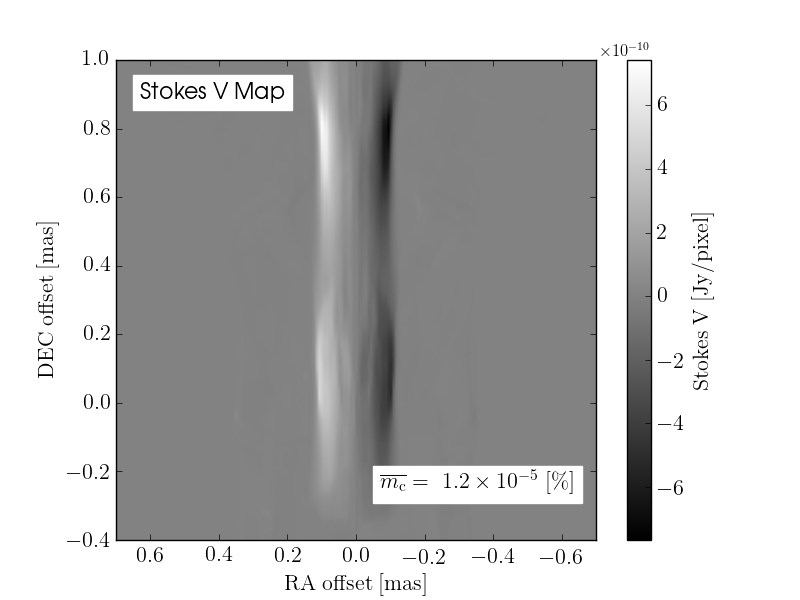}};

	\draw[semithick] (Fig9.south east) -- 
	node[anchor=north,rotate=90] {\emph{(iii)} Toroidal} (Fig9.north east) --
	(Fig6.south east) --
	node[anchor=north,rotate=90] {\emph{(ii)} Helical} (Fig6.north east) --
	(Fig3.south east) --
	node[anchor=north,rotate=90] {\emph{(i)} Poloidal} (Fig3.north east);
	
	\end{tikzpicture}
	
	\caption[Synthetic polarized emission maps: Total intensity, linearly 
	polarized intensity, and circular polarization for three magnetic field 
	morphologies]{Ray-traced images of our jet propagating from top to bottom 
		in 
		total
		intensity (left column), in linearly polarized intensity (middle column), 
		and	circular polarization (right column). The pictures
		illustrate similar epochs in the jet's evolution of each 3D RMHD simulation at $86\,$GHz. The jet carries 
		a purely poloidal magnetic field (top row), a helical magnetic field
		(middle row), and a purely toroidal magnetic field (bottom row). Integrated values of the fractional linear and circular polarization are listed to the lower right in the middle and right columns. The purely 
		toroidal field jet appears edge-brightened in contrast to the poloidal jet 
		which is brightest along the spine.}
	\label{38a}
\end{figure*}

We visualize the 
polarized synchrotron
emission of the jet for three different magnetic field morphologies, i.e.,
purely poloidal, helical, and purely toroidal. To begin with, the images in
\fig{38a} show the total intensity of the jet's emission in the left column, the 
linearly
polarized intensity ($=\sqrt{Q^2 + U^2}\,\,$) including electric vector position angles (EVPAs $= 0.5\arctan{[U/Q]}$ ) in the
middle column and the circular polarization in the right column. Moreover, the
different rows represent the three magnetic field morphologies introduced in
\sect{mag}. The top row depicts the poloidal magnetic field, the middle row
the helical field, and the bottom row the purely toroidal magnetic field. All ray-traced images are viewed at $45^\circ$ to the jet-axis in \fig{38a}.

We can see that the emission for the purely poloidal magnetic field is
concentrated in the inner part of the jet, and is brightest within the standing
recollimation shock. The EVPAs, shown as white line segments in the middle column 
of
\fig{38a}, are predominantly perpendicular to the magnetic field orientation, in the ideal
case. As the poloidal field is streaming in the direction of the jet the EVPAs
accurately convey the field orientation within our simulations. In addition, the circular polarization has only positive values, unlike the purely toroidal magnetic field.

The toroidal magnetic field (bottom row of \fig{38a}), clearly produces emission
that is centered along the edges of the jet, as we see an edge-brightened jet 
in
all our images along the bottom row. In addition, we can see both positive and negative circular polarization highlighting the changing orientation of the jet's magnetic field with respect to our line-of-sight.

The helical magnetic field, illustrated in the middle row in \fig{38a}, exhibits a mixture of the emission/polarization morphologies present in the
toroidal/poloidal cases. The emission is concentrated on the right side of the relativistic jet which stresses the structure of the helical magnetic field lines.  

Additionally, we compute integrated levels of fractional polarization. These are flux weighted averages of the Stokes parameters across the entire jet
emission region in each set of images (listed 
to the lower right in the linearly polarized and circularly polarized images of \fig{38a}, $\bar{m_l} \equiv \left(\bar{Q}^2+\bar{U}^2\right)^{1/2}/\bar{I}$ \& $\bar{m_c}\equiv-\bar{V}/\bar{I}$). The fractional linear polarization decreases from the poloidal, to the toroidal magnetic field model ( from $\sim 7.1,\%$ to $\sim 1.8\,\%$) and drops for the helical magnetic field morphology ($\sim 1.6\,\%$). The fractional circular polarization decreases as well from the poloidal to the helical magnetic field structure. The calculated value for the toroidal field jet changed the sign and is several orders of magnitude smaller (from $\sim-4.8\times 10^{-1}\,\%$ to $\sim 1.2\times10^{-5}\,\%$).

\begin{figure*}
	\centering
	\begin{tikzpicture}
	\small
	\node[anchor=south west,inner sep=0] (Fig1) at (0,0) 
	{\includegraphics[width=0.31\textwidth]{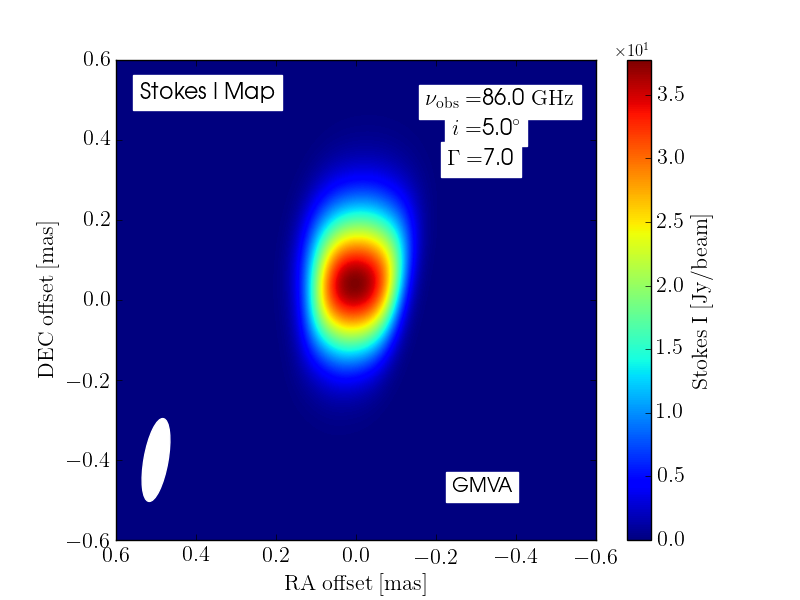}};
	\node[anchor=south west, inner sep=0] (Fig2) at (Fig1.south east) 
	{\includegraphics[width=0.31\textwidth]{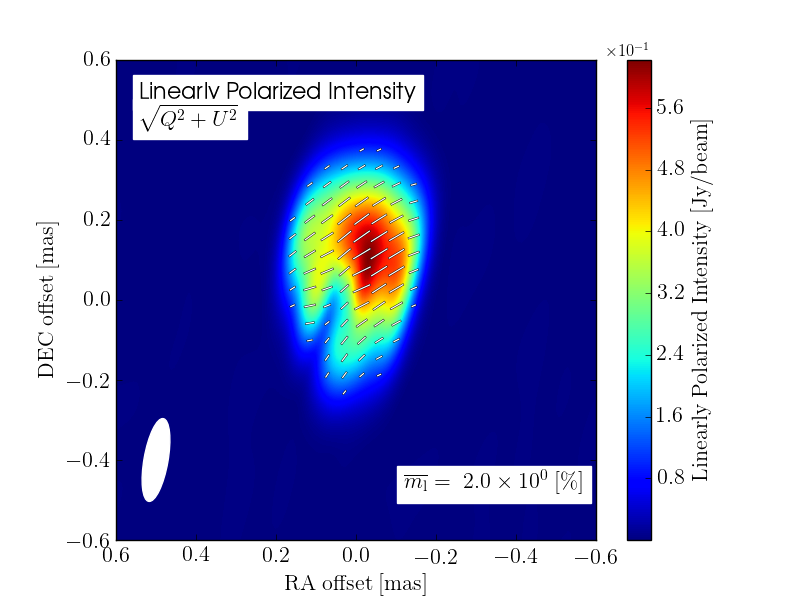}};
	\node[anchor=south west, inner sep=0] (Fig3) at (Fig2.south east) 
	{\includegraphics[width=0.31\textwidth]{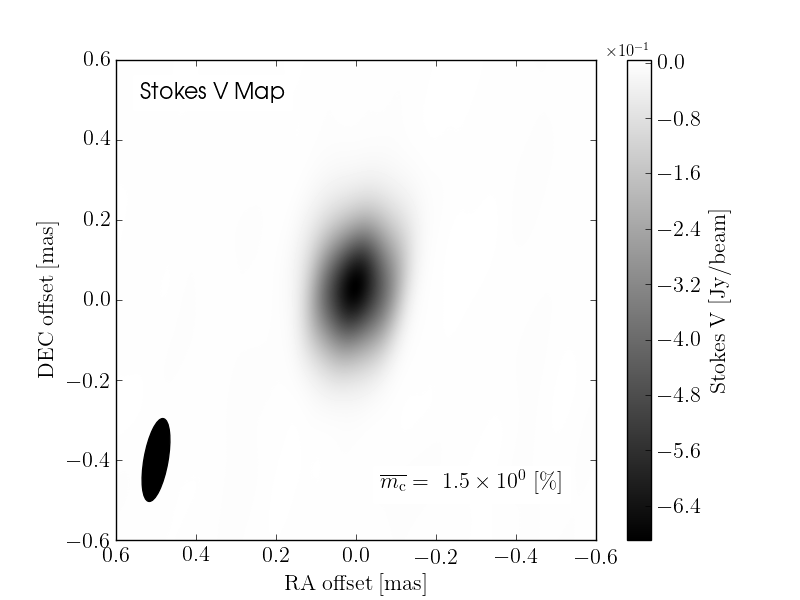}};
	
	\node[anchor=north west, inner sep=0] (Fig4) at (Fig1.south west) 
	{\includegraphics[width=0.31\textwidth]{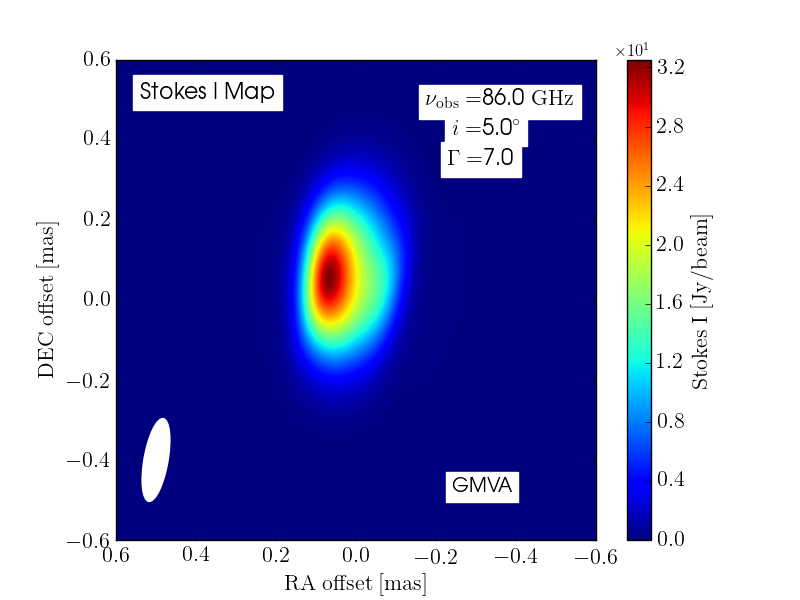}};
	\node[anchor=south west, inner sep=0] (Fig5) at (Fig4.south east) 
	{\includegraphics[width=0.31\textwidth]{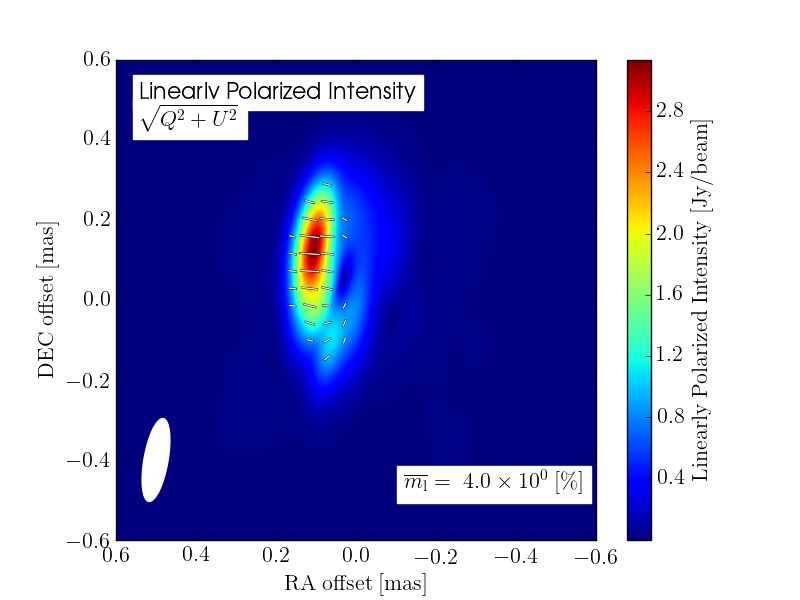}};
	\node[anchor=south west, inner sep=0] (Fig6) at (Fig5.south east) 
	{\includegraphics[width=0.31\textwidth]{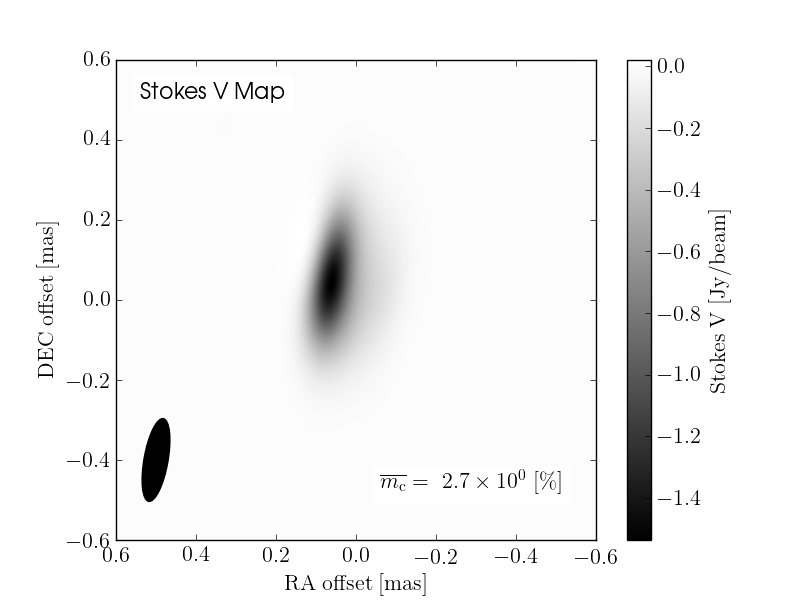}};
	
	\node[anchor=north west, inner sep=0] (Fig7) at (Fig4.south west) 
	{\includegraphics[width=0.31\textwidth]{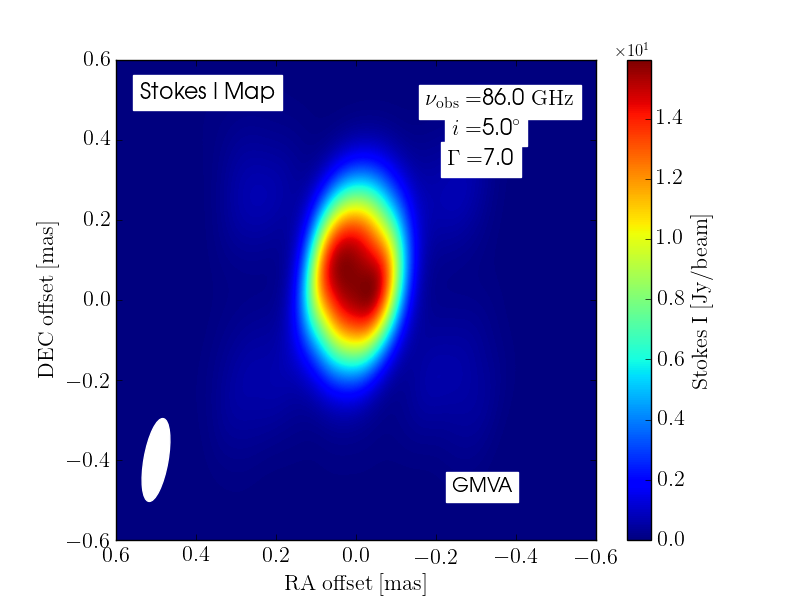}};
	\node[anchor=south west, inner sep=0] (Fig8) at (Fig7.south east) 
	{\includegraphics[width=0.31\textwidth]{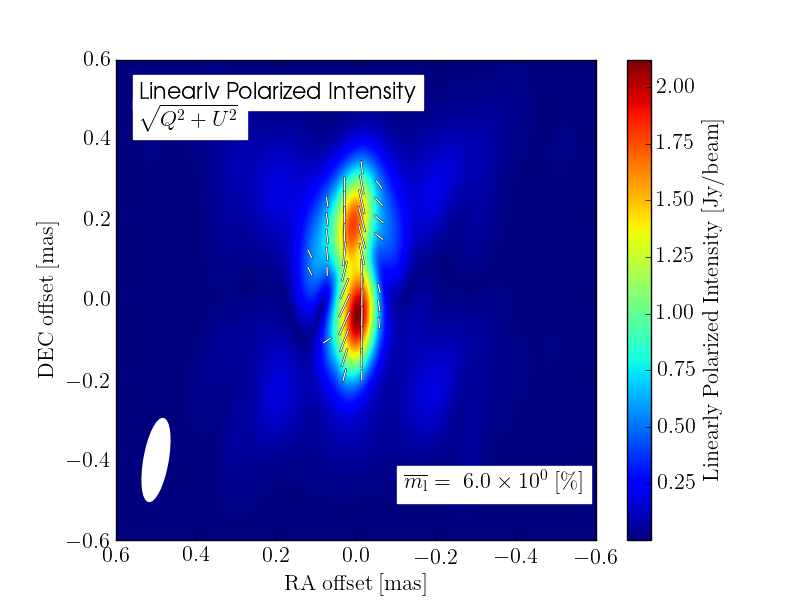}};
	\node[anchor=south west, inner sep=0] (Fig9) at (Fig8.south east) 
	{\includegraphics[width=0.31\textwidth]{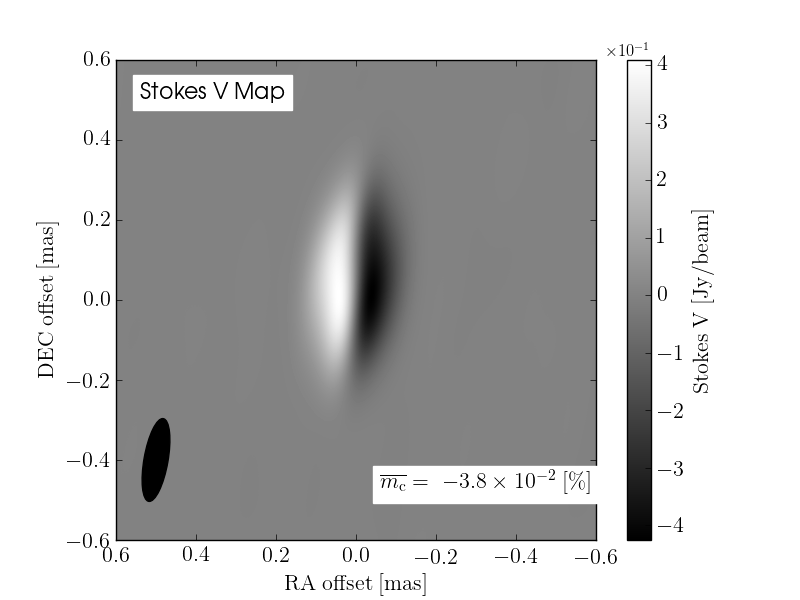}};
	
	\draw[semithick] (Fig9.south east) -- 
	node[anchor=north,rotate=90] {\emph{(iii)} Toroidal} (Fig9.north east) --
	(Fig6.south east) --
	node[anchor=north,rotate=90] {\emph{(ii)} Helical} (Fig6.north east) --
	(Fig3.south east) --
	node[anchor=north,rotate=90] {\emph{(i)} Poloidal} (Fig3.north east);
	
	\end{tikzpicture}
	
	\caption[Synthetic polarized emission maps: Total intensity, linearly 
	polarized intensity, and circular polarization for three magnetic field 
	morphologies]{Ray-traced images of RMHD jet simulations in total intensity (left), linearly polarized intensity (middle), and circular polarization (right) when each jet is viewed edge-on to the jet-axis. The images highlight similar epochs in the jet's evolution of each 3D RMHD simulation at $86\,$GHz. The jet carries 
		a purely poloidal magnetic field (top row), a helical magnetic field
		(middle row), and a purely toroidal magnetic field (bottom row). The ray-traced images are convolved with a Gaussian beam indicative of the resolution of the GMVA and with a Gaussian noise floor of $10^{-4}\,\text{Jy/beam}$. }
	\label{38b}
\end{figure*}

The images presented in \fig{38a} display a resolved RMHD jet observed at $45^\circ$. To further simulate the emission of a blazar radio core we: \emph{(i)} alter the viewing angle to $5^\circ$, \emph{(ii)} convolve our resultant images with a Gaussian beam indicative of the resolution of the GMVA, and \emph{(iii)} mimic  VLBI array sensitivity by introducing a Gaussian noise floor of $10^{-4}\,\text{Jy/beam}$. This results in a dynamic range of $\sim 1:100000$ in our synthetic images.
\Fig{38b} shows the total intensity of the jet's emission in the left column, the 
linearly
polarized intensity including EVPAs in the
middle column and the circular polarization in the right column. Again, the
different rows represent the three magnetic field morphologies. The top row depicts purely poloidal magnetic field, the middle one
the helical field, and the bottom one the purely toroidal magnetic field. All simulations are viewed edge-on to the jet axis in \fig{38b}.

In total intensity the poloidal field case exhibits a bright central radio core whereas, in contrast, the helical and toroidal field cases exhibit emission peaks offset from the central shock. Also, the helical and toroidal field cases exhibit two signs in circular polarization whereas the poloidal field case exhibits only one. In all cases, the linear polarized emission peaks are offset from the total intensity peaks which is commonly seen in blazars.

Again we computed integrated levels of fractional linear and circular polarization ($\bar{m_l}$ \& $\bar{m_c}$, listed 
to the lower right in the linear polarization and circular polarization images of \fig{38b}). In case of $i=5^\circ$ and in contrast to the calculated values at $45^\circ$, the fractional linear polarization increases from the poloidal, to the
helical, to the toroidal magnetic field model ( from $\sim 2.0\,\%$ to $\sim 6.0\,\%$). The fractional circular polarization decreases and changes sign from the poloidal to toroidal magnetic field model (from $\sim 1.5\,\%$ to $\sim -3.8\times10^{-2}\,\%$). The fractional circular polarization peaks for the helical magnetic field at $\sim 2.7\,\%$.

Here, the most striking result is that \figs{38a}{38b} demonstrate that resolved circular polarization imaging of relativistic jets has the potential to distinguish between a purely poloidal or purely toroidal magnetic field configuration within standing/recollimation shocks.

\section{Emission recipe study}\label{4}
\subsection{Emission Recipe Prescriptions}
We now shift our focus to better understanding the impact that 
different 
electron scaling relations have on the synchrotron 
polarization produced in our RMHD jet simulations.

In particular, we explore three additional scaling relations to account for the jet's micro physics by adopting the methods presented in~\cite{Porth2011}. In particular, we set the 
NTEs energy distribution proportional to: \emph{(i)} the fluid density, 
\emph{(ii)} the thermal pressure, and \emph{(iii)} the magnetic energy density:
\begin{align}\label{41}
\begin{split}
&\int_{\gamma_\text{min}}^{\gamma_\text{max}} \text{d}\gamma n_e(\gamma) = 
\frac{\rho}{m_p},\\
&\int_{\gamma_\text{min}}^{\gamma_\text{max}}\text{d}\gamma n_e(\gamma) \gamma m_ec^2 
=  \frac{p}{\left(\hat{\gamma}-1\right)},\\
&\int_{\gamma_\text{min}}^{\gamma_\text{max}} \text{d}\gamma n_e(\gamma) \gamma 
m_e 
c^2 = \epsilon_B \frac{B^2}{8\pi\cdot\left(\hat{\gamma}-1\right)}.
\end{split}
\end{align}
In this treatment we set the conversion factors $\zeta_e = \epsilon_e = \epsilon_B = 1$ (see \sect{emission1}), where $\epsilon_B$ is the equipartition fraction.
Again, $n_e(\gamma)$ takes the form of a power-law, see \eq{ne}.

Instead of solving the terms in \eq{41} for two unknowns, we assume 
the bounds of the electron power-law ($\gamma_\text{min}$ \& 
$\gamma_\text{max}$) to be fixed. The lower cutoff for injected NTEs is 
set to $\gamma_\text{min}=10$ and the upper limit to: 
$\gamma_\text{max}=10^6\cdot\gamma_\text{min}$ (see~\eq{ne}).We solve each 
equation in \eq{41} for $n_0$ as a function of either $p$, $\rho$, or $B$:

\begin{align}\label{234}
\begin{split}
n_0 &= \frac{\rho 
	(1-s)}{m_p\gamma_\text{min}\left(\left(10^6\right)^{1-s}-1\right)}.\\
n_0 &= \frac{p(2-s)}{\left(\hat{\gamma}-1\right)\gamma_\text{min}^2m_e c^2
	\left(\left(10^6\right)^{2-s}-1\right)}\\
n_0 &= \frac{\epsilon_B B^2(2-s)}{8\pi\left(\hat{\gamma}-1\right)
	\gamma_\text{min}^2m_e c^2\left(\left(10^6\right)^{2-s}-1\right)}.
\end{split}
\end{align}
Again, we set the electron power-law index to $s=2.3$ (hence  $\alpha=0.65$).

\subsection{Results}\label{sec:resporth}
\begin{figure*}
	\centering
	\begin{tikzpicture}
	\small
	\node[anchor=south west,inner sep=0] (Fig1) at (0,0) 
	{\includegraphics[width=0.31\textwidth]{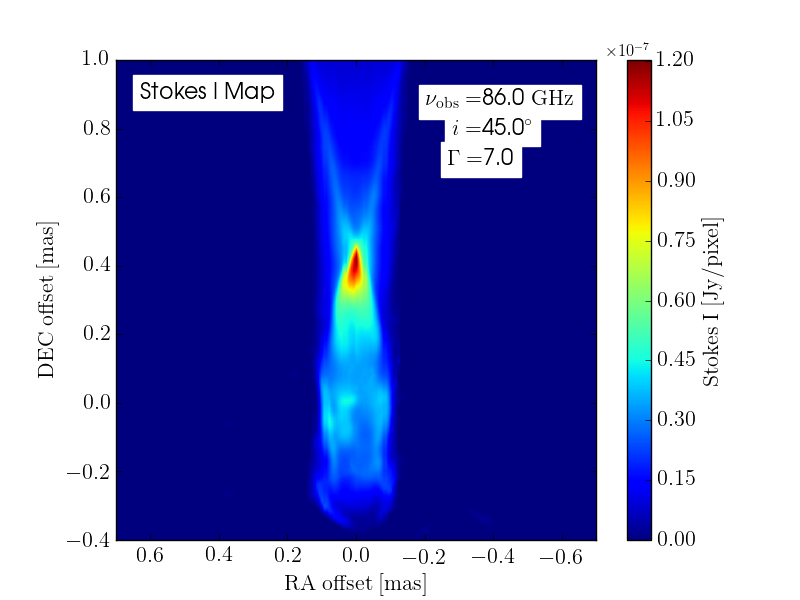}};
	\node[anchor=south west, inner sep=0] (Fig2) at (Fig1.south east) 
	{\includegraphics[width=0.31\textwidth]{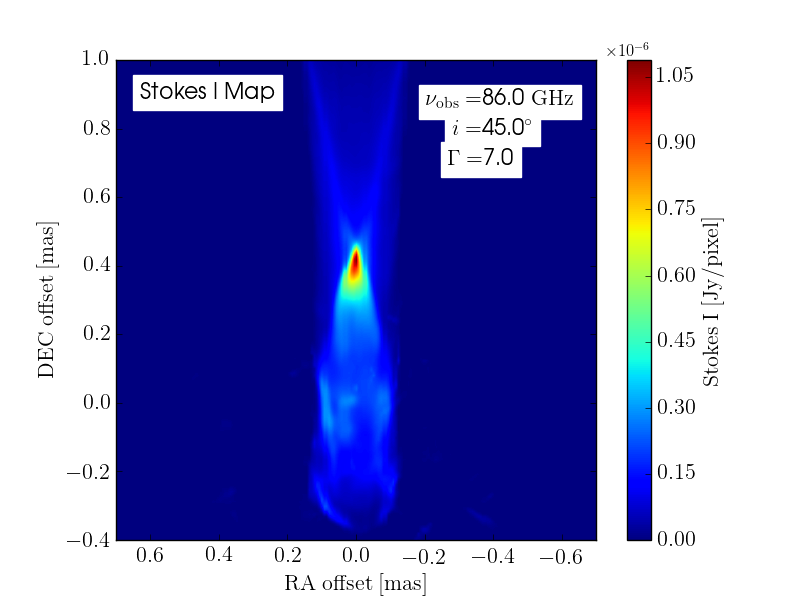}};
	\node[anchor=south west, inner sep=0] (Fig3) at (Fig2.south east) 
	{\includegraphics[width=0.31\textwidth]{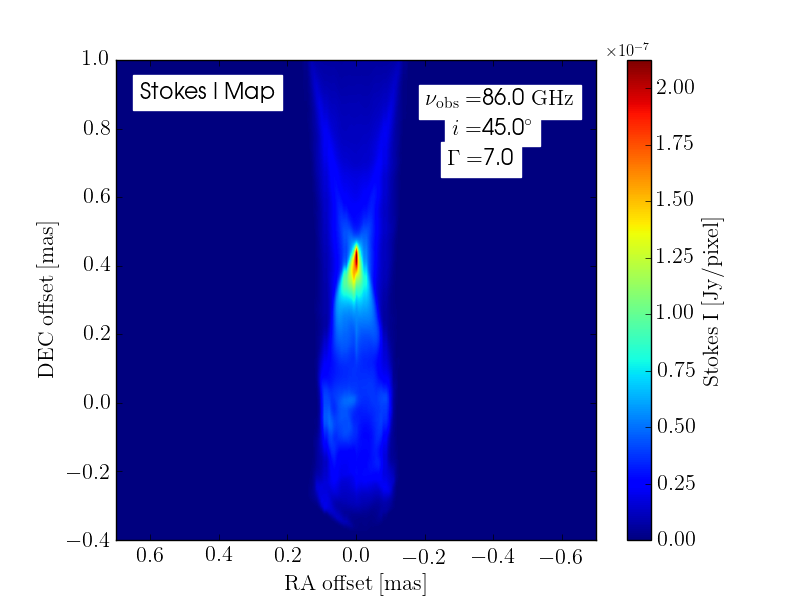}};
	
	\node[anchor=north west, inner sep=0] (Fig4) at (Fig1.south west) 
	{\includegraphics[width=0.31\textwidth]{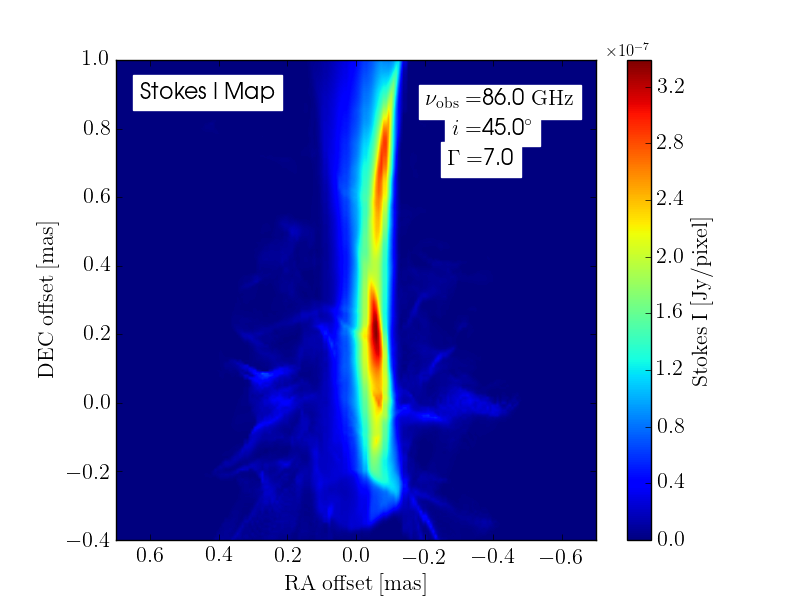}};
	\node[anchor=south west, inner sep=0] (Fig5) at (Fig4.south east) 
	{\includegraphics[width=0.31\textwidth]{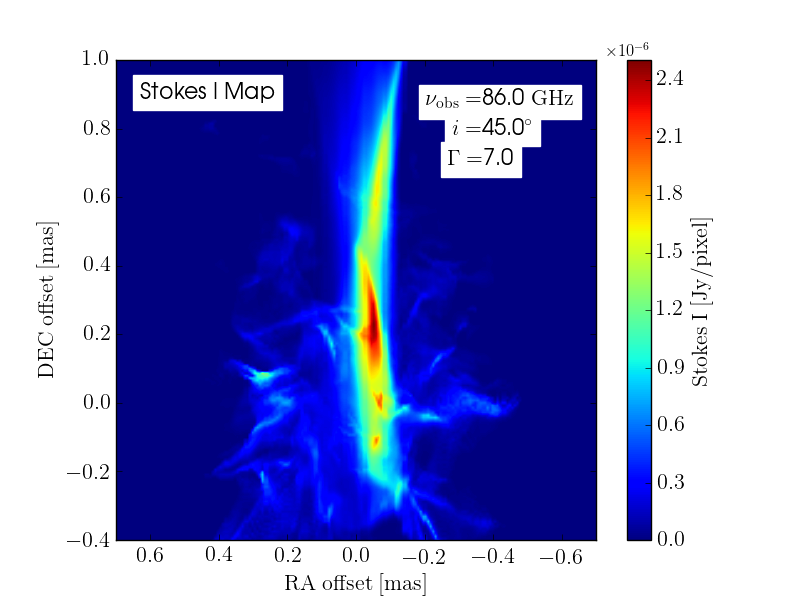}};
	\node[anchor=south west, inner sep=0] (Fig6) at (Fig5.south east) 
	{\includegraphics[width=0.31\textwidth]{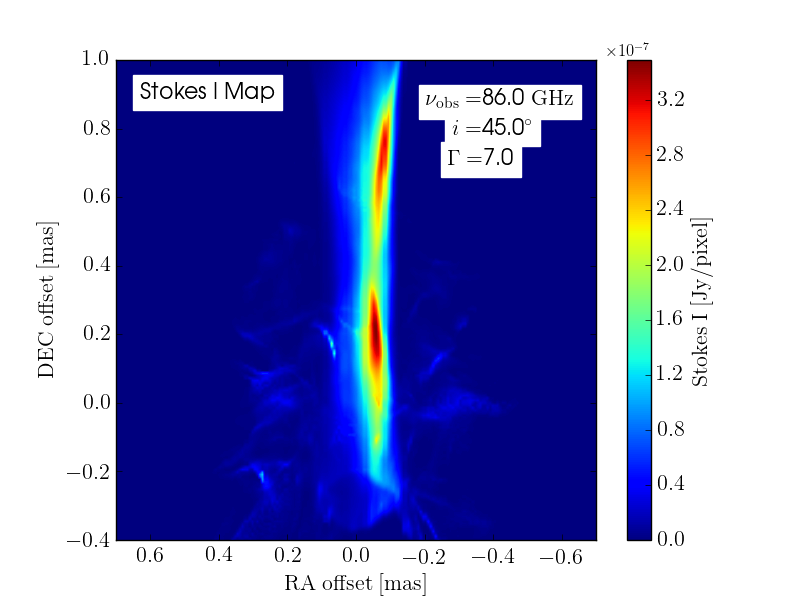}};
	
	\node[anchor=north west, inner sep=0] (Fig7) at (Fig4.south west) 
	{\includegraphics[width=0.31\textwidth]{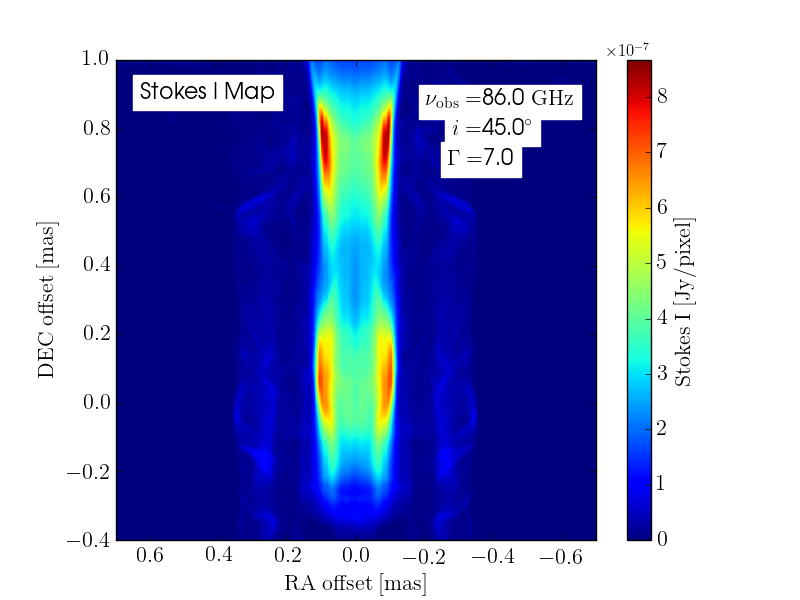}};
	\node[anchor=south west, inner sep=0] (Fig8) at (Fig7.south east) 
	{\includegraphics[width=0.31\textwidth]{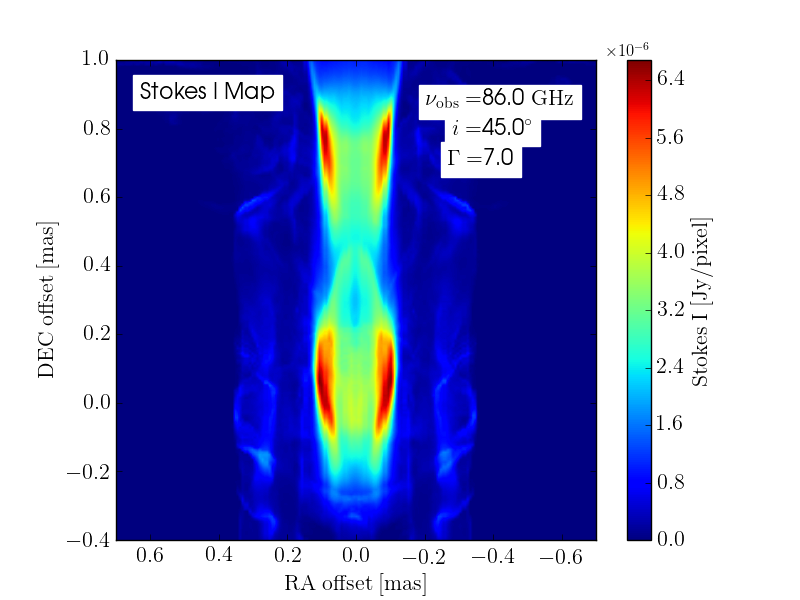}};
	\node[anchor=south west, inner sep=0] (Fig9) at (Fig8.south east) 
	{\includegraphics[width=0.31\textwidth]{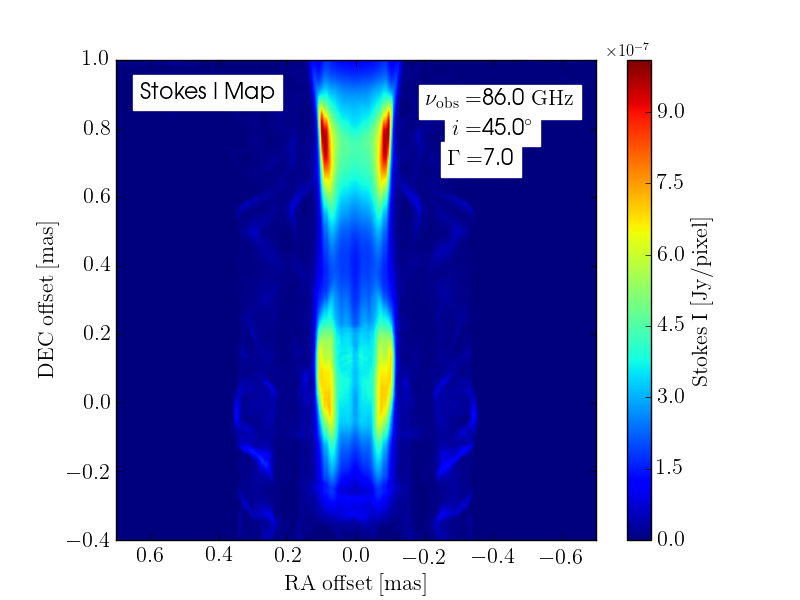}};
	
	\draw[semithick] (Fig9.south east) -- 
	node[anchor=north,rotate=90] {\emph{(iii)} Toroidal} (Fig9.north east) --
	(Fig6.south east) --
	node[anchor=north,rotate=90] {\emph{(ii)} Helical} (Fig6.north east) --
	(Fig3.south east) --
	node[anchor=north,rotate=90] {\emph{(i)} Poloidal} (Fig3.north east);
	\draw[semithick] ([yshift=-6pt]Fig7.south west) -- 
	node[below, xshift=-6pt] {(1) $n_e(\gamma) \propto \rho$} 
	([yshift=-6pt]Fig7.south east) --
	([yshift=-6pt]Fig8.south west) -- 
	node[below, xshift=-6pt] {(2) $n_e(\gamma) \propto p$} 
	([yshift=-6pt]Fig8.south east) -- 
	([yshift=-6pt]Fig9.south west) -- 
	node[below, xshift=-6pt] {(3) $n_e(\gamma) \propto B^2$} 
	([yshift=-6pt]Fig9.south east);
	
	\end{tikzpicture}
	
	\caption[Synthetic emission maps: Total intensity for three 
	different magnetic fields and emission recipes]{Total intensity maps 
		of our 3D RMHD jet simulations using 
		different electron scaling relations. The jet is viewed at $i=45^\circ$ and 
		propagates from top to bottom in each frame. The images highlight similar epochs in the jet's evolution of each 3D RMHD simulation at $86\,$GHz. From left to right: 
		proportionality of the NTEs to the 
		fluid's 
		density, internal energy density, and magnetic energy density. From top 
		to bottom: 
		purely 
		poloidal magnetic field, helical magnetic field, and purely toroidal 
		magnetic field. }
	\label{fig:41a}
\end{figure*}

Based on the solutions presented in \eq{234} we generate synthetic maps of total 
intensity, linearly polarized intensity, circular polarization, and calculate integrated fractional 
levels of polarization.
\Fig{fig:41a} presents resolved total intensity images computed using the three emission 
recipes (see \eq{41}). The top row illustrates the poloidal magnetic field, the middle row 
illustrates
the helical field, and the bottom row illustrates the toroidal magnetic field. 
Here,
we analyze the dependence on the electron scaling relations while viewing the jet at $i=45^\circ$. The left column shows images produced using the density scaling relation, the middle column show images produced using the pressure scaling relation, and the right column shows images produced using the magnetic energy density scaling relation.
We see that the three different scaling relations do
not have a drastic impact on the resultant jet emission when using a jet tracer to exclude the ambient medium.
Similar to our findings in \sect{resfromm}, we see an edge-brightened jet in the toroidal field case and a spine/shock-brightened jet in the poloidal field case for all three electron scaling relations.

\begin{figure*}
	\centering
	\begin{tikzpicture}
	\small
	\node[anchor=south west,inner sep=0] (Fig1) at (0,0) 
	{\includegraphics[width=0.31\textwidth]{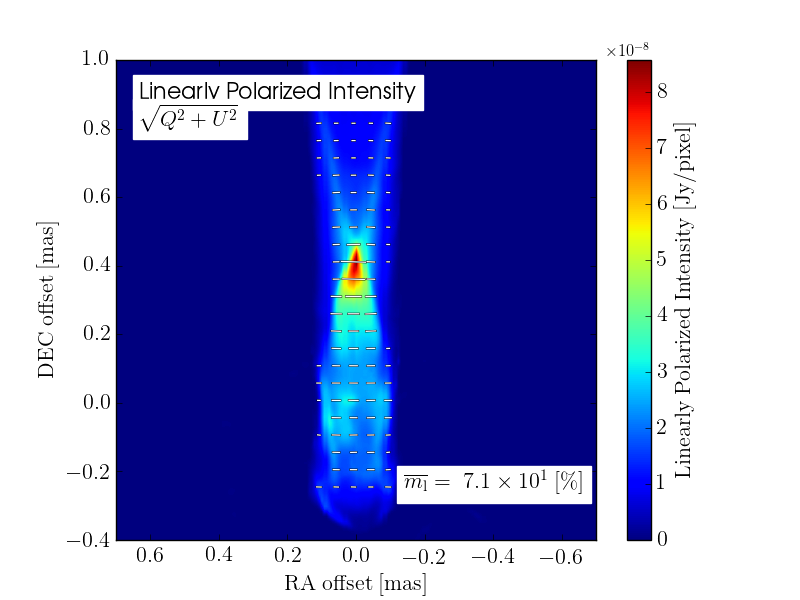}};
	\node[anchor=south west, inner sep=0] (Fig2) at (Fig1.south east) 
	{\includegraphics[width=0.31\textwidth]{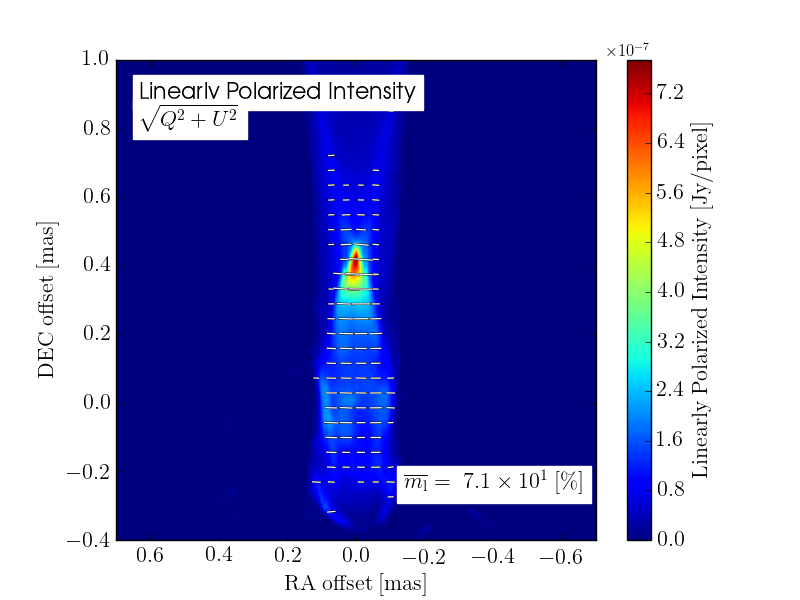}};
	\node[anchor=south west, inner sep=0] (Fig3) at (Fig2.south east) 
	{\includegraphics[width=0.31\textwidth]{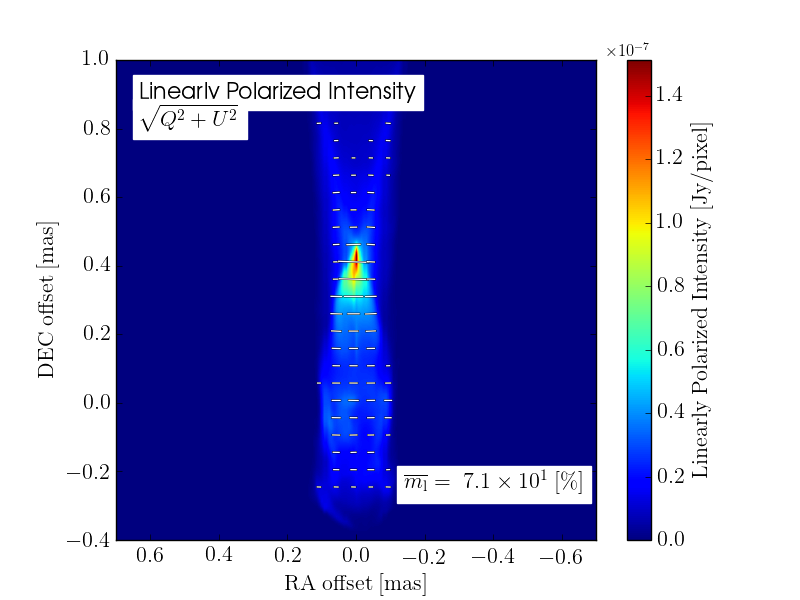}};
	
	\node[anchor=north west, inner sep=0] (Fig4) at (Fig1.south west) 
	{\includegraphics[width=0.31\textwidth]{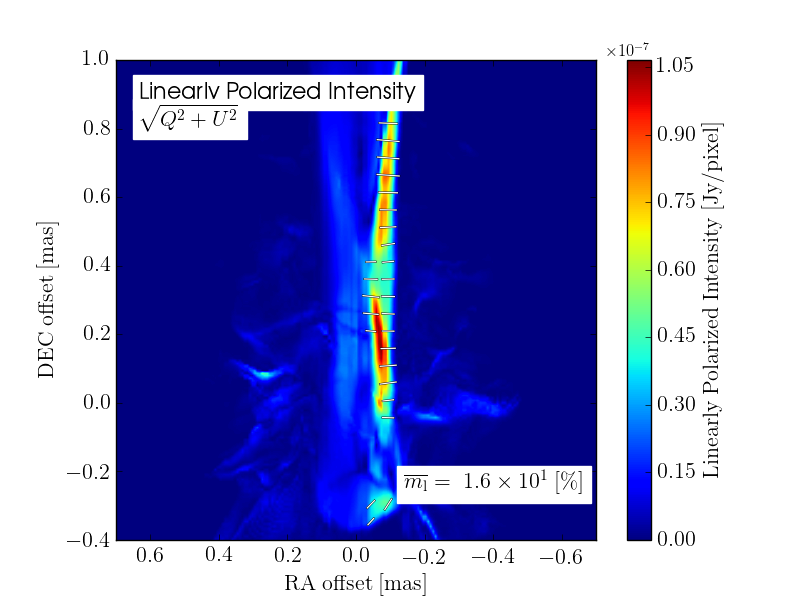}};
	\node[anchor=south west, inner sep=0] (Fig5) at (Fig4.south east) 
	{\includegraphics[width=0.31\textwidth]{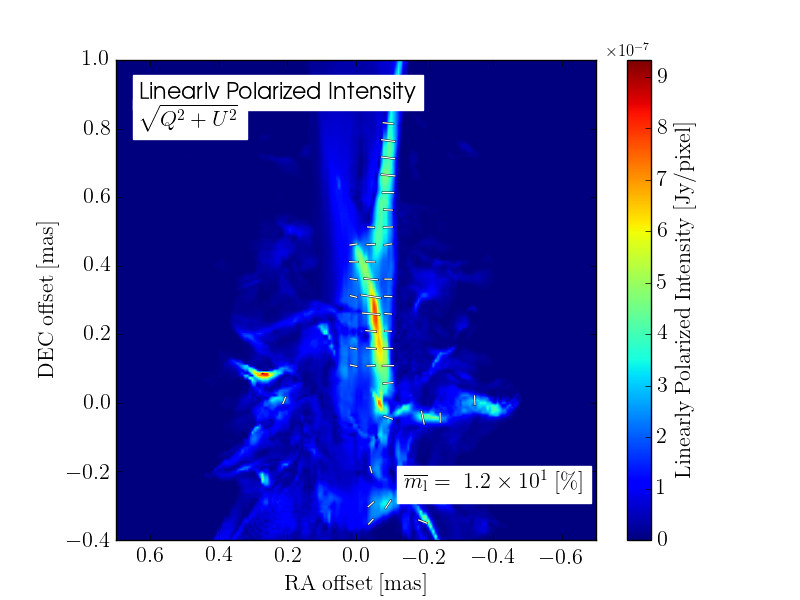}};
	\node[anchor=south west, inner sep=0] (Fig6) at (Fig5.south east) 
	{\includegraphics[width=0.31\textwidth]{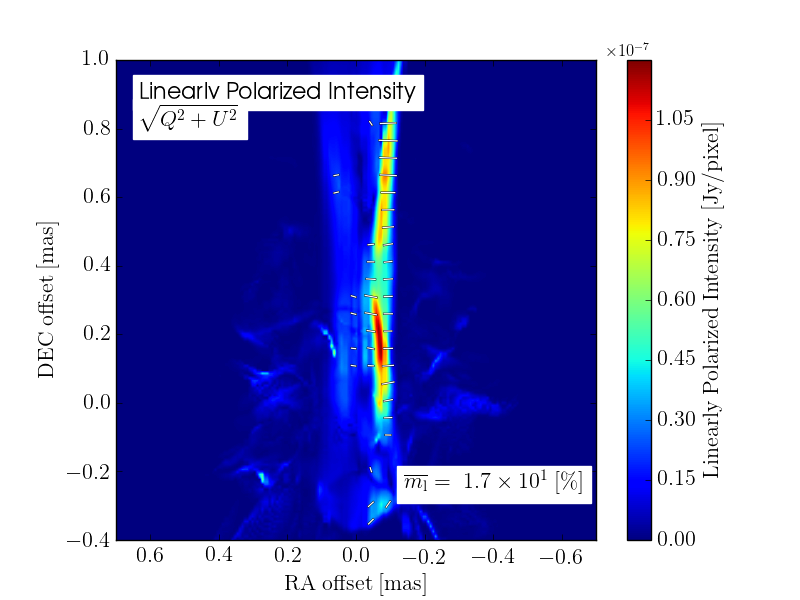}};
	
	\node[anchor=north west, inner sep=0] (Fig7) at (Fig4.south west) 
	{\includegraphics[width=0.31\textwidth]{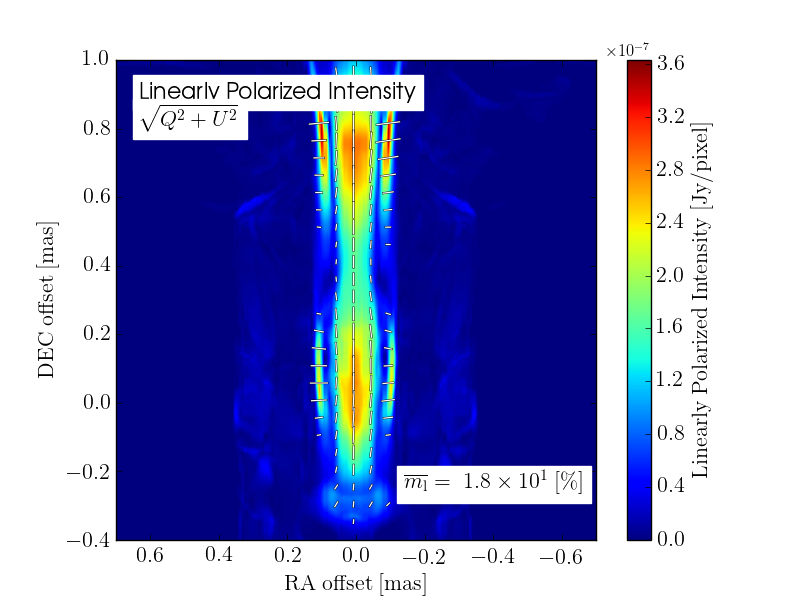}};
	\node[anchor=south west, inner sep=0] (Fig8) at (Fig7.south east) 
	{\includegraphics[width=0.31\textwidth]{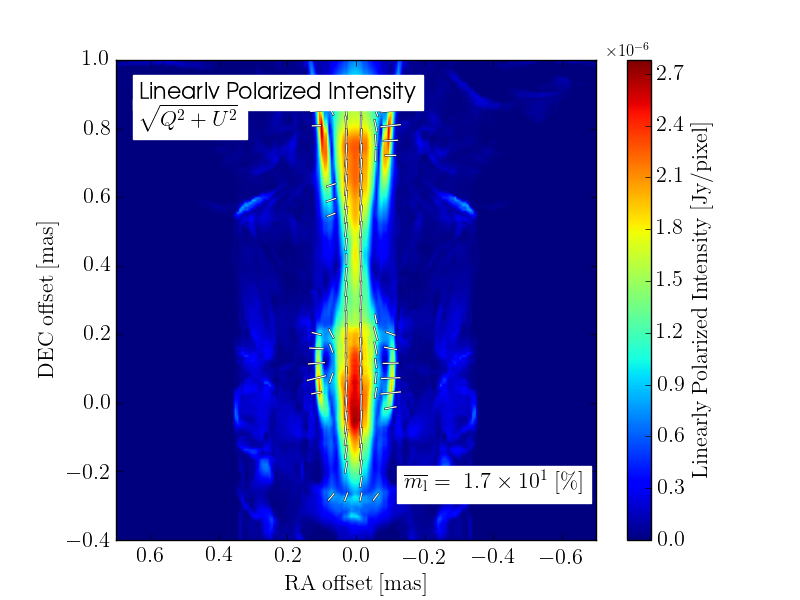}};
	\node[anchor=south west, inner sep=0] (Fig9) at (Fig8.south east) 
	{\includegraphics[width=0.31\textwidth]{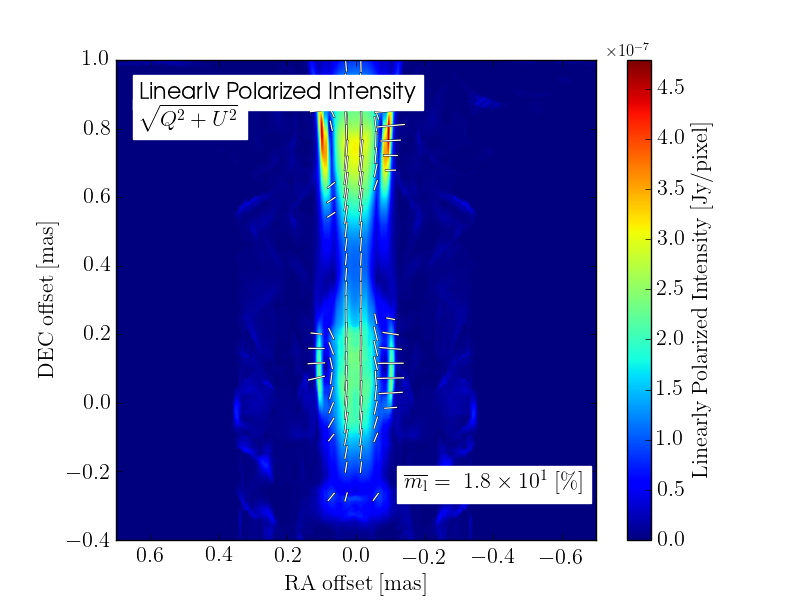}};
	
	\draw[semithick] (Fig9.south east) -- 
	node[anchor=north,rotate=90] {\emph{(iii)} Toroidal} (Fig9.north east) --
	(Fig6.south east) --
	node[anchor=north,rotate=90] {\emph{(ii)} Helical} (Fig6.north east) --
	(Fig3.south east) --
	node[anchor=north,rotate=90] {\emph{(i)} Poloidal} (Fig3.north east);
	\draw[semithick] ([yshift=-6pt]Fig7.south west) -- 
	node[below, xshift=-6pt] {(1) $n_e(\gamma) \propto \rho$} 
	([yshift=-6pt]Fig7.south east) --
	([yshift=-6pt]Fig8.south west) -- 
	node[below, xshift=-6pt] {(2) $n_e(\gamma) \propto p$} 
	([yshift=-6pt]Fig8.south east) -- 
	([yshift=-6pt]Fig9.south west) -- 
	node[below, xshift=-6pt] {(3) $n_e(\gamma) \propto B^2$} 
	([yshift=-6pt]Fig9.south east);
	
	\end{tikzpicture}
	
	\caption[Synthetic emission maps: Total intensity for three 
	different magnetic fields and emission recipes]{Linearly polarized emission maps 
		of our 3D RMHD jet simulations using 
		different emission electron scaling relations. The jet is viewed at $i=45^\circ$ and 
		propagates from top to bottom in each frame. The images highlight similar epochs in the jet's evolution of each 3D RMHD simulation at $86\,$GHz. From left to right: 
		proportionality of the NTEs to the 
		fluid's 
		density, internal energy density, and magnetic energy density. From top 
		to bottom: 
		purely 
		poloidal magnetic field, helical magnetic field, and purely toroidal 
		magnetic field. }
	\label{fig:41c}
\end{figure*}

\Fig{fig:41c} shows
the corresponding linear polarization maps with EVPAs overplotted. The EVPA orientations are similar to the orientations illustrated in
\sect{resfromm}. 

\begin{figure*}
	\centering
	\begin{tikzpicture}
	\small
	\node[anchor=south west,inner sep=0] (Fig1) at (0,0) 
	{\includegraphics[width=0.31\textwidth]{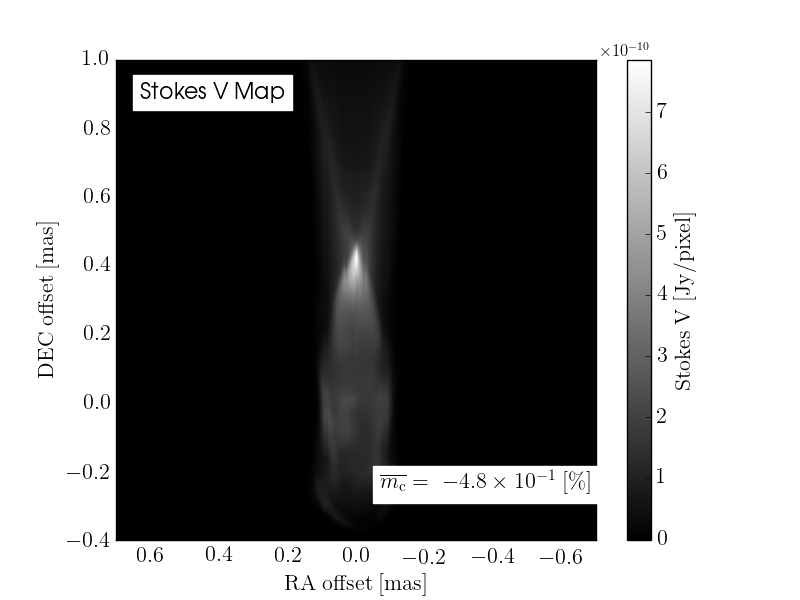}};
	\node[anchor=south west, inner sep=0] (Fig2) at (Fig1.south east) 
	{\includegraphics[width=0.31\textwidth]{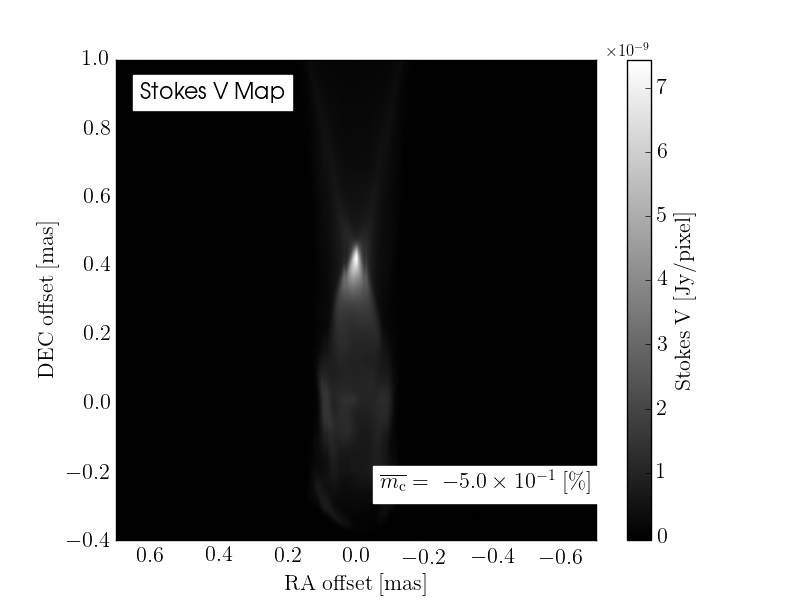}};
	\node[anchor=south west, inner sep=0] (Fig3) at (Fig2.south east) 
	{\includegraphics[width=0.31\textwidth]{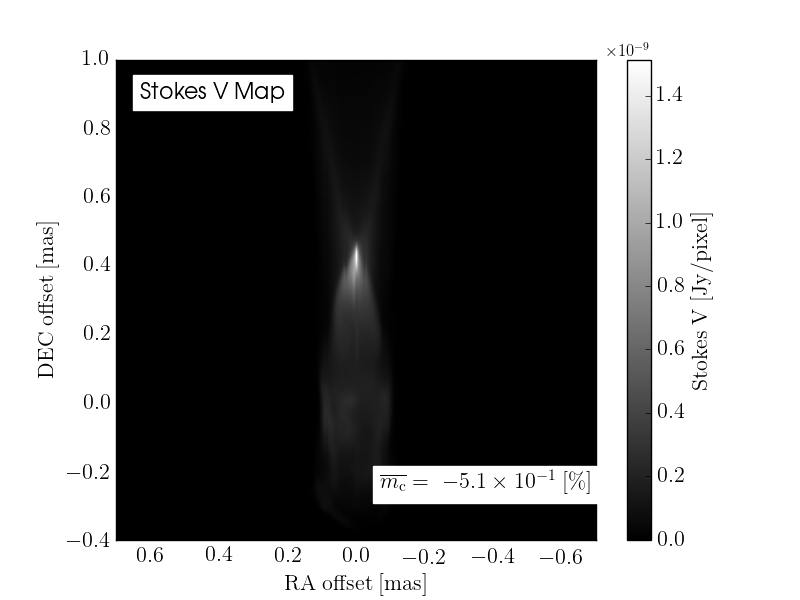}};
	
	\node[anchor=north west, inner sep=0] (Fig4) at (Fig1.south west) 
	{\includegraphics[width=0.31\textwidth]{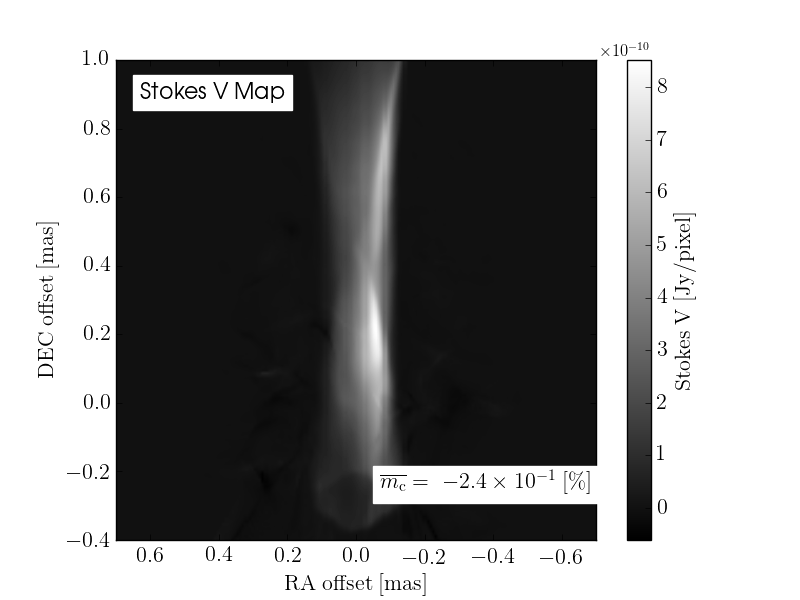}};
	\node[anchor=south west, inner sep=0] (Fig5) at (Fig4.south east) 
	{\includegraphics[width=0.31\textwidth]{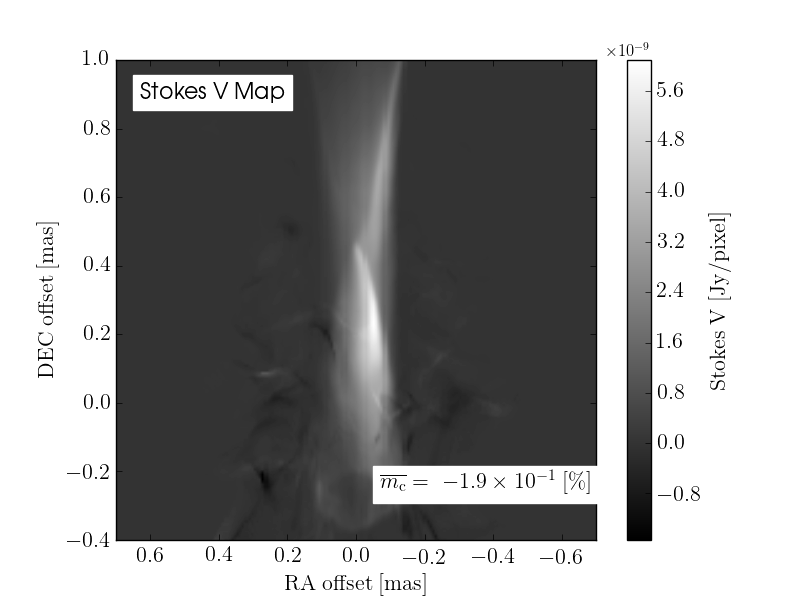}};
	\node[anchor=south west, inner sep=0] (Fig6) at (Fig5.south east) 
	{\includegraphics[width=0.31\textwidth]{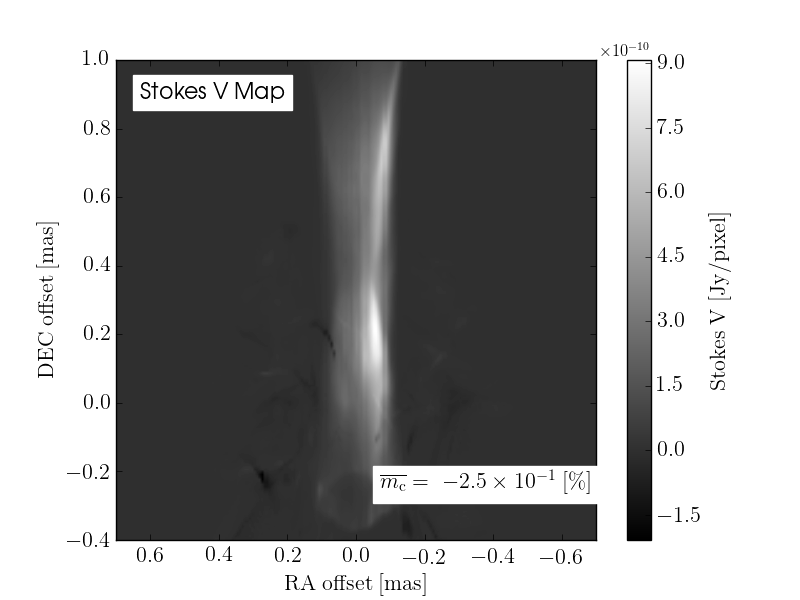}};
	
	\node[anchor=north west, inner sep=0] (Fig7) at (Fig4.south west) 
	{\includegraphics[width=0.31\textwidth]{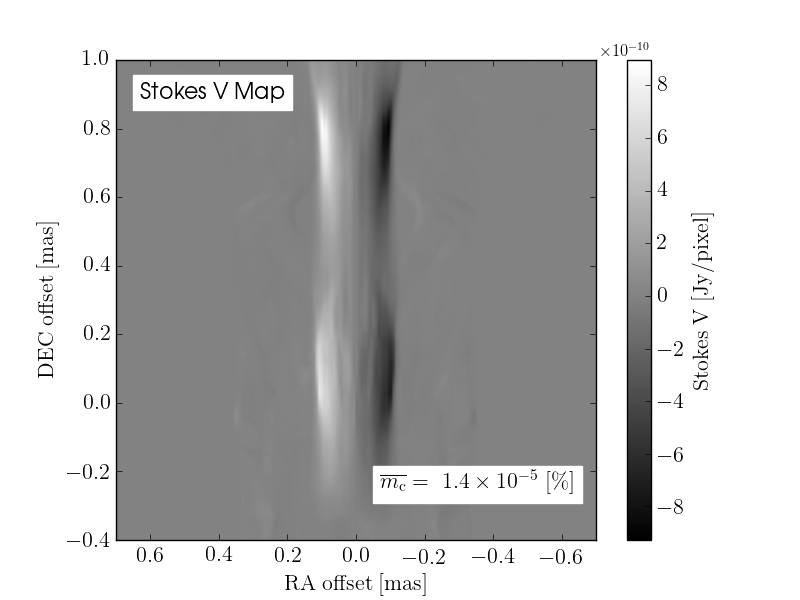}};
	\node[anchor=south west, inner sep=0] (Fig8) at (Fig7.south east) 
	{\includegraphics[width=0.31\textwidth]{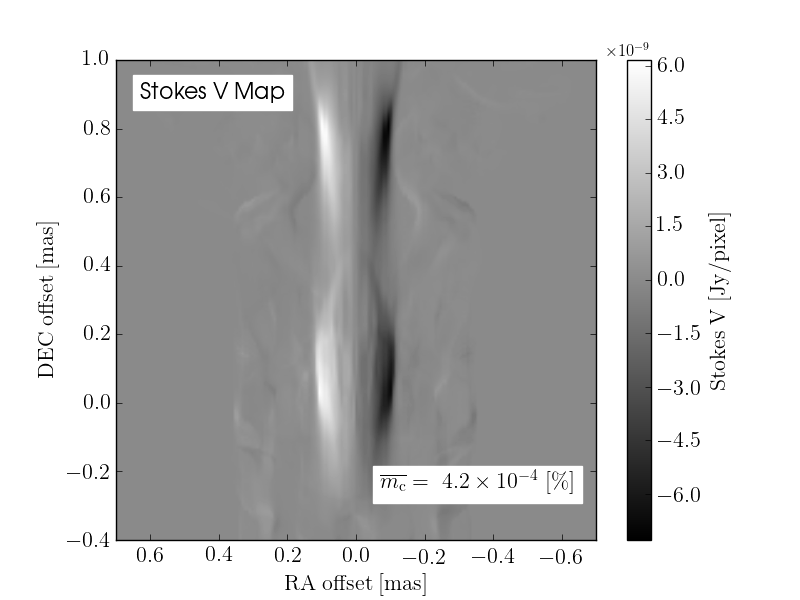}};
	\node[anchor=south west, inner sep=0] (Fig9) at (Fig8.south east) 
	{\includegraphics[width=0.31\textwidth]{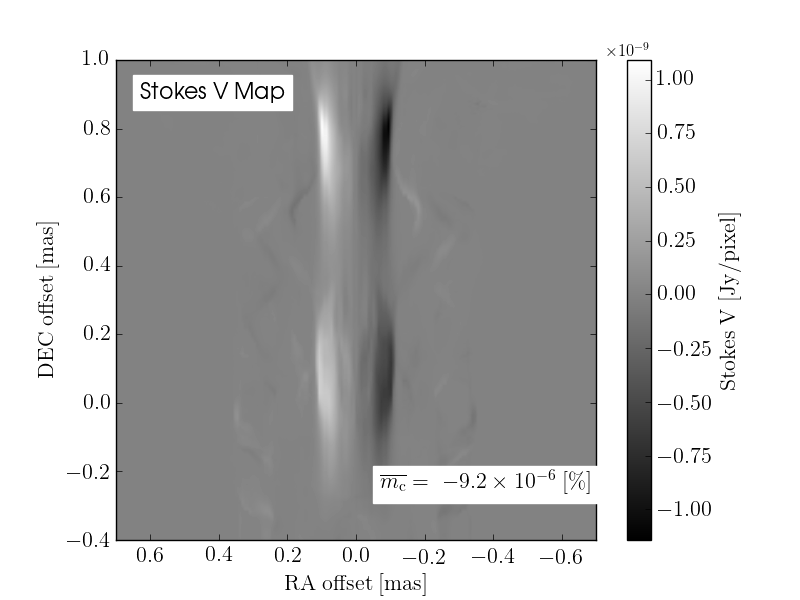}};
	
	\draw[semithick] (Fig9.south east) -- 
	node[anchor=north,rotate=90] {\emph{(iii)} Toroidal} (Fig9.north east) --
	(Fig6.south east) --
	node[anchor=north,rotate=90] {\emph{(ii)} Helical} (Fig6.north east) --
	(Fig3.south east) --
	node[anchor=north,rotate=90] {\emph{(i)} Poloidal} (Fig3.north east);
	\draw[semithick] ([yshift=-6pt]Fig7.south west) -- 
	node[below, xshift=-6pt] {(1) $n_e(\gamma) \propto \rho$} 
	([yshift=-6pt]Fig7.south east) --
	([yshift=-6pt]Fig8.south west) -- 
	node[below, xshift=-6pt] {(2) $n_e(\gamma) \propto p$} 
	([yshift=-6pt]Fig8.south east) -- 
	([yshift=-6pt]Fig9.south west) -- 
	node[below, xshift=-6pt] {(3) $n_e(\gamma) \propto B^2$} 
	([yshift=-6pt]Fig9.south east);
	
	\end{tikzpicture}
	
	\caption[Synthetic emission maps: Total intensity for three 
	different magnetic fields and emission recipes]{Circular polarization maps 
		of our 3D RMHD jet simulations using 
		different emission electron scaling relations. The jet is viewed at $i=45^\circ$ and 
		propagates from top to bottom in each frame. The images highlight similar epochs in the jet's evolution of each 3D RMHD simulation at $86\,$GHz. From left to right: 
		proportionality of the NTEs to the 
		fluid's 
		density, internal energy density, and magnetic energy density. From top 
		to bottom: 
		purely 
		poloidal magnetic field, helical magnetic field, and purely toroidal 
		magnetic field. }
	\label{fig:41b}
\end{figure*}

\Fig{fig:41b} shows the resulting circular polarization maps of the RMHD jet 
simulations for 
the three different electron scaling relations. The arrangement is the same as in \fig{fig:41a}. All three magnetic field morphologies show again only minor differences.
The polarized synchrotron emission is focused in the center 
of the jet, i.e., highlighting the recollimation shock. 
The helical magnetic field exhibits positive circular polarization on the right side of the jet. 
For the toroidal magnetic field, the circular polarized emission shows positive values on the left part of the 
jet and
negative values on the right (i.e., left \& right-handed). This is a reflection of the pitch angle present in the helical field case.

\begin{figure*}
	\centering
	\begin{tikzpicture}
	\small
	\node[anchor=south west,inner sep=0] (Fig1) at (0,0) 
	{\includegraphics[width=0.31\textwidth]{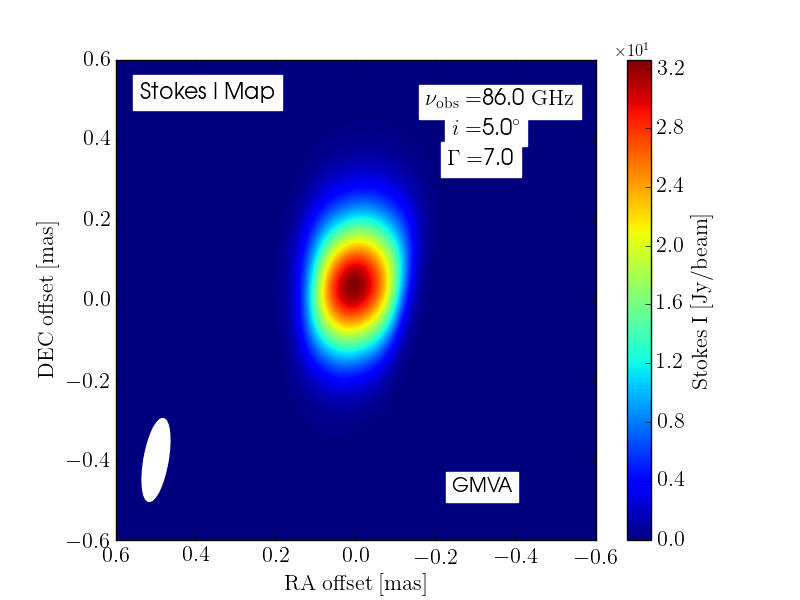}};
	\node[anchor=south west, inner sep=0] (Fig2) at (Fig1.south east) 
	{\includegraphics[width=0.31\textwidth]{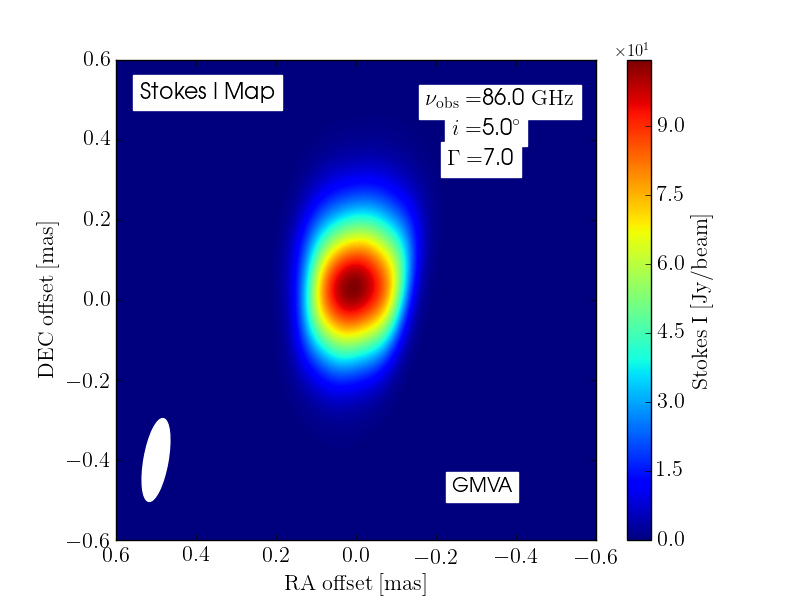}};
	\node[anchor=south west, inner sep=0] (Fig3) at (Fig2.south east) 
	{\includegraphics[width=0.31\textwidth]{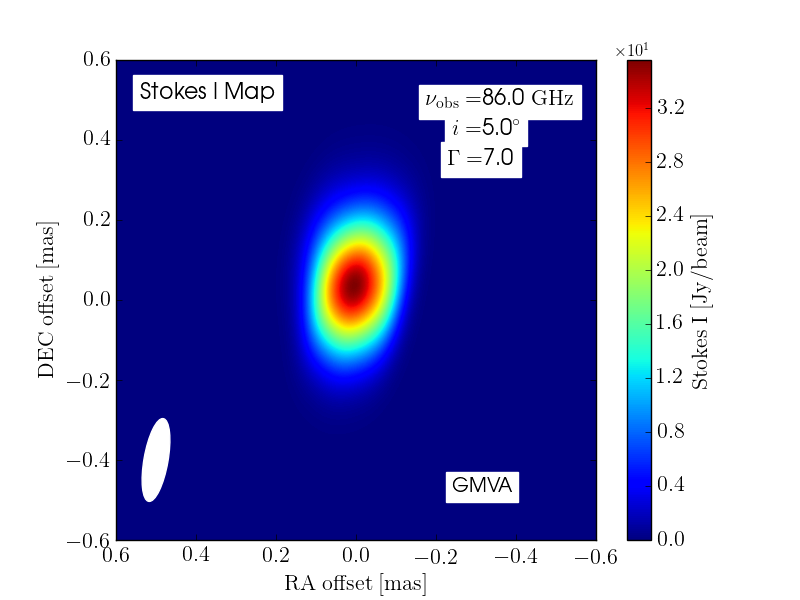}};
	
	\node[anchor=north west, inner sep=0] (Fig4) at (Fig1.south west) 
	{\includegraphics[width=0.31\textwidth]{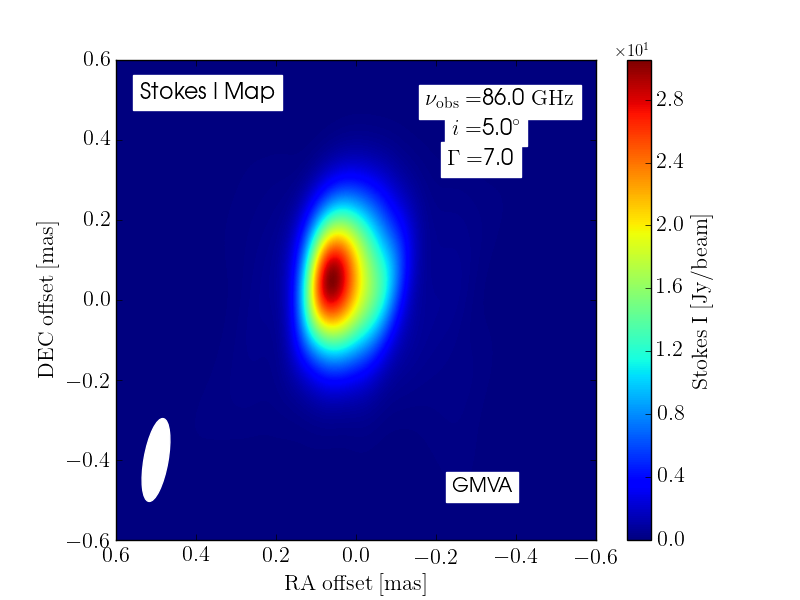}};
	\node[anchor=south west, inner sep=0] (Fig5) at (Fig4.south east) 
	{\includegraphics[width=0.31\textwidth]{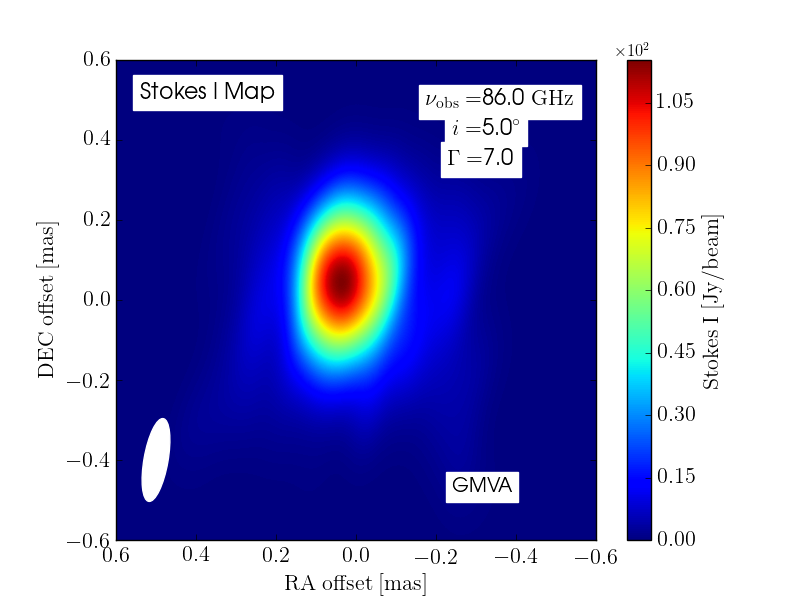}};
	\node[anchor=south west, inner sep=0] (Fig6) at (Fig5.south east) 
	{\includegraphics[width=0.31\textwidth]{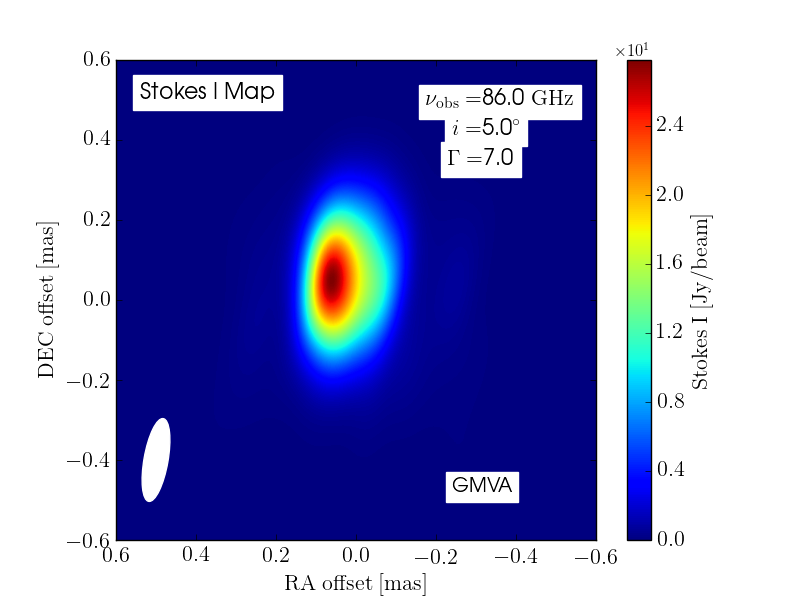}};
	
	\node[anchor=north west, inner sep=0] (Fig7) at (Fig4.south west) 
	{\includegraphics[width=0.31\textwidth]{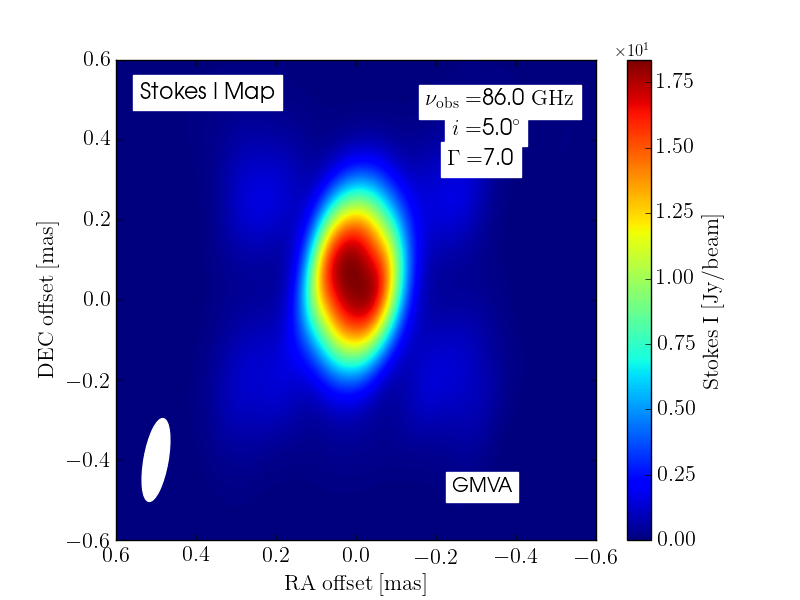}};
	\node[anchor=south west, inner sep=0] (Fig8) at (Fig7.south east) 
	{\includegraphics[width=0.31\textwidth]{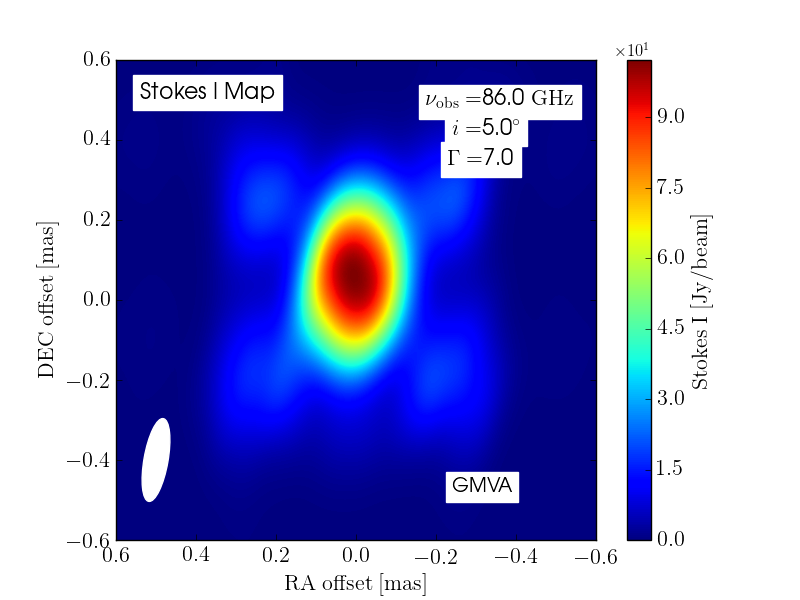}};
	\node[anchor=south west, inner sep=0] (Fig9) at (Fig8.south east) 
	{\includegraphics[width=0.31\textwidth]{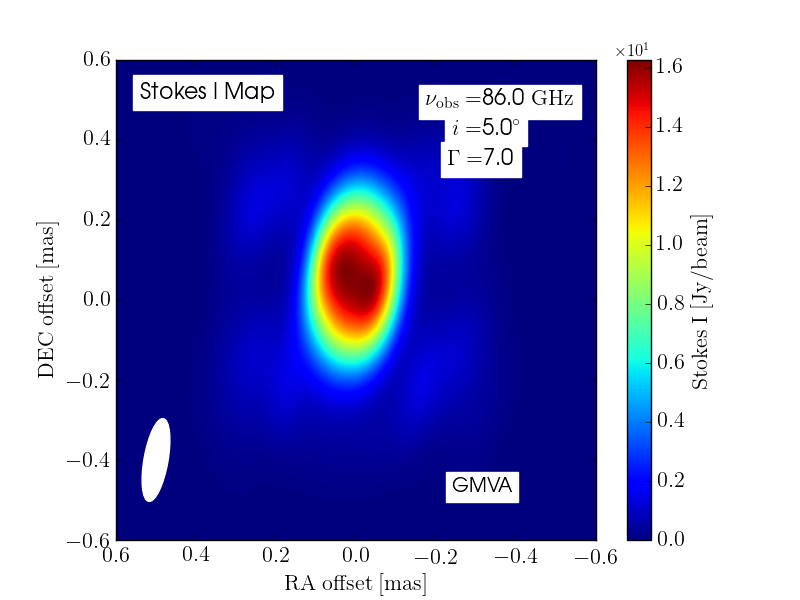}};
	
	\draw[semithick] (Fig9.south east) -- 
	node[anchor=north,rotate=90] {\emph{(iii)} Toroidal} (Fig9.north east) --
	(Fig6.south east) --
	node[anchor=north,rotate=90] {\emph{(ii)} Helical} (Fig6.north east) --
	(Fig3.south east) --
	node[anchor=north,rotate=90] {\emph{(i)} Poloidal} (Fig3.north east);
	\draw[semithick] ([yshift=-6pt]Fig7.south west) -- 
	node[below, xshift=-6pt] {(1) $n_e(\gamma) \propto \rho$} 
	([yshift=-6pt]Fig7.south east) --
	([yshift=-6pt]Fig8.south west) -- 
	node[below, xshift=-6pt] {(2) $n_e(\gamma) \propto p$} 
	([yshift=-6pt]Fig8.south east) -- 
	([yshift=-6pt]Fig9.south west) -- 
	node[below, xshift=-6pt] {(3) $n_e(\gamma) \propto B^2$} 
	([yshift=-6pt]Fig9.south east);
	
	\end{tikzpicture}
	
	\caption[Synthetic emission maps: Total intensity for three 
	different magnetic fields and emission recipes]{Ray-tracing images of our jets in total intensity when each jet is viewed edge-on. The images highlight similar epochs in the jet's evolution of each 3D RMHD simulation at $86\,$GHz. From left to right: 
		proportionality of the NTEs to the 
		fluid's 
		density, internal energy density, and magnetic energy density. From top 
		to bottom: 
		purely 
		poloidal magnetic field, helical magnetic field, and purely toroidal 
		magnetic field. The ray-traced images are convolved with a Gaussian beam indicative of the resolution of the GMVA and with a Gaussian noise floor of $10^{-4}\,\text{Jy/beam}$. }
	\label{fig:43a}
\end{figure*}

\begin{figure*}
	\centering
	\begin{tikzpicture}
	\small
	\node[anchor=south west,inner sep=0] (Fig1) at (0,0) 
	{\includegraphics[width=0.31\textwidth]{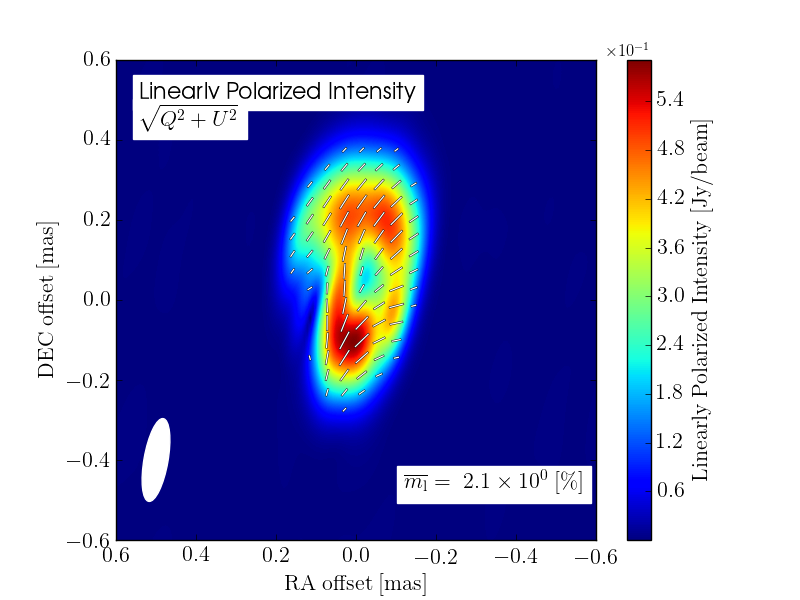}};
	\node[anchor=south west, inner sep=0] (Fig2) at (Fig1.south east) 
	{\includegraphics[width=0.31\textwidth]{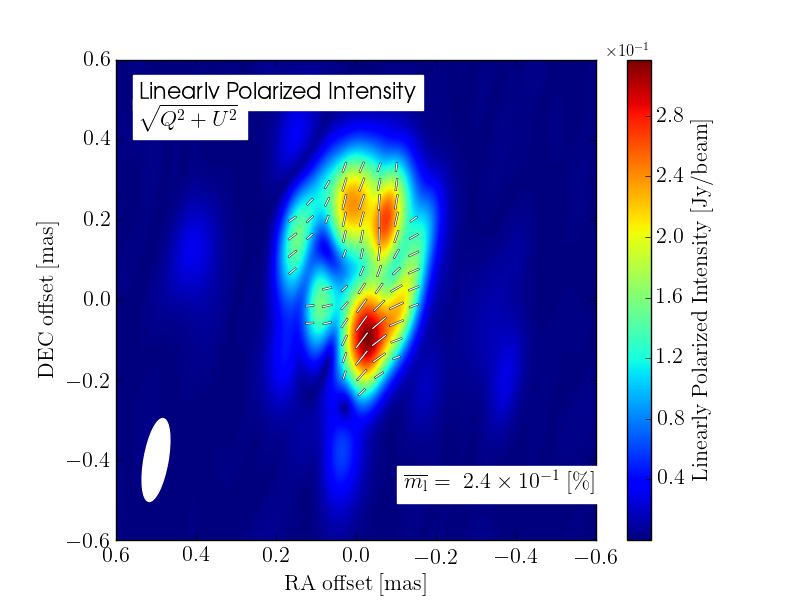}};
	\node[anchor=south west, inner sep=0] (Fig3) at (Fig2.south east) 
	{\includegraphics[width=0.31\textwidth]{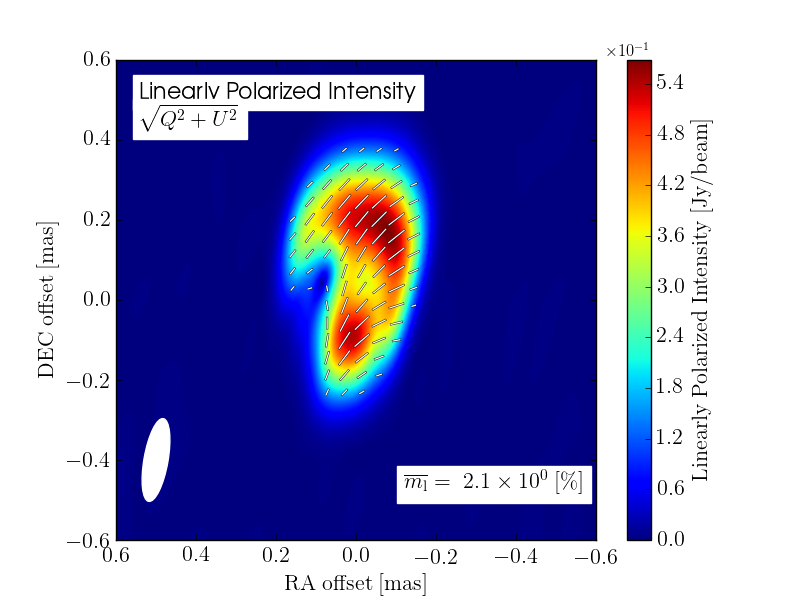}};
	
	\node[anchor=north west, inner sep=0] (Fig4) at (Fig1.south west) 
	{\includegraphics[width=0.31\textwidth]{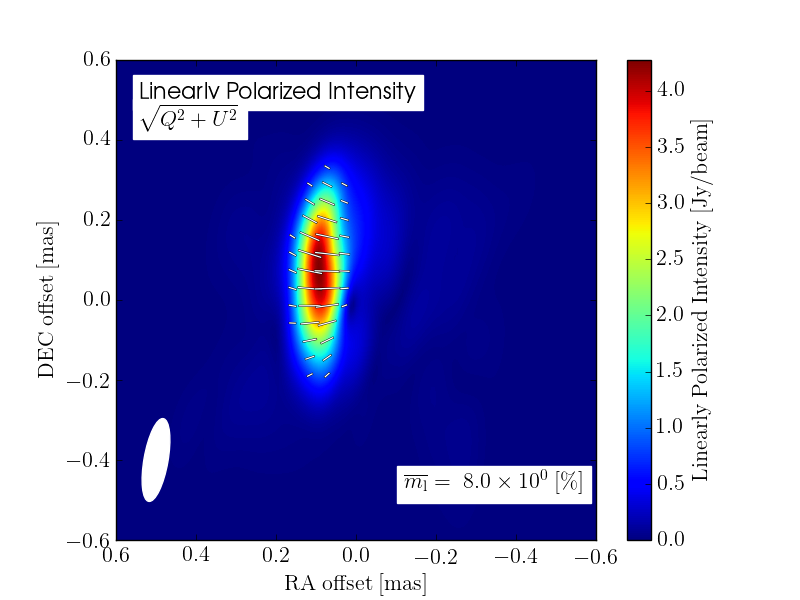}};
	\node[anchor=south west, inner sep=0] (Fig5) at (Fig4.south east) 
	{\includegraphics[width=0.31\textwidth]{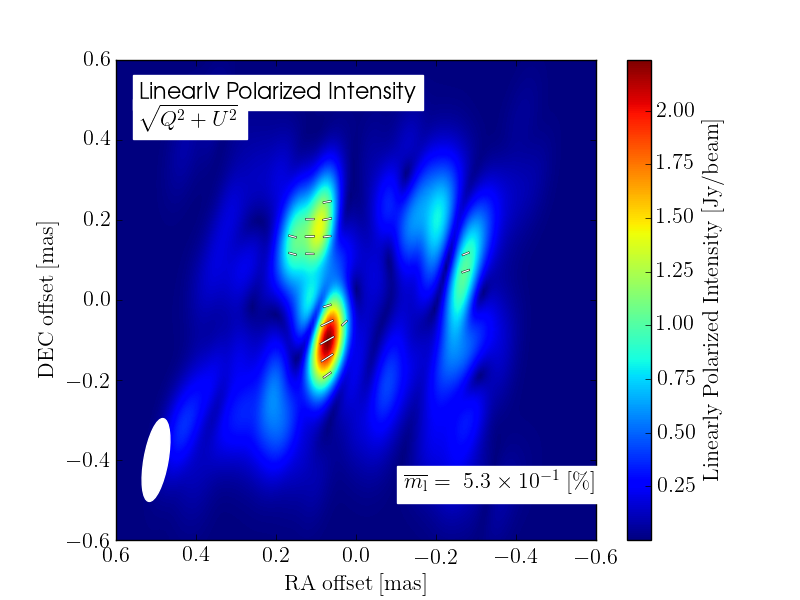}};
	\node[anchor=south west, inner sep=0] (Fig6) at (Fig5.south east) 
	{\includegraphics[width=0.31\textwidth]{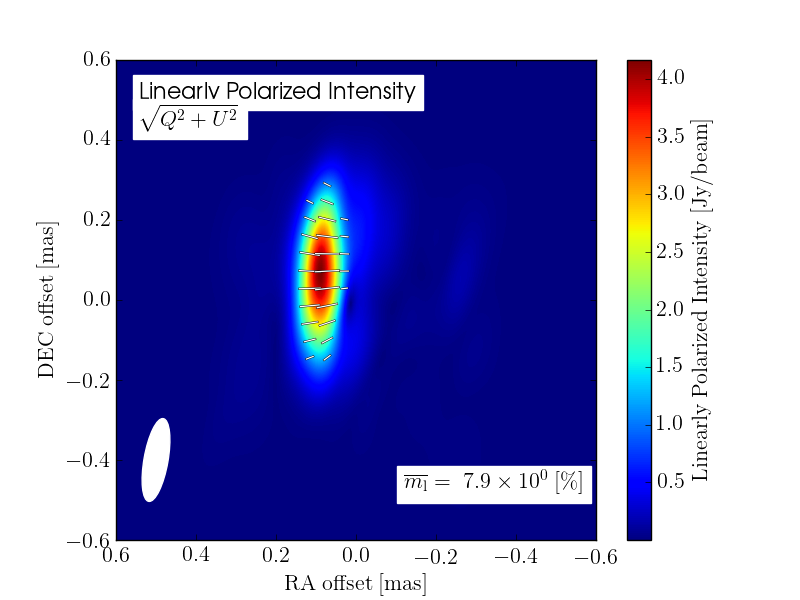}};
	
	\node[anchor=north west, inner sep=0] (Fig7) at (Fig4.south west) 
	{\includegraphics[width=0.31\textwidth]{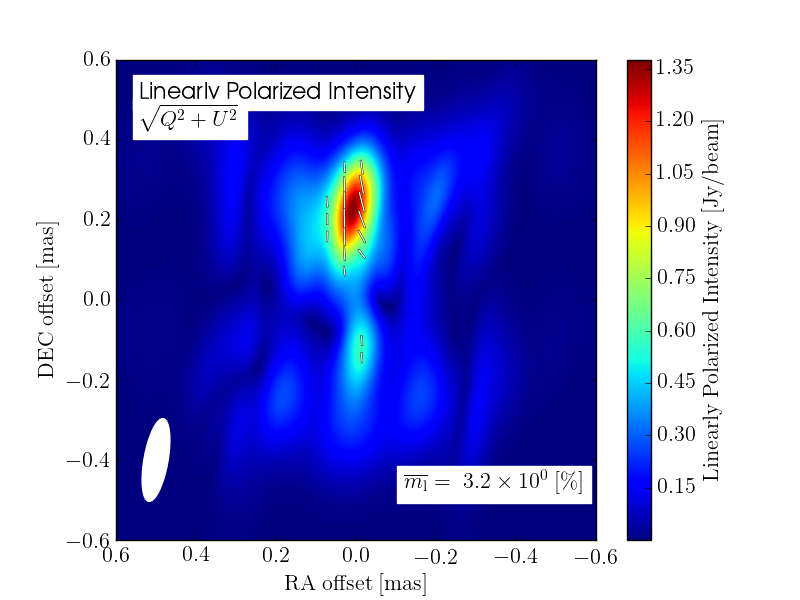}};
	\node[anchor=south west, inner sep=0] (Fig8) at (Fig7.south east) 
	{\includegraphics[width=0.31\textwidth]{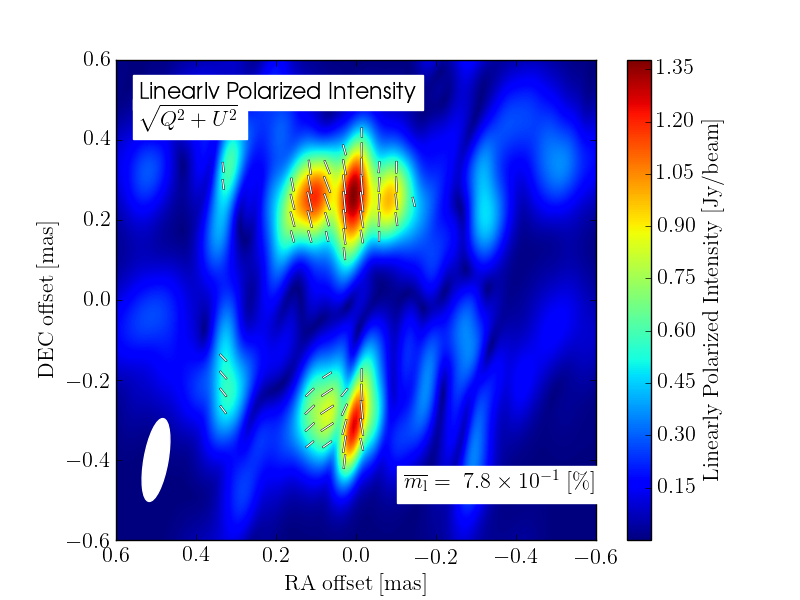}};
	\node[anchor=south west, inner sep=0] (Fig9) at (Fig8.south east) 
	{\includegraphics[width=0.31\textwidth]{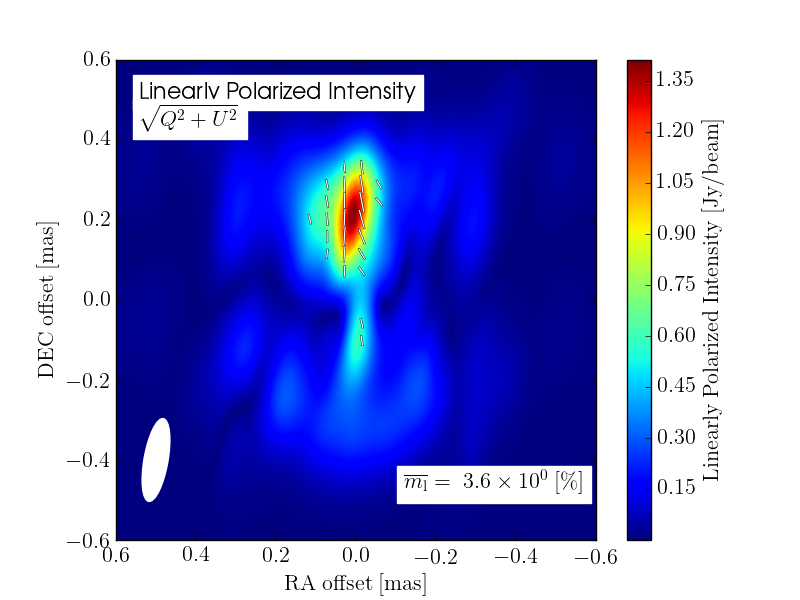}};
	
	\draw[semithick] (Fig9.south east) -- 
	node[anchor=north,rotate=90] {\emph{(iii)} Toroidal} (Fig9.north east) --
	(Fig6.south east) --
	node[anchor=north,rotate=90] {\emph{(ii)} Helical} (Fig6.north east) --
	(Fig3.south east) --
	node[anchor=north,rotate=90] {\emph{(i)} Poloidal} (Fig3.north east);
	\draw[semithick] ([yshift=-6pt]Fig7.south west) -- 
	node[below, xshift=-6pt] {(1) $n_e(\gamma) \propto \rho$} 
	([yshift=-6pt]Fig7.south east) --
	([yshift=-6pt]Fig8.south west) -- 
	node[below, xshift=-6pt] {(2) $n_e(\gamma) \propto p$} 
	([yshift=-6pt]Fig8.south east) -- 
	([yshift=-6pt]Fig9.south west) -- 
	node[below, xshift=-6pt] {(3) $n_e(\gamma) \propto B^2$} 
	([yshift=-6pt]Fig9.south east);
	
	\end{tikzpicture}
	
	\caption[Synthetic emission maps: Total intensity for three 
	different magnetic fields and emission recipes]{Ray-tracing images of our jets in linearly polarized intensity when each jet is viewed edge-on. The images highlight similar epochs in the jet's evolution of each 3D RMHD simulation at $86\,$GHz. From left to right: 
		proportionality of the NTEs to the 
		fluid's 
		density, internal energy density, and magnetic energy density. From top 
		to bottom: 
		purely 
		poloidal magnetic field, helical magnetic field, and purely toroidal 
		magnetic field. The ray-traced images are convolved with a Gaussian beam indicative of the resolution of the GMVA and with a Gaussian noise floor of $10^{-4}\,\text{Jy/beam}$.  }
	\label{fig:43c}
\end{figure*}

\begin{figure*}
	\centering
	\begin{tikzpicture}
	\small
	\node[anchor=south west,inner sep=0] (Fig1) at (0,0) 
	{\includegraphics[width=0.31\textwidth]{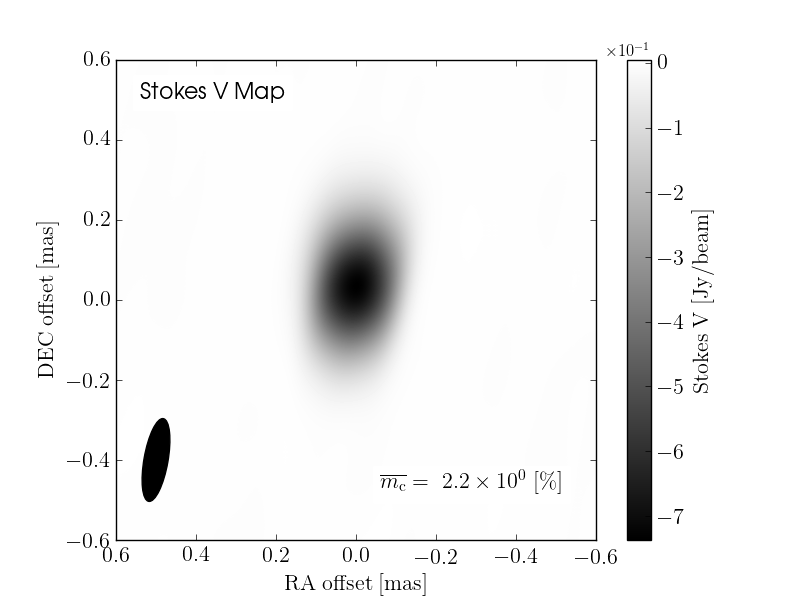}};
	\node[anchor=south west, inner sep=0] (Fig2) at (Fig1.south east) 
	{\includegraphics[width=0.31\textwidth]{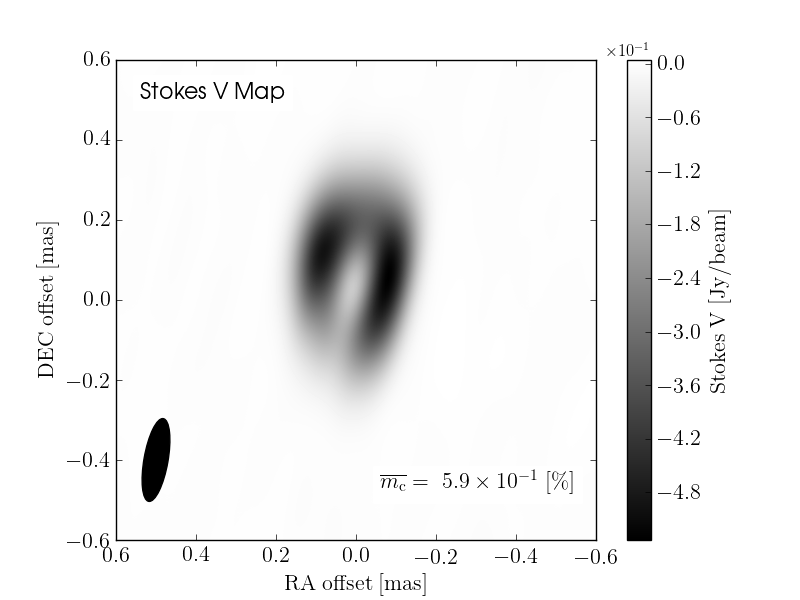}};
	\node[anchor=south west, inner sep=0] (Fig3) at (Fig2.south east) 
	{\includegraphics[width=0.31\textwidth]{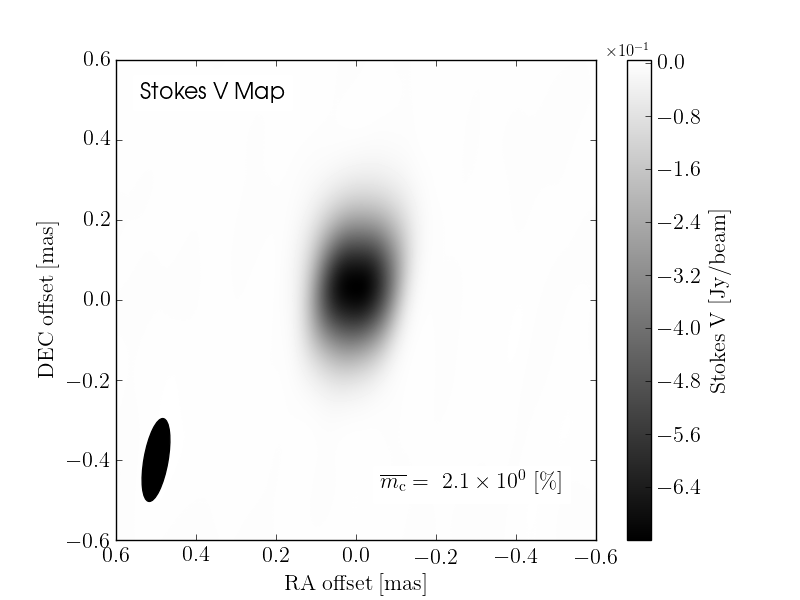}};
	
	\node[anchor=north west, inner sep=0] (Fig4) at (Fig1.south west) 
	{\includegraphics[width=0.31\textwidth]{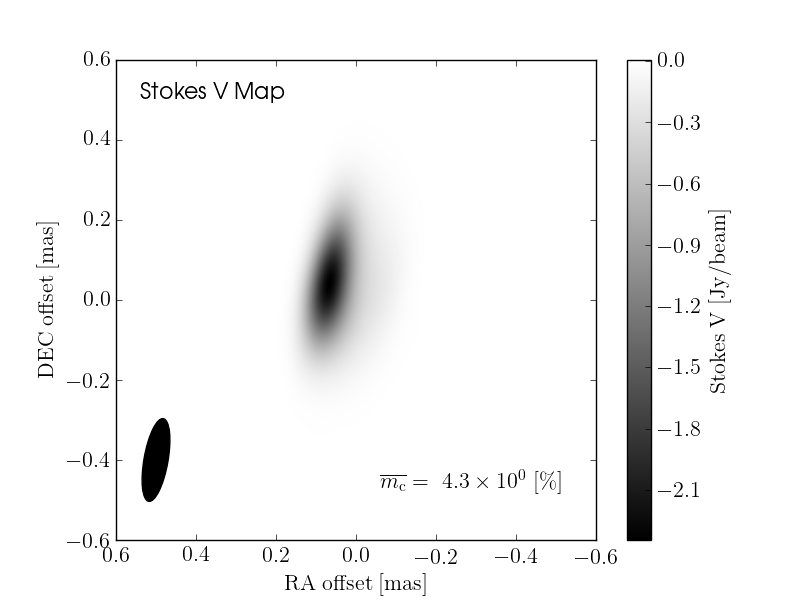}};
	\node[anchor=south west, inner sep=0] (Fig5) at (Fig4.south east) 
	{\includegraphics[width=0.31\textwidth]{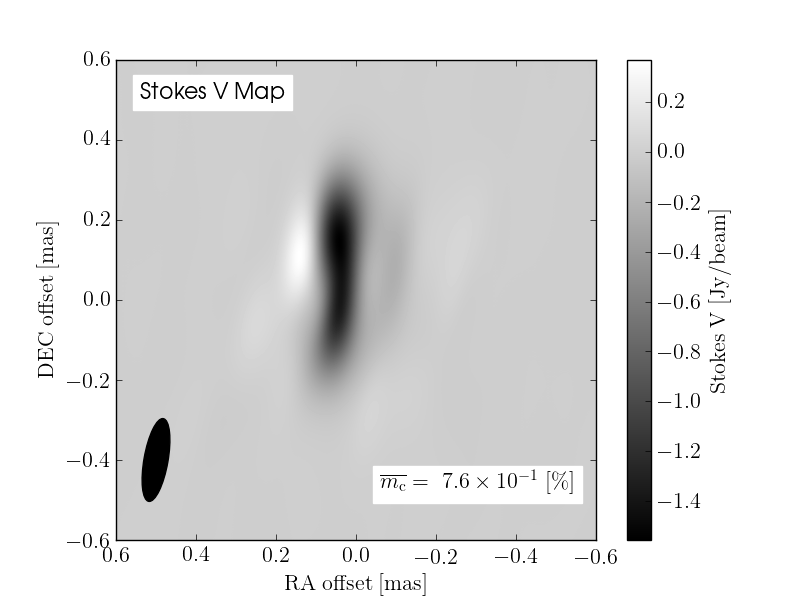}};
	\node[anchor=south west, inner sep=0] (Fig6) at (Fig5.south east) 
	{\includegraphics[width=0.31\textwidth]{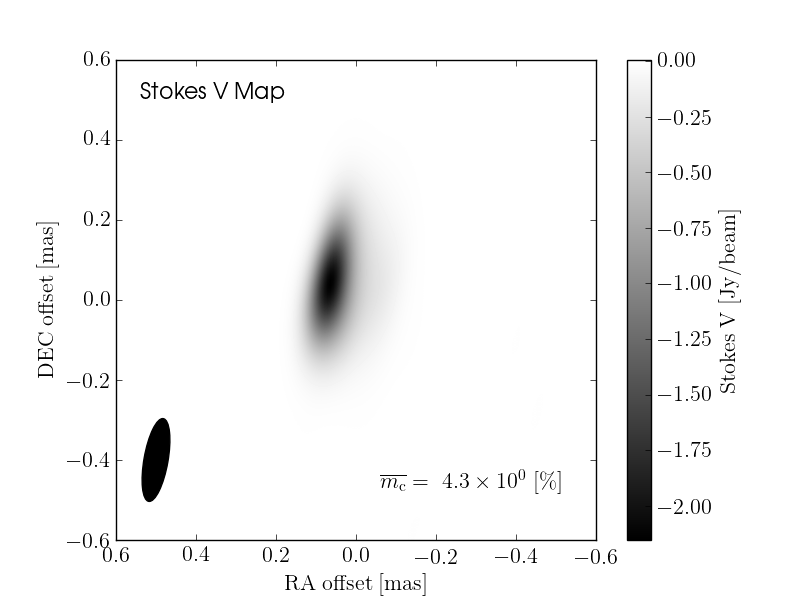}};
	
	\node[anchor=north west, inner sep=0] (Fig7) at (Fig4.south west) 
	{\includegraphics[width=0.31\textwidth]{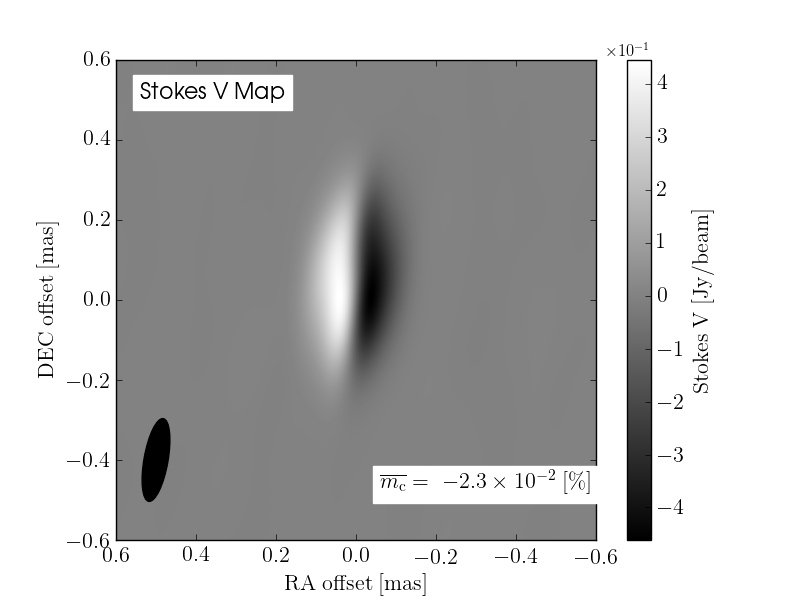}};
	\node[anchor=south west, inner sep=0] (Fig8) at (Fig7.south east) 
	{\includegraphics[width=0.31\textwidth]{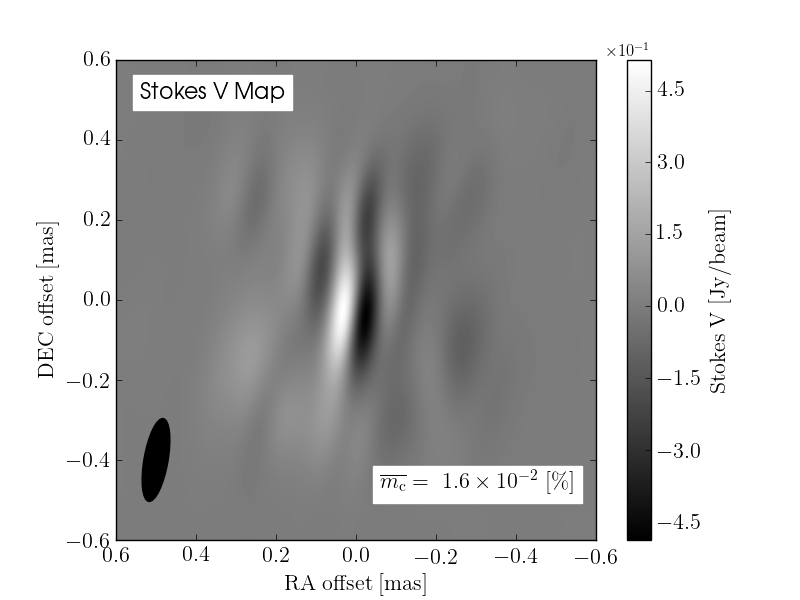}};
	\node[anchor=south west, inner sep=0] (Fig9) at (Fig8.south east) 
	{\includegraphics[width=0.31\textwidth]{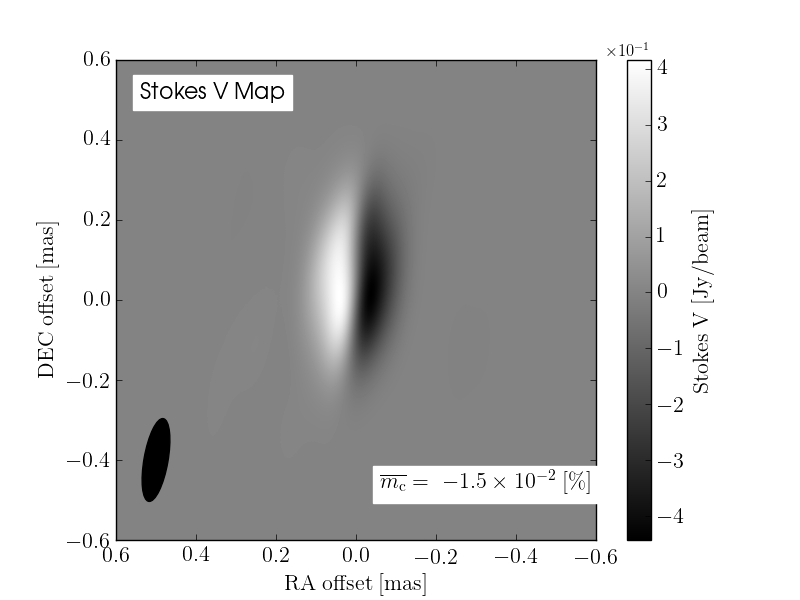}};
	
	\draw[semithick] (Fig9.south east) -- 
	node[anchor=north,rotate=90] {\emph{(iii)} Toroidal} (Fig9.north east) --
	(Fig6.south east) --
	node[anchor=north,rotate=90] {\emph{(ii)} Helical} (Fig6.north east) --
	(Fig3.south east) --
	node[anchor=north,rotate=90] {\emph{(i)} Poloidal} (Fig3.north east);
	\draw[semithick] ([yshift=-6pt]Fig7.south west) -- 
	node[below, xshift=-6pt] {(1) $n_e(\gamma) \propto \rho$} 
	([yshift=-6pt]Fig7.south east) --
	([yshift=-6pt]Fig8.south west) -- 
	node[below, xshift=-6pt] {(2) $n_e(\gamma) \propto p$} 
	([yshift=-6pt]Fig8.south east) -- 
	([yshift=-6pt]Fig9.south west) -- 
	node[below, xshift=-6pt] {(3) $n_e(\gamma) \propto B^2$} 
	([yshift=-6pt]Fig9.south east);
	
	\end{tikzpicture}
	
	\caption[Synthetic emission maps: Total intensity for three 
	different magnetic fields and emission recipes]{Ray-tracing images of our jets in circular polarization when each jet is viewed edge-on. The images highlight similar epochs in the jet's evolution of each 3D RMHD simulation at $86\,$GHz. From left to right: 
		proportionality of the NTEs to the 
		fluid's 
		density, internal energy density, and magnetic energy density. From top 
		to bottom: 
		purely 
		poloidal magnetic field, helical magnetic field, and purely toroidal 
		magnetic field. The ray-traced images are convolved with a Gaussian beam indicative of the resolution of the GMVA and with a Gaussian noise floor of $10^{-4}\,\text{Jy/beam}$.  }
	\label{fig:43b}
\end{figure*}

Figures~\ref{fig:43a} through \ref{fig:43b} show the same sequence of images for each scaling relations except that the jets are now viewed at an inclination of $i=5^\circ$ to the jet axis. In particular, the proportionality to the thermal pressure, visible in the middle column of \fig{fig:43c}, highlights both the recollimation shock in the central region of the jet and \label{key}the pressurized regions of the ambient plasma due to interactions with the surrounding medium.

The integrated levels of fractional linear and fractional circular polarization (shown to the lower right in Figures~\ref{fig:41c},~\ref{fig:41b},~\ref{fig:43c}, and~\ref{fig:43b}) exhibit the same level and behavior as discussed in \sect{resfromm} and show no major dependency on the chosen electron scaling relation.

\section{Electron Emission Scaling Relation Variations}\label{variations}
There are two additional variations of the electron emission scaling relations presented in~\sect{4}. In the first variation we exclude the use of a jet tracer (which is a \emph{PLUTO} variable) and consider the impact the toroidal magnetic field morphology and various emission scaling relations have on the ray-traced image when including the surrounding ambient medium. In a second variation we keep the jet tracer but replace the assumption of a constant lower cutoff in the electron power-law spectrum and instead compute $\gamma_\text{min}$ as a function related to the thermal pressure and density.
\subsection{No Jet Tracer}
In contrast to~\sect{sec:resporth}, we no longer make use of a jet tracer to exclude the surrounding ambient medium in our ray-tracing images. \Fig{fig:45a} highlights the resultant emission for the toroidal magnetic field case for the three electron scaling relations with the additional component of the ambient medium. For simplicity we choose to show the total intensity where we see the most striking difference to the results presented in~\sect{sec:resporth}.
As before, the left column shows images produced using the density scaling relation, the middle column show images produced using the pressure scaling relation, and the right column shows images produced using the magnetic energy density scaling relation. In contrast to~\sect{sec:resporth}, we clearly see an impact of the various emission scaling relations on the jet emission when assuming a toroidal magnetic field morphology and including the ambient medium. The ambient medium is most visible for the density scaling relation and the jet structure itself is largely hidden behind the intervening ambient plasma (which is not radiatively cooled in our simulations). Clearly, the second two recipes (which are proportional to thermal pressure and magnetic energy density) are better at highlighting the jet through the intervening ambient plasma. A solution to the arbitrary use of a jet tracer, moving forward, will be to properly include the effects of synchrotron cooling and diffusive shock acceleration - this is planned for a future paper.
\subsection{Lower Cutoff for Injected NTEs}
\begin{figure*}
	\centering
	\begin{tikzpicture}
	\small
	\node[anchor=south west,inner sep=0] (Fig1) at (0,0) 
	{\includegraphics[width=0.31\textwidth]{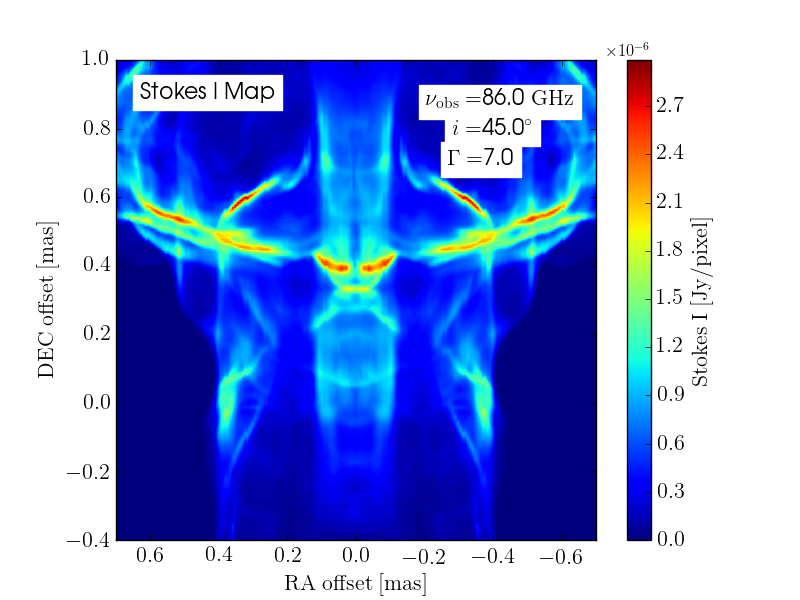}};
	\node[anchor=south west, inner sep=0] (Fig2) at (Fig1.south east) 
	{\includegraphics[width=0.31\textwidth]{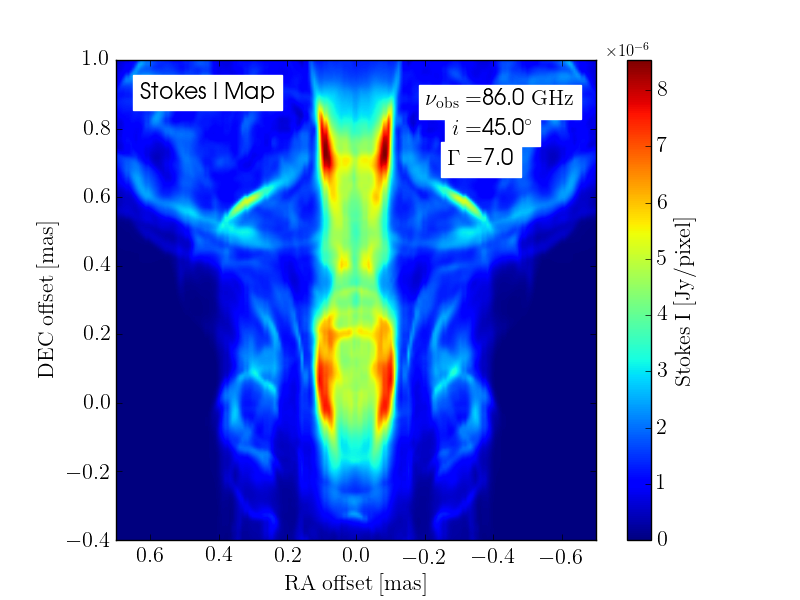}};
	\node[anchor=south west, inner sep=0] (Fig3) at (Fig2.south east) 
	{\includegraphics[width=0.31\textwidth]{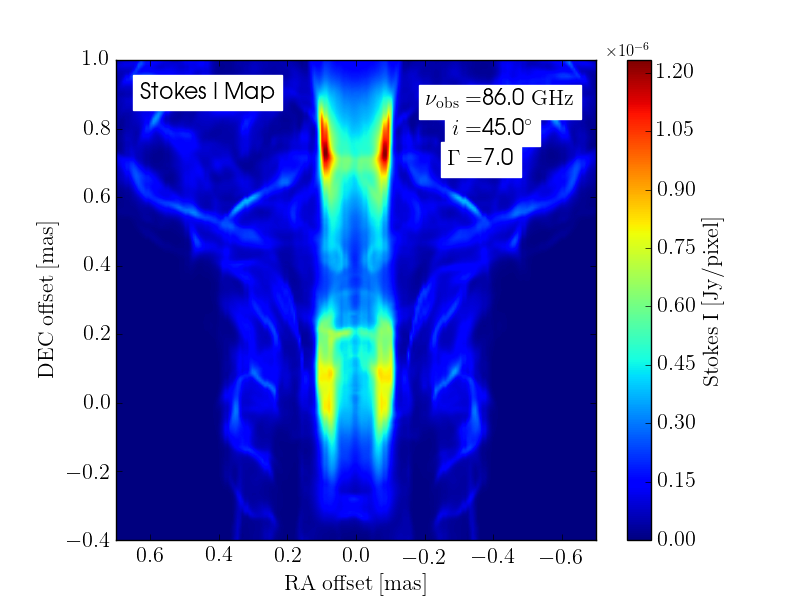}};
	
\draw[semithick] (Fig3.south east) -- 
node[anchor=north,rotate=90] {Toroidal} (Fig3.north east);
\draw[semithick] ([yshift=-6pt]Fig1.south west) -- 
node[below, xshift=-6pt] {(1) $n_e(\gamma) \propto \rho$} 
([yshift=-6pt]Fig1.south east) --
([yshift=-6pt]Fig2.south west) -- 
node[below, xshift=-6pt] {(2) $n_e(\gamma) \propto p$} 
([yshift=-6pt]Fig2.south east) -- 
([yshift=-6pt]Fig3.south west) -- 
node[below, xshift=-6pt] {(3) $n_e(\gamma) \propto B^2$} 
([yshift=-6pt]Fig3.south east);
	
	\end{tikzpicture}
	
	\caption[Synthetic emission maps: Total intensity for three 
	different magnetic fields and emission recipes]{Total intensity maps 
		of our 3D RMHD jet simulations using 
		different electron scaling relations and a purely toroidal magnetic field morphology. The images highlight similar epochs in the jet's evolution of each 3D RMHD simulation at $86\,$GHz. The images, in contrast to \fig{fig:41a}, are generated without the use of a jet tracer and illustrate both the emission of the ambient medium and the hidden jet structure. The jet is viewed at $i=45^\circ$ and 
		propagates from top to bottom in each frame. From left to right: 
		proportionality of the NTEs to the 
		fluid's 
		density, internal energy density, and magnetic energy density.}
	\label{fig:45a}
\end{figure*}

\begin{figure*}
	\centering
	\begin{tikzpicture}
	\small
	\node[anchor=south west,inner sep=0] (Fig1) at (0,0) 
	{\includegraphics[width=0.31\textwidth]{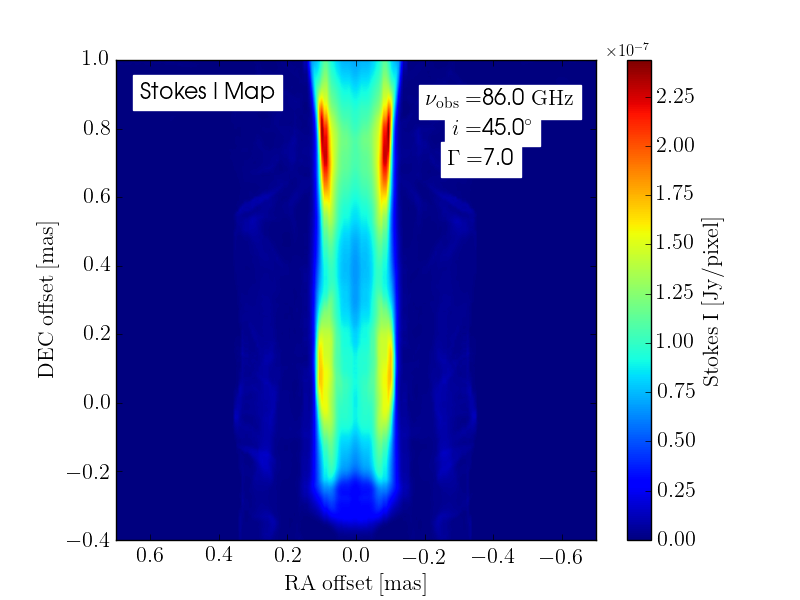}};
	\node[anchor=south west, inner sep=0] (Fig2) at (Fig1.south east) 
	{\includegraphics[width=0.31\textwidth]{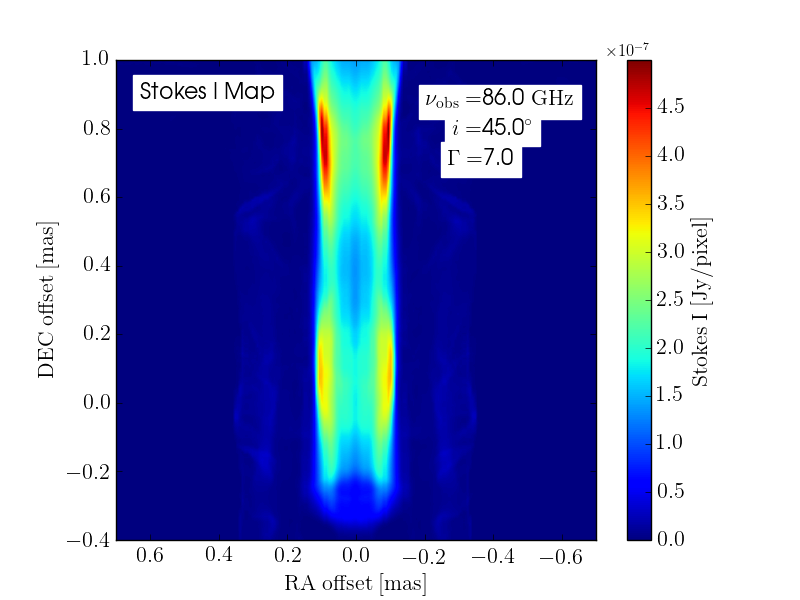}};
	\node[anchor=south west, inner sep=0] (Fig3) at (Fig2.south east) 
	{\includegraphics[width=0.31\textwidth]{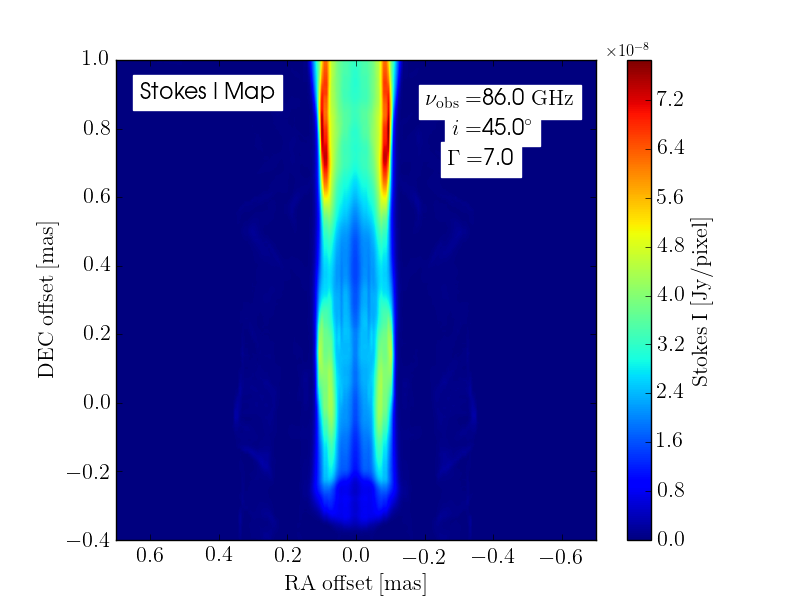}};
	
	\draw[semithick] (Fig3.south east) -- 
	node[anchor=north,rotate=90] {Toroidal} (Fig3.north east);
	\draw[semithick] ([yshift=-6pt]Fig1.south west) -- 
	node[below, xshift=-6pt] {(1) $n_e(\gamma) \propto \rho$} 
	([yshift=-6pt]Fig1.south east) --
	([yshift=-6pt]Fig2.south west) -- 
	node[below, xshift=-6pt] {(2) $n_e(\gamma) \propto p$} 
	([yshift=-6pt]Fig2.south east) -- 
	([yshift=-6pt]Fig3.south west) -- 
	node[below, xshift=-6pt] {(3) $n_e(\gamma) \propto B^2$} 
	([yshift=-6pt]Fig3.south east);
	
	\end{tikzpicture}
	
	\caption[Synthetic emission maps: Total intensity for three 
	different magnetic fields and emission recipes]{Total intensity maps 
		of our 3D RMHD jet simulations using 
		different electron scaling relations and a purely toroidal magnetic field morphology. The images highlight similar epochs in the jet's evolution of each 3D RMHD simulation at $86\,$GHz. The images are created with a jet tracer which excludes the obscuring ambient medium.
		The images are comparable to the lower panels of \fig{fig:41a} in which, in contrast to computing $\gamma_\text{min}$  from the ratio of pressure to density, it is fixed to 10.
		The jet is viewed at $i=45^\circ$ and 
		propagates from top to bottom in each frame. From left to right: 
		proportionality of the NTEs to the 
		fluid's 
		density, internal energy density, and magnetic energy density. The jet structure does not show considerable differences although the jet is most edge-brightened for the most right proportionality.  }
	\label{fig:45b}
\end{figure*}

In the second variation, we keep the jet tracer but instead compute the constant lower cutoff for injected NTEs ($\gamma_\text{min}=10$ for~\sect{4}) with a prescription based on the ratio between thermal pressure and density (similar to~\cite{Porth2011}). In particular, we compute (cell-to-cell):
\begin{align}\label{gammin}
	\gamma_\text{min} = 1 + \frac{p}{2\rho c^2}\frac{m_p}{m_c}.
\end{align}
\Fig{fig:45b} shows the resulting jet emission in total intensity. The arrangement is the same as in~\fig{fig:45a}. 
The proportionality to: \emph{(i)} the density, \emph{(ii)} pressure, and \emph{(iii)} magnetic energy density does not show any major differences when compared with the lower panels of~\fig{fig:41a} (in which we set $\gamma_\text{min}=10$ ). The jet remains edge-brightened for the toroidal magnetic field as seen in~\sect{sec:resporth}.

\section{Discussion}\label{discuss}
A wealth of new polarimetric data has been amassed of relativistic jets over the last decade, i.e., maps of AGN in both linearly polarized intensity and circular polarization, maps of Faraday rotation measure (RM) gradients along the jet, and analysis of EVPA orientation along the jet, all of which help to probe the underlying magnetic field geometry. 
In this section we make further comparisons between our ray-traced emission maps and observations to  to further understand and interpret the different polarimetric features we observe in our simulations.

Faraday rotation measure (RM) gradients observed transverse to the jet axis hint at the existence of helical magnetic fields~\citep[e.g., in 0133+479, see][]{Gabuzda2018, Gabuzda2019}. In a future work we plan to experiment with generating synthetic RM maps of our three jet simulations to explore how robust RM is as a metric of the jet's internal magnetic field structure.

Helical/toroidal magnetic field morphologies have been invoked to explain an increase in linear polarization towards the edges of the jet~\citep{Pushkarev2005, Lyutikov2005}. This edge brightened emission morphology is clearly seen in our simulations. 

As discussed in~\cite{Gabuzda2020}, a purely toroidal magnetic field should result in symmetric edge brightened emission across the jet whereas, in contrast, a helical magnetic field should result in asymmetric emission along the jet edges. These distinct emission morphologies are present within our simulations (see, e.g., \fig{38a}).

In our poloidal field simulation the jet spine/shock dominates the emission. This emission morphology has been observed in the jet of NGC$\,$1052~\citep[][]{Baczko}. In contrast, in the helical and toroidal field simulations the outer sheath is edge brightened and dominates the emission. This emission morphology has been observed in the jet of 3C$\,$84~\citep[see, e.g.,][]{Giovannini2018, Kim3C, Paraschos}.

Bi-modal EVPA patterns have been observed in a number of jets~\citep[][]{OSullivan2009}  in which the EVPAs preferentially align with the jet axis in the spine and, in contrast, appear predominantly perpendicular to the jet axis in the sheath. We see this bi-modal pattern in our toroidal field simulation (see, e.g., lower middle panel of \fig{38a}).

The majority of blazars in which CP is detected tend to exhibit one sign/handedness of circular polarization~\citep[see, e.g.,][]{Homan2005, Homan2004}. A small number of sources, however, exhibit both negative and positive CP in the radio core region~\citep[see, e.g.,][]{Vasili2008}. As illustrated in the right column of~\fig{38b}, we find that a poloidal field produces only one sign of CP in the radio core whereas, in contrast, the the toroidal field produces both signs of CP in the core. This highlights the potential of combining linear and circular polarization maps to make a more robust determination of the magnetic field orientation within the jet.

\section{Conclusion}\label{con}

We have carried out a systematic survey of full Stokes radiative transfer calculations exploring how: \emph{(i)} the jet's magnetic field morphology and \emph{(ii)} how various electron scaling relations effect the resultant linear and circular polarized emission.\\
In particular, we find:
\begin{itemize}
	\item[$\bullet$] Resolved circular polarization imaging has the potential to discriminate between a purely poloidal or a purely toroidal magnetic field morphology within the jet. \\
	\item[$\bullet$] When the jet is resolved (i.e., \fig{38a}), toroidal magnetic fields result in edge-brightened jets whereas poloidal magnetic fields seem to highlight the jet spine/recollimation shock.\\
	\item[$\bullet$] The integrated levels of fractional linear $\bar{m}_l$ 
	and 
	circular polarization $\bar{m}_c$ are only midly sensitive to the choice of electron scaling relation. The integrated fractional circular polarization of the toroidal jet, however, is found to be several orders of magnitude smaller than the poloidal and helical jets. \\
	\item[$\bullet$] Scaling the electron number density to the thermal (fluid) density $\rho$, pressure $p$, or magnetic energy density $B^2$ while fixing the bounds of the electron power-law distribution ($\gamma_\text{min}\, \&\, \gamma_\text{max}$) does not seem to have an appreciable effect on the morphology of the linear and circular polarized emission within the shock when the ambient medium is excluded from the ray-tracing. However, when the ambient medium is included, the magnetic energy density recipe best highlights the jet emission through the intervening plasma.
\end{itemize}

In an effort to further compare simulations to observations, our numerical RMHD jet models have formed the basis of an accepted VLBA proposal to conduct deep full Stokes imaging of a number of blazar jets. This will allow us to make further comparisons between our numerical models and CP observations in an attempt to better understand the nature of the intrinsic magnetic field morphologies of relativistic jets.

In future, we plan to incorporate the effects of synchrotron cooling and diffusive shock acceleration in our ray-tracing calculations~\citep[see, e.g.,][]{Vaidya2018}. 

\vspace{20pt}
\begin{acknowledgements}
	This research was supported through a PhD grant from the International Max Planck Research School (IMPRS) for Astronomy and Astrophysics at the Universities of Bonn and Cologne. 
	The three-dimensional jet simulations presented in this paper are computed with the \emph{PLUTO} Code. The ray-tracing software \emph{RADMC-3D} produced the polarized images of the synchrotron emission.
	The authors are grateful to E. Ros for feedback regarding VLBI, M. Perucho for helpful discussions on the physics of RMHD jet simulations, and to the referee, Philip Hughes, for a thorough review of this manuscript.
	
\end{acknowledgements}

\bibliographystyle{bibtex/aa}
\bibliography{bib/thesis_refs}


\begin{appendix}
	\appendix
	\section{Persistence of the Injected Magnetic Field Morphology}\label{appendix}
	\begin{figure}
		\centering
		\begin{tikzpicture}
		\small
		\node[anchor=south west,inner sep=0] (Fig1) at (5,5) 
		{\includegraphics[width=0.25\textwidth]{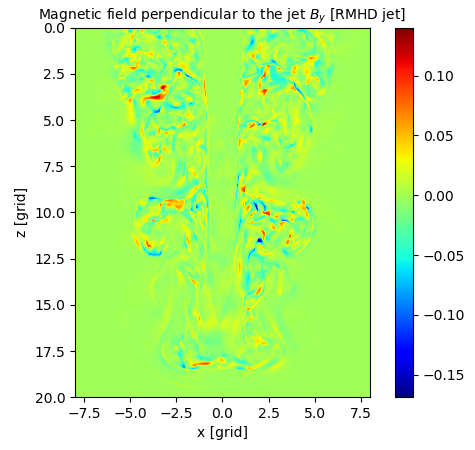}};
		\node[anchor=south west, inner sep=0] (Fig2) at (Fig1.south east) 
		{\includegraphics[width=0.25\textwidth]{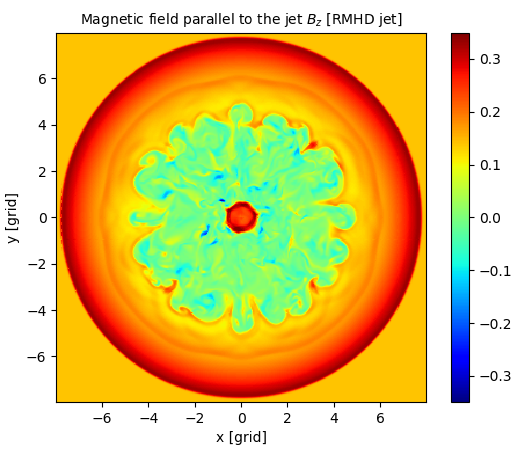}};
		
		\node[anchor=north west, inner sep=0] (Fig4) at (Fig1.south west) 
		{\includegraphics[width=0.25\textwidth]{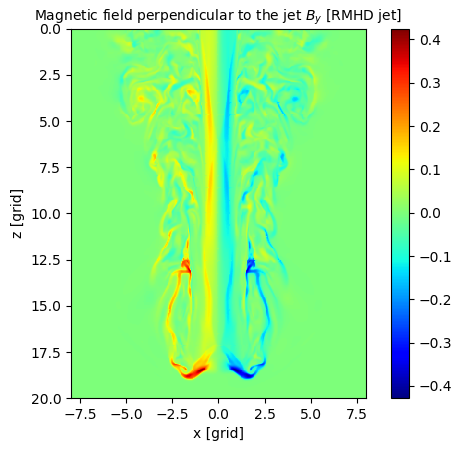}};
		\node[anchor=south west, inner sep=0] (Fig5) at (Fig4.south east) 
		{\includegraphics[width=0.25\textwidth]{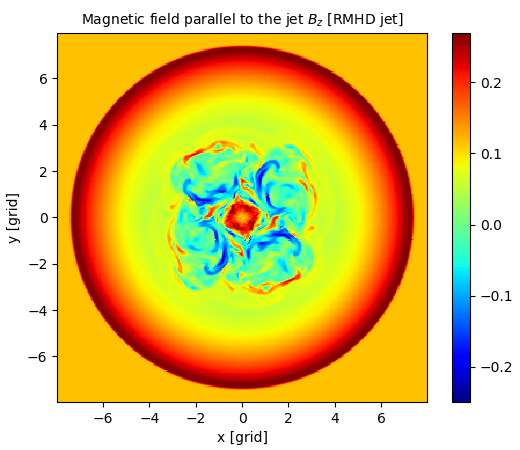}};
		
		\node[anchor=north west, inner sep=0] (Fig7) at (Fig4.south west) 
		{\includegraphics[width=0.25\textwidth]{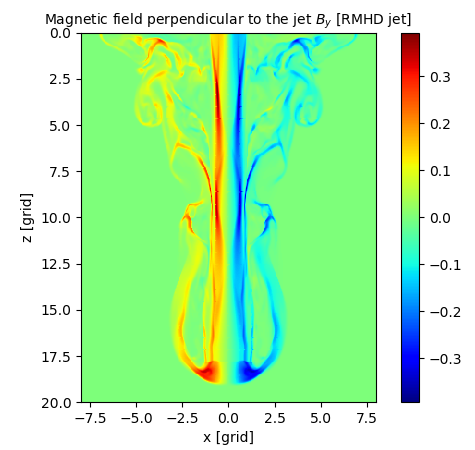}};
		\node[anchor=south west, inner sep=0] (Fig8) at (Fig7.south east) 
		{\includegraphics[width=0.25\textwidth]{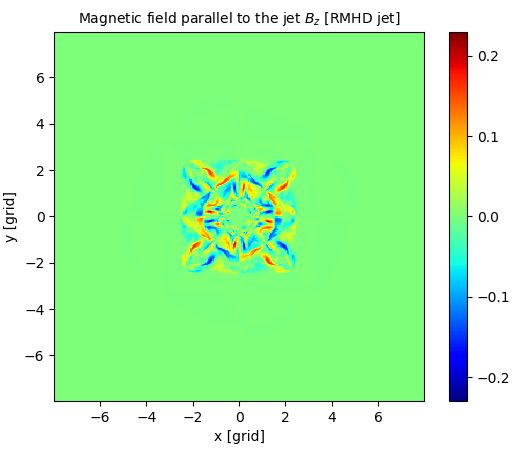}};
		
		\draw[semithick] (Fig8.south east) -- 
		node[anchor=north,rotate=90] {\emph{(iii)} Toroidal} (Fig8.north east) --
		(Fig5.south east) --
		node[anchor=north,rotate=90] {\emph{(ii)} Helical} (Fig5.north east) --
		(Fig2.south east) --
		node[anchor=north,rotate=90] {\emph{(i)} Poloidal} (Fig2.north east);
		\draw[semithick] ([yshift=-6pt]Fig7.south west) -- 
		node[below, xshift=-6pt] {$B_y$} 
		([yshift=-6pt]Fig7.south east) --
		([yshift=-6pt]Fig8.south west) -- 
		node[below, xshift=-6pt] {$B_z$} 
		([yshift=-6pt]Fig8.south east);
		
		\draw[dashed, semithick, red] (5.72,2.930) -- 
		node[anchor=north, rotate=90] {} (8.56,2.930);
		\draw[dashed, semithick, red] (5.75,7.31) -- 
		node[anchor=north, rotate=90] {} (8.59,7.31);
		\draw[dashed, semithick, red] (5.78,-1.485) -- 
		node[anchor=north, rotate=90] {} (8.62,-1.485);
		\end{tikzpicture}
		
		\captionof{figure}{These images illustrate 2D slices through our 3D RMHD jets and are color coded according to the magnetic field component perpendicular to the jet axis - $B_y$ (left column), and the field component parallel to the jet axis - $B_z$ (right column) through the jet's mid-plane (see dashed line in the left column). From top to bottom: poloidal, helical, and toroidal magnetic field simulations.}
		\label{Appendix}
	\end{figure}
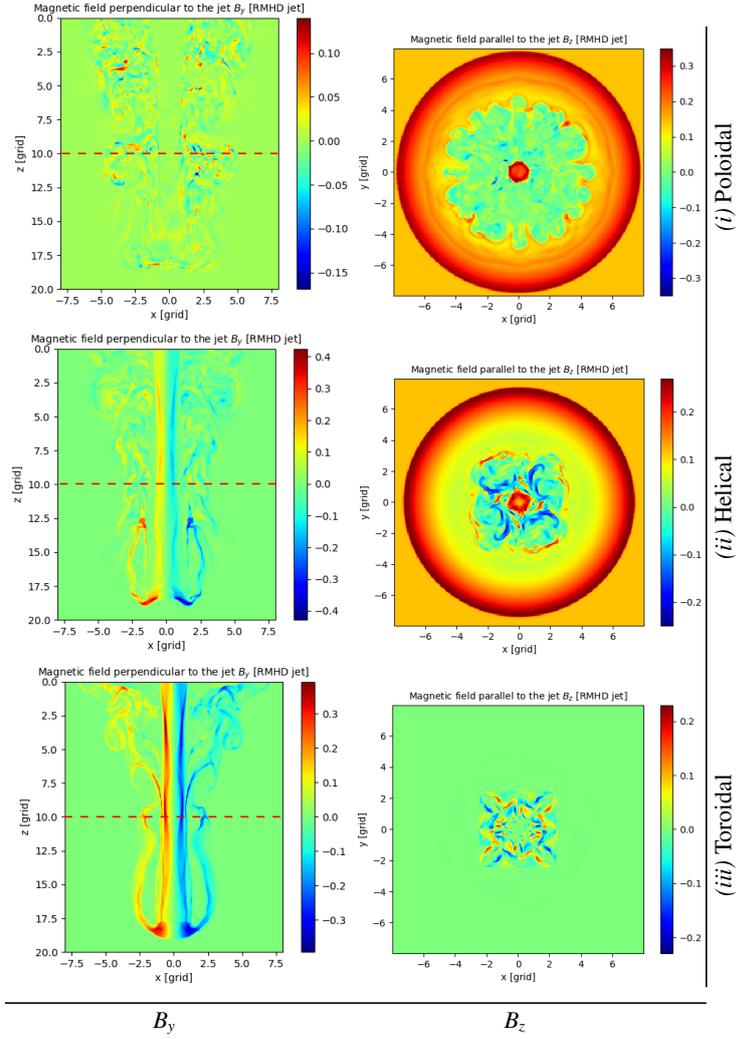
	
	\noindent
	The underlying magnetic field included in the RMHD jet simulations exhibits distinct characteristics for each simulation (i.e., poloidal, helical, toroidal).
	We implement a poloidal component ($B_z$, see \eq{37}) and toroidal components ($B_x$ and $B_y$, see \eq{36}). The magnetization parameters ($\sigma_z$ and $\sigma_\phi$ - see \sect{mag}) set the overall morphology of the field (i.e., by varying the ratio between the poloidal and toroidal components).
	
	\Fig{Appendix} illustrates how the magnetic field morphologies injected at the jet orifice persist down the jet. The left column shows 2D slices through the mid-plane of each simulation with the color scheme highlighting the component of the magnetic field ($B_y$) which is perpendicular to the jet axis. In contrast, the right column shows 2D slices through the jet's mid-plane (see dashed line in left column), respectively, with the color scheme highlighting the component of the magnetic field ($B_z$) which is parallel to the jet direction. The upper, middle, and lower rows correspond to the poloidal, helical, and toroidal field simulations, respectively. 
	
	The slices presented in Fig. A1 are made through the same simulation epochs used in our ray-tracing calculations above. Despite the existence of turbulent eddies and jet shear along each jet axis, the injected magnetic field morphologies largely persist down the jet (be it poloidal, helical, or toroidal).

\end{appendix}



\end{document}